\documentclass[bibyear]{aa}
\usepackage[varg]{txfonts}
\usepackage{graphicx}
\usepackage{longtable}
\usepackage{lscape}
\newcommand{\gam}{$\gamma$~Vel}
\newcommand{\e}{et al.\ }
\newcommand{\ha}{H$\alpha$}
\newcommand{\teff}{T$_{eff}$}
\newcommand{\logg}{$\log g$}
\newcommand{\fe}{$[Fe/H]$}
\begin{document}

\title{Gaia-ESO Survey:
Empirical classification of VLT/Giraffe stellar spectra in the wavelength range 
6440-6810 \AA\ in the $\gamma$~Vel cluster,
and calibration of spectral indices.
\thanks{
Based on observations collected with the FLAMES spectrograph at VLT/UT2
telescope (Paranal Observatory, ESO, Chile), for the Gaia-ESO Large
Public Survey (program 188.B-3002).
Table~\ref{table2} is only available
in electronic form at the CDS via anonymous ftp to cdsarc.u-strasbg.fr
(130.79.128.5) or via http://cdsweb.u-strasbg.fr/cgi-bin/qcat?J/A$+$A/}
}

\date{Received date / Accepted date}

\author{F. Damiani\inst{1}, L. Prisinzano\inst{1}, G. Micela\inst{1},\\
S. Randich\inst{2}, G. Gilmore\inst{3},
J.~E. Drew\inst{4}, R.~D. Jeffries\inst{5},
Y. Fr\'emat\inst{6},
E.~J. Alfaro\inst{7}, T. Bensby\inst{8}, A. Bragaglia\inst{9},
E. Flaccomio\inst{1},
A.~C. Lanzafame\inst{10},
E. Pancino\inst{9,11}, A. Recio-Blanco\inst{12}, G.~G. Sacco\inst{2},
R. Smiljanic\inst{13,14},
R.~J. Jackson\inst{5}, P. de Laverny\inst{12}, L. Morbidelli\inst{2},
C.~C. Worley\inst{3}, A. Hourihane\inst{3}, M.~T. Costado\inst{7},
P. Jofr\'e\inst{3}, K. Lind\inst{3}, E. Maiorca\inst{2}
}
\institute{INAF - Osservatorio Astronomico di Palermo G.S.Vaiana,
Piazza del Parlamento 1, I-90134 Palermo,
Italy. \email{damiani@astropa.inaf.it}
\and INAF - Osservatorio Astrofisico di Arcetri, Largo E. Fermi 5,
50125, Florence, Italy
\and Institute of Astronomy, University of Cambridge, Madingley
Road, Cambridge CB3 0HA, United Kingdom
\and Centre for Astrophysics Research, STRI, University of
Hertfordshire, College Lane Campus, Hatfield AL10 9AB, United Kingdom
\and Astrophysics Group, Research Institute for the Environment,
Physical Sciences and Applied Mathematics, Keele University, Keele,
Staffordshire ST5 5BG, United Kingdom
\and Royal Observatory of Belgium, 3 avenue circulaire, 1180
Brussels, Belgium
\and Instituto de Astrof\'{i}sica de Andaluc\'{i}a, Camino Bajo de
Hu\'etor 50, 18008, Granada, Spain
\and Lund Observatory, Department of Astronomy and Theoretical
Physics, Box 43, SE-221 00 Lund, Sweden
\and INAF - Osservatorio Astronomico di Bologna, via Ranzani 1,
40127, Bologna, Italy
\and Dipartimento di Fisica e Astronomia, Sezione Astrofisica,
Universit\`{a} di Catania, via S. Sofia 78, 95123, Catania, Italy
\and ASI Science Data Center, Via del Politecnico SNC, 00133 Roma, Italy
\and Laboratoire Lagrange (UMR7293), Universit\'e de Nice Sophia
Antipolis, CNRS, Observatoire de la C\^ote d'Azur, BP 4229,F-06304 Nice
cedex 4, France
\and European Southern Observatory, Karl-Schwarzschild-Str. 2, 85748
Garching bei M\"unchen, Germany
\and Department for Astrophysics, Nicolaus Copernicus Astronomical Center,
ul. Rabia\'nska 8, 87-100 Toru\'n, Poland
}

\abstract{
We present a study of spectral diagnostics available from optical
spectra with R=17000 obtained with the VLT/Giraffe HR15n setup, using
observations from the Gaia-ESO Survey, on the \object{$\gamma$~Vel} young
cluster, with the purpose of classifying these stars and finding their
fundamental parameters.
We define several spectroscopic indices, sampling the amplitude of TiO
bands, the \ha\ line core and wings, and temperature- and
gravity-sensitive sets of lines, each useful as a \teff\ or \logg\
indicator over a limited range of stellar spectral types. \ha\ line
indices are also useful as
chromospheric activity or accretion indicators. Furthermore, we use
all indices to define additional global \teff- and \logg-sensitive
indices $\tau$ and $\gamma$, valid for the entire range of types in the
observed sample. We find a clear difference between
gravity indices of main-sequence and pre-main-sequence stars,
as well as a much larger difference between these and giant stars.
The potentially great usefulness of the $(\gamma,\tau)$ diagram as a
distance-independent age measurement tool for young clusters is discussed.
We discuss the effect on the defined indices of classical T-Tauri star
veiling, which is however detected in only a few stars in the
present sample.
Then, we present tests and calibrations of these indices, on the basis of
both photometry and literature reference spectra, from the UVES
Paranal Observatory Project and the ELODIE 3.1 Library.  The known
properties of these stars, spanning a wide range of stellar parameters,
enable us to obtain a good understanding of the performances of our
new spectral indices. For non-peculiar stars with known temperature, gravity,
and metallicity, we are able to calibrate quantitatively our indices,
and derive stellar parameters for a wide range of stellar types.
To this aim, a new composite index is defined,
providing a good metallicity indicator.
The ability of our indices to select peculiar, or
otherwise rare classes of stars is also established.
For pre-main-sequence stars outside the parameter range of the ELODIE dataset,
index calibration relies on model isochrones.
We check our calibrations against current Gaia-ESO UVES results, plus a
number of Survey benchmark stars, and
also against Gaia-ESO observations of young clusters, which
contribute to establishing the good performance of our method across a wide
range of stellar parameters.
Our gravity determination for late-type PMS stars is found to be accurate
enough to let us obtain gravity-based age estimates for PMS clusters.
Finally, our gravity determinations support the existence
of an older pre-main-sequence population in the \gam\ sky region,
in agreement with evidence obtained from the lithium depletion pattern
of the same stars.
}
%{context}
%{aims}
%{methods}
%{results}
%{}


\titlerunning{Classification of Gaia-ESO Survey spectra of $\gamma$ Vel cluster}
\authorrunning{Damiani et al.}

\maketitle

\section{Introduction}
\label{intro}

The \gam\ cluster was discovered in X-rays by Pozzo \e (2000), in a ROSAT
observation of the Wolf-Rayet binary $\gamma^2$ Vel itself.
Its age and distance have been recently estimated as about 5-10~Myr and
356$\pm 11$~pc by Jeffries \e (2009), in a study based
on deep optical photometry and XMM-Newton X-ray observations.
Star formation appears to have ceased long ago, as there is no sign of
ongoing formation events (Hernandez \e 2008); the cluster
pre-main-sequence (PMS)
stars also show only rare evidence of discs (Hernandez \e 2008), and most of
them are therefore weak-line T Tauri stars (WTTS), with no accretion of
circumstellar material on the stellar surface.

The \gam\ cluster was observed in the framework of the Gaia-ESO
Survey (Gilmore \e 2012), consisting of VLT/FLAMES
observations of 80-100 open clusters (of all ages, except embedded ones),
and selected Milky Way field star samples, with 300 nights allocated over 5
years. Survey observations started on 31 December 2011, and \gam\ 
was one of the earliest observed clusters.
The FLAMES fiber positioner was used to feed both UVES and Giraffe
spectrographs, for the brighter and fainter cluster stars respectively.
We focus here on the Giraffe observations, all made using the HR15n
setup, with R$\sim$ 17000, in the band $\lambda\lambda6440-6815$ \AA.
This setup has the advantage of observing both the \ha\ line and the Li~I
resonance line at 6707.7~\AA, of great interest when studying young stars, and
PMS stars in particular. On the other hand, this has not been the band of
choice for deriving stellar fundamental parameters such as
effective temperature \teff, gravity \logg, or metal abundance $z$.
This wavelength range is barely mentioned in stellar classification
textbooks such as Jaschek and Jaschek (1990), or Gray and
Corbally (2009). Curve-of-growth methods, often based on the study
of a multitude of Fe~I/Fe~II lines, are also not applicable here, essentially
because of the paucity and weakness of Fe~II lines in this range.
Therefore, it is important to investigate whether stellar spectra
obtained with this instrumental setup do contain indicators useful to
derive these fundamental parameters, and with what degree of
uncertainty.
Since all stars later than F spectral type in both
young and old clusters in the Gaia-ESO Survey will be observed using the
HR15n Giraffe setup, the solution to this problem has a general interest
for the entire Survey. As we discuss below, the Gaia-ESO Survey observations of
the \gam\ cluster offer several advantages for this type of investigations.

We initially consider likely members of the \gam\ cluster, which
have a homogeneous near-solar metallicity, as
recently determined by Spina et al.\ (submitted) using Gaia-ESO UVES
data. This also agrees with existing studies of metallicity in other
nearby star-formation regions (Padgett 1996, Santos et al.\ 2008, D'Orazi et
al.\ 2009a, Biazzo et al.\ 2011).
Many stars in the stellar sample considered here are however not
cluster members, and are therefore of unknown metallicity.
On the basis of the \gam\ dataset spectra we are able to define temperature-
and gravity-sensitive indices, as well as indices probing
pre-main-sequence accretion and veiling.

Then, we consider the issue of calibrating quantitatively the
temperature, gravity, and metallicity dependence of the newly defined indices
on the basis of additional data. The indices are
optimized to take advantage of the HR15n setup resolution ($R \sim
17000$), and there are not many public sets of stellar spectra with high
enough spectral resolution, in the same wavelength range, useful to our
purposes. We have nevertheless been able to
calibrate indices on the basis of two sets of public
spectral libraries; these results were then checked for consistency against
(1) current Gaia-ESO Survey results obtained from UVES spectra of several
hundreds stars, and a few tens common UVES-Giraffe targets,
(2) about two dozen Gaia-ESO Survey benchmark stars with
accurately known properties, and (3) Gaia-ESO Survey HR15n datasets of
young clusters.

This paper is structured as follows: in Sect.~\ref{obs-sample} the
observed sample is described, while in Sect.~\ref{prepare} we
describe the preliminary data preparation; then in Sect.~\ref{all-diag} we study
several spectroscopic diagnostics available from these observations,
ranging from molecular bands, to \ha, to lesser-known lines in the HR15n
range, and we define sets of new spectral indices, especially suited as
\teff\ and \logg\ diagnostics.
Section~\ref{calibr} presents
an initial photometric calibration of our new \teff\ indicator.
In Section~\ref{refdata} we discuss the choice of the reference
spectra for general calibration, their merits and disadvantages.
In Section~\ref{results} we
describe the calibration of indices made from each of these spectral
sets, and the relevant consistency checks.
In Sect.~\ref{discuss} we discuss all results obtained, how they can
be applied to other observations, and also what can be possible future
developments of this work.
A summary is provided in Sect.~\ref{concl}.

\section{Observed sample composition in the \gam\ dataset}
\label{obs-sample}

\begin{figure}
\resizebox{\hsize}{!}{
\includegraphics[bb=20 10 465 475]{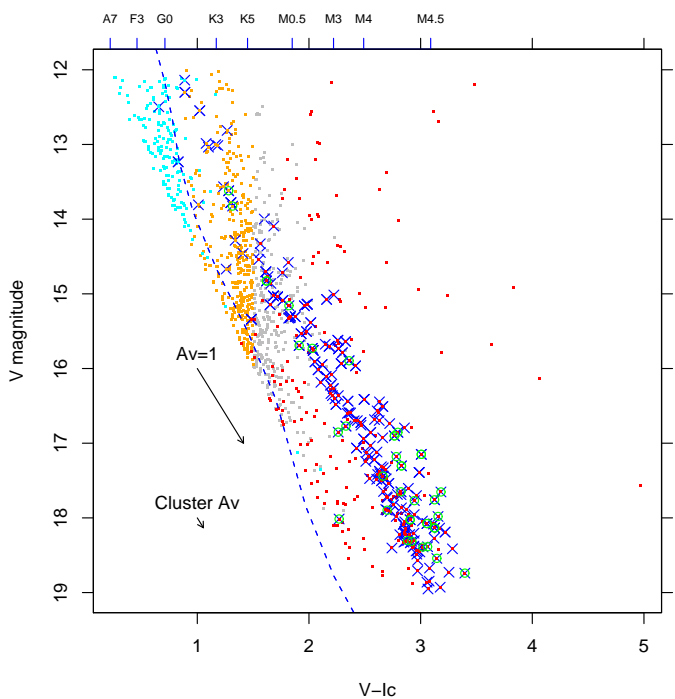}}
\caption{The $(V,V-I_c)$ color-magnitude diagram of stars
in the \gam\ cluster region observed with Giraffe and with S/N$>$15.
Here and in all following figures color-coded symbols are consistently
used: blue crosses are candidate cluster members, cyan dots are A to early-G
stars, red dots are late-K and M stars, and orange and gray dots are the
remaining stars; see Section~\ref{mol-bands} for details.
Stars with wide \ha\ {\rm emission} are indicated with green circles.
The unreddened ZAMS from Siess et al.\ (2000) at the cluster distance is
shown as a blue dashed line.
The large arrow indicate a reddening vector for $A_V=1$, while the small
arrow indicates the low estimated reddening for the cluster.
In the top axis, spectral types corresponding to unreddened main-sequence
stars are indicated.
\label{v-vi-2}}
\end{figure}

The VLT/FLAMES/Giraffe fiber-fed spectrograph allows
multi-object spectroscopy of up to $\sim$130 objects at a time.
Simultaneous to the observations described here, FLAMES/UVES observations
were also performed, but are not discussed in this work.
The target sample was chosen to be inclusive of all possible candidate
cluster members (Bragaglia et al., in preparation),
on the basis of the $(V,V-I_c)$ color-magnitude diagram
(CMD), using the photometric catalog from Jeffries \e (2009).  The CMD of
Figure~\ref{v-vi-2} shows all \gam\ stars observed with Giraffe in the
Survey with signal/noise ratio (S/N) $>15$ on individual spectra, down
to the Survey limiting magnitude $V=19$.
Throughout this paper we use consistent color/symbol codes for scatter plots:
red dots are stars exhibiting molecular bands, cyan dots
early-type stars with wide \ha\ absorption, orange and gray dots the
remaining (mostly GK) stars; see Sections~\ref{mol-bands} and
\ref{halpha} for details.
Blue crosses indicate bona-fide cluster members, selected by the
presence of {\rm at least two indicators} among: 
X-ray emission (from Jeffries \e 2009), lithium $EW>150$~m\AA, radial
velocity (RV) in the range 10-22~km/s
(the latter two membership indicators for \gam\ stars are studied
by Jeffries et al.\ 2014, and Prisinzano et al., in preparation).
To these membership criteria we have added the presence of {\rm wide} \ha\
emission (green circles, see Sect.~\ref{halpha}), considered sufficient
by itself (as distinctive of Classical
T Tauri Stars - CTTS). Each of the other membership indicators above,
taken alone, did instead not seem to guarantee a clean enough candidate
member list. This member selection cannot be considered either complete
(e.g., the X-ray data do not cover the whole cluster), nor entirely free of
contaminants (like slightly older lithium-rich G stars), but is found
to be accurate enough for the purposes of this work. A much more
detailed membership study of the cluster will be presented by Prisinzano
\e (in preparation).

The putative members (labeled 'M' in Table~\ref{table2})
define the cluster locus, which for types later than
about G0 lies distinctly above the ZAMS (blue dashed line) at the
cluster distance (356~pc, Jeffries \e 2009), indicating their PMS nature
corresponding to the cluster young age. ZAMS colors are taken from Siess
\e (2000), with color calibration from Kenyon and Hartmann (1995).
The absence of stars below a
cutoff line in the CMD is caused by the sample selection for the Survey
observation (discussed by Bragaglia et al., in preparation),
while the complete photometric database (from Jeffries \e
2009, their Fig.1) shows many stars below that line.
In \gam, the parental molecular cloud appears to have dissipated,
with no excess dust and gas remaining in the cluster region. This is
also entirely consistent with the low value of extinction found toward
cluster stars by Jeffries \e (2009), namely $E(B-V)=0.038 \pm
0.016$~mag, or $A_V=0.13 \pm 0.055$~mag (and $E(V-I)=0.055 \pm 0.023$),
also shown in Fig.~\ref{v-vi-2}.
The reddening of foreground stars is expected to be even lower than this value.

The fact that $E(V-I)$ is so small for cluster members, and even smaller
for main-sequence (MS) foreground stars,
together with the fact that nearly all cluster
members have normal photospheric spectra (without veiling or strong
emission lines), make the $V-I_c$ color index a good proxy
for \teff, once the MS/PMS status of a star is ascertained.
The relatively small distance has enabled cluster stars as late as mid-M
to be observed with Giraffe. The predominance of WTTS over
CTTS is also of advantage, since it permits a study of stellar
properties, uncomplicated by accretion phenomena. At the same time,
the study of the few CTTS present enables to check whether methods used
to study WTTS may be extended to accreting stars as well. 

Above the observational cutoff line, Survey targets were selected randomly,
and therefore include a substantial number of field stars.
The spectral-type range expected for un-reddened stars (thus including MS/PMS
stars) is late-A to mid-M.
Fig.~\ref{v-vi-2} shows that the selection limit for the Survey corresponds
roughly to the unreddened ZAMS at the cluster distance: therefore,
low-reddening ZAMS stars can belong to the observed sample only up to a
distance of 710~pc for late-A stars, and of 280~pc for mid-M stars. Thus,
this main-sequence subsample of all observed stars is not expected to
have reddening higher than the cluster itself (or only a little more
for stars earlier than G0), i.e. {\rm the bulk of observed MS or PMS stars has
negligible or little reddening, $A_V<0.2$~mag.}
Early-type stars, which are rare objects and are thus a priori expected
to make only a small fraction of the observed sample, can enter it even
if they are MS stars found at high distances and reddening. Apart from
MS/PMS stars, the observed sample certainly comprises many giant stars
(suggested from the excess of data points in the CMD of
Fig.~\ref{v-vi-2} between $1.3<V-I_c<2$),
observable up to large distances.
Moreover, most giant stars are expected to belong to the red clump, in
which they spend a relatively long part of their evolved-star lifetime,
several magnitudes above the main sequence: this
implies that most giants in the observed sample
are likely to be much more distant than MS/PMS sample stars.
For this reason {\rm most giant stars in the observed sample are expected
to be substantially reddened}. Therefore,
MS/PMS stars and giants are expected to make two mostly distinct,
reddening-segregated subsamples.

Overall, the \gam\ Giraffe observed sample comprises 1802 
spectra of 1242 distinct stars; 439 stars (35\% of the sample)
were observed more than once. Duplicated spectra of the same star are
here analyzed individually.
Such a large sample of spectra permits a study of spectral diagnostics
with very good statistics, for the different populations contained in
the sample.
Tasks within the Gaia-ESO Survey consortium are distributed among
several distinct Working Groups (WG). In particular,
the initial reduction of Giraffe spectra is made by
the Cambridge Astronomical Survey Unit (CASU) within WG7, and provides
wavelength-calibrated, sky-subtracted, barycentrically-corrected
spectra (Lewis et al., in preparation). Radial-velocity (RV)
measurements, $v \sin i$ and other directly-measured parameters
are also provided for Giraffe data by CASU (Koposov et al., in preparation),
as part of WG8 activities.
These spectra and RVs, from data release {\rm GESiDR1Final},
form the basis for the work discussed in this paper.
The spectral range is 6444-6816~\AA, and the pixel size is 0.05~\AA.
Results presented here contribute to the activities of WG12, whose task is
spectrum analysis for young and PMS clusters, WG10, dealing with
analysis of FGK star spectra of field and older cluster stars, and WG14,
dealing with peculiar stars.

\begin{figure}
\resizebox{\hsize}{!}{
\includegraphics[bb=20 10 465 475]{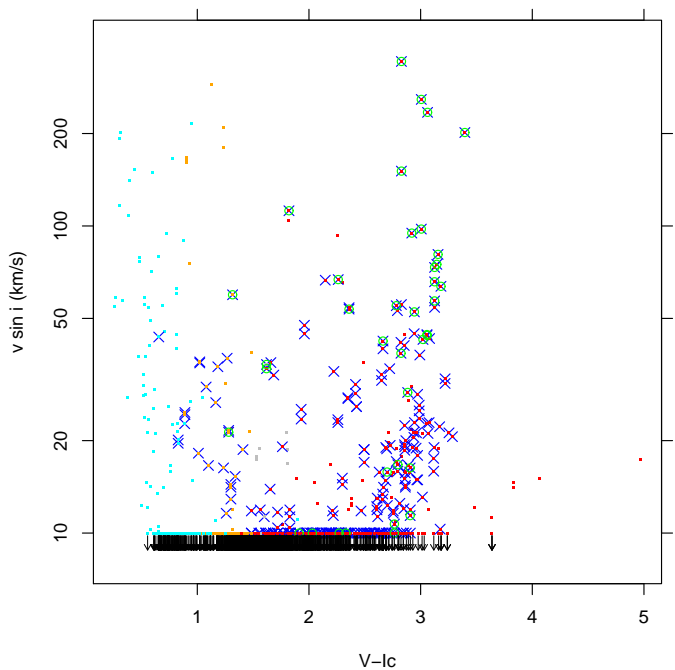}}
\caption{Rotational velocity $v \sin i$ vs.\ $V-I_c$.
A significant number of cluster candidate members are
fast-rotating stars, especially at the latest spectral types.
Symbols are as in Fig.~\ref{v-vi-2}.
Sources with released $v \sin i$ values lower than 10 km/s are marked as
upper limits (arrow symbols).
\label{vrot-vi}}
\end{figure}

\section{Data preparation}
\label{prepare}

In this work we will define several spectral indices, computed as in
several previous studies over fixed wavelength ranges comprising
crucial line features.
In the following, we will repeatedly remark the effect of fast rotation
on the spectral indices we define, which is particularly relevant for some 
of the \gam\ probable members (young and fast-rotating
stars, mostly of late type). Therefore, we show for general reference
a $(V-I_c, v \sin i)$ diagram in Figure~\ref{vrot-vi}, made using the
{\rm GESiDR1Final} released
$v \sin i$ values\footnote{In several cases, the released $v
\sin i$ values are exceedingly low, e.g., 1~km/s or less, considering
the instrumental resolution: these are here considered as
upper limits at 10~km/s, consistently with the work by Jeffries et al.
(2014), on the same data.}.

The continuum shape has impact on the
spectral indices which will be defined in the next Section, both because
asymmetric\footnote{Namely, with a blue-red asymmetry in the used
wavelength regions.}
(molecular band-strength) indices are sensitive to continuum
gradients, and also because indices which are symmetric, but computed over a
relatively large wavelength region (e.g., the \ha\ wings), are sensitive
to second derivatives in the continuum. We have therefore normalized the
spectra ourselves, using a fifth-degree polynomial. The choice of such a
low polynomial order is important, since it ensures that wide features
in the spectrum, such as broad molecular bands and \ha\ wings are not
"fitted" by the continuum normalization procedure, but their shape is
preserved. In order to avoid problems arising from both strong \ha\
emission or wide absorption, we did not use in the fitting procedure a
40~\AA\ wide region centered on \ha.
Thus, this normalization step is rather different from, e.g.,
pseudo-continuum recovery in M~stars, which is meant to trace the
detailed shape of molecular bands, over which individual spectral lines
are found superimposed.
We remark that the adopted normalization recipe deliberately sets to unity the
local average spectrum intensity, but not necessarily the true continuum level:
this is irrelevant for our indices definition.
We emphasize that the adopted normalization procedure is an integral
part of the procedure described in the following sections: with a
different normalization recipe results are not guaranteed to remain
unchanged.

\begin{figure}
\resizebox{\hsize}{!}{
\includegraphics[bb=20 10 465 475]{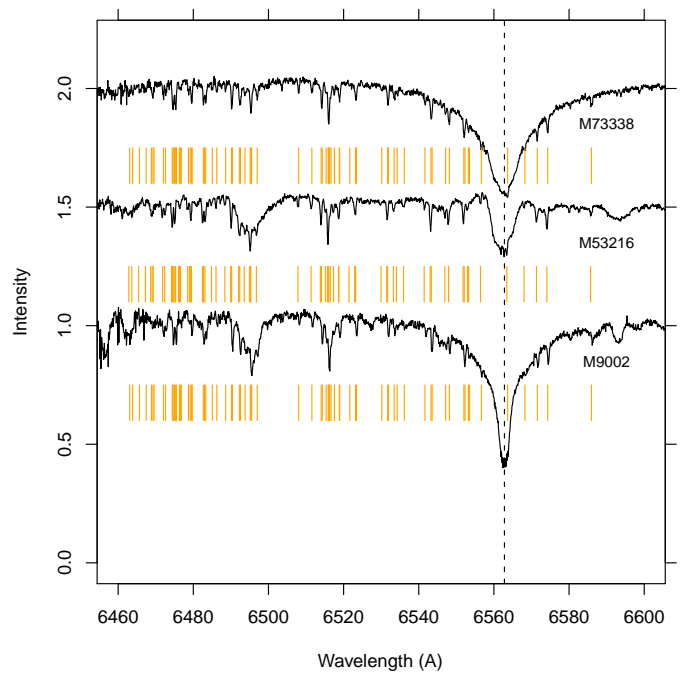}}
\caption{Three examples of fast-rotating early-type star spectra with
evident telluric-line contamination (orange segments).
The dashed line indicates the \ha\ line wavelength.
Spectra are labeled with the sequential star number from Jeffries \e (2009).
\label{telluric}}
\end{figure}

The released spectra are not corrected for telluric lines. Therefore, in
the cases where telluric contamination was strong, we did the correction
ourselves.  Identification of lines as telluric was made
by considering the spectra of (possibly fast-rotating) early-type stars,
and looking at narrow lines which were fixed in (observed) wavelength,
regardless of the star RV (most early-type stars do not belong to the
cluster, and their RV are broadly distributed).  An average, template
telluric-line spectrum was thus computed, which yielded good results
when subtracted from all massive stars, with the only difference of a
scaling factor.  For a given Observing Block (OB), all stars are observed
simultaneously and the scaling factor for the telluric absorption spectrum
is the same for all of them:
therefore, the template telluric-line spectrum, with the
OB-dependent scaling factor, was divided off from all stars in the same
OB. Only a few OBs were lacking early-type stars to use this method;
in these cases, the stars with the highest S/N and the fastest rotation were
used to determine the scaling factor of the template telluric spectrum.
The contaminating telluric lines were found only in the short-wavelength
side of the HR15n range, between 6460-6587~\AA, including \ha\ and other
very useful features which will be introduced below. The red part of the HR15n
range (including the lithium line) is unaffected. Examples of the
strongest telluric-contaminated spectra are shown in Figure~\ref{telluric}.

Finally, released RV values for about 80 stars did not appear to be
correct and we derived them independently
from the cross-correlation function (CCF) with ten different templates,
chosen from among slowly-rotating, high-S/N stars in the \gam\ dataset,
without emission lines, and spanning the entire range of (visually
defined) spectral types. Our redetermined RV values have a likely
larger error ($\sim 1-2$ km/s) compared to that typical of
{\rm GESiDR1Final} released RVs (Jeffries et al.\ 2014), but
this is not critical for the present work.
In most cases, the incorrect RV values were due to SB2 nature, and
are being revised in forthcoming data releases.
After inspection of all spectra (and their CCFs), we discard all the SB2
systems (labeled in Table~\ref{table2}) from our
analysis\footnote{Values of spectral indices for SB2 systems are however
included in Table~\ref{table2} since they may be of some use,
with due care, to infer some rough average properties of the components;
this is also true of spectra with low S/N: also in this case, the choice
of the most appropriate threshold for rejection of noisy results is left to the
prospective user, being dependent on their specific needs.}.

\section{Definition of spectral indices}
\label{all-diag}

The range of spectral types expected for observed stars is so wide
that, in order to classify them with some level of detail, a
hierarchical classification is necessary, to distinguish in the
first place between early-type, intermediate, and
late-type stars. Then we proceed to consider if more refined
indicators of, e.g., luminosity class, stellar activity, accretion,
are available from the observed spectra.
In this section we introduce such a classification scheme, first using
molecular bands and the \ha\ line, and subsequently turning our attention
to other groups of interesting spectral lines.

\begin{figure}
\resizebox{\hsize}{!}{
\includegraphics[bb=20 10 465 475]{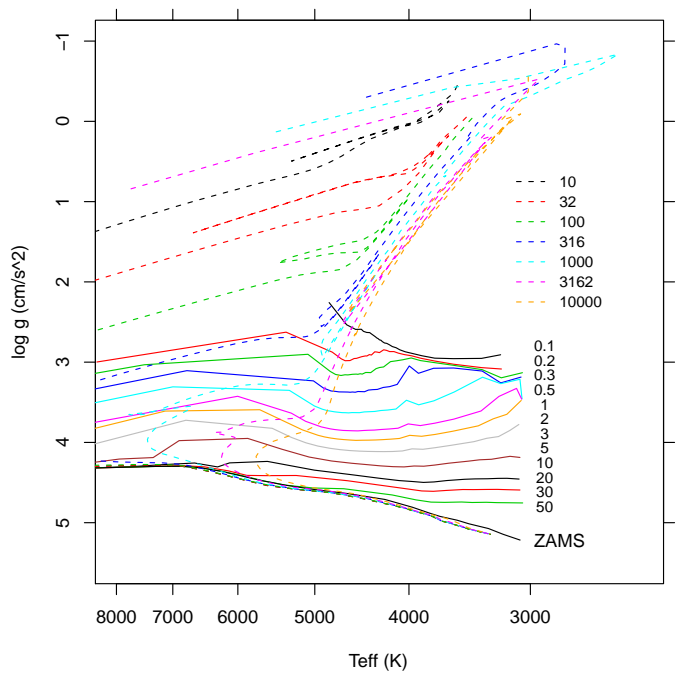}}
\caption{Model isochrones in the $(T_{eff},\log g)$ plane, for both PMS
stars in the age range 0.1-50~Myr and ZAMS stars (solid lines, from
Siess \e 2000), and evolved stars in the age range 10~Myr-10~Gyr (dashed lines,
from Marigo \e 2008). Isochrone ages are labeled in Myr units.
\label{teff-g-isochr}}
\end{figure}

It may be useful to consider the expected locations of stars of
different ages/masses in the $(T_{eff},\log g)$ plane. 
Figure~\ref{teff-g-isochr} shows model isochrones in such a plane, for
both PMS stars in the age range 0.1-50~Myr and ZAMS stars (solid lines, from
Siess \e 2000), and evolved stars in the age range 10~Myr-10~Gyr
and Z=0.019
(dashed lines, from Padova group, Marigo \e 2008 with corrections from
Case A in Girardi et al. 2010\footnote{Isochrones available
through http://stev.oapd.inaf.it/cgi-bin/cmd.}).
The bulk of field giants are expected
to be comprised between the 1-10~Gyr isochrones. The 5-10~Myr
isochrones, where we expect to find \gam\ cluster members, cross the
giant-star region for $T_{eff}>5500$~K, but become increasingly
separated from giants at lower \teff\ values. At the same time, \logg\ in
such cold cluster members is at least 0.5~dex lower than ZAMS stars of
same \teff. Therefore, the latest-type stars are the most promising
place where to expect differences in gravity-sensitive features between
giants and PMS stars in particular, and to a lesser extent between PMS
and ZAMS stars. At \teff\ above 5500-6000~K, the \logg\ of PMS stars
is similar to that of giants, making their discrimination impossible, while
remaining slightly above ZAMS \logg\ values.

\subsection{Molecular bands}
\label{mol-bands}

Molecular bands, especially from TiO, are an unmistakable characteristic of M
stars. The strongest TiO molecular bands fall at redder wavelengths than the
Giraffe HR15n range (e.g., $\lambda>7000$ \AA), but a few bands are
clearly seen in the HR15n range as well. Several spectral indices based on
molecular bands were extensively studied by e.g.,
Kirkpatrick \e (1991),
Reid \e (1995),
Brice\~{n}o, Hartmann and Mart\'{\i}n (1998),
Jeffries \e (2003, for NGC~2547),
Lyo \e (2004, 2008, for $\eta$~Cha and $\epsilon$~Cha),
Riddick \e (2007, and references therein),
Lawson \e (2009),
Shkolnik \e (2009).
In nearly all these studies, with the exception of the "TiO 1" index
defined by Reid \e (1995), the spectral region used to define
interesting indices is outside the HR15n range. Again in Reid \e (1995),
the "CaH 2" index is defined using a spectral window (6814-6846~\AA)
just beyond the red limit of the HR15n band; some
CaH lines (reported to be a gravity indicator) fall inside the HR15n
band as well, longward of 6750~\AA\ (Kirkpatrick \e 1991).
By studying the HR15n band, we are therefore going to explore 
poorly studied molecular indicators, which are the subject of this section.

\begin{figure}
\resizebox{\hsize}{!}{
\includegraphics[bb=20 10 465 475]{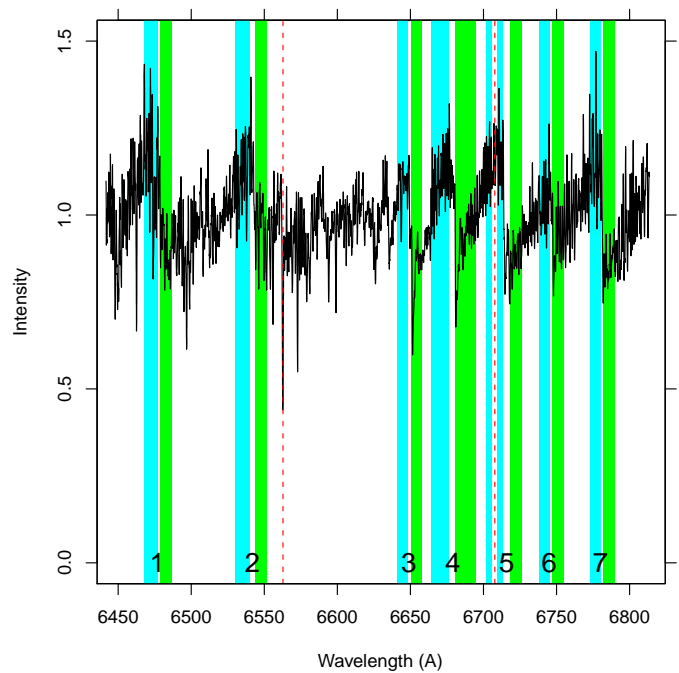}}
\caption{HR15n spectrum of a mid-M star in the \gam\ dataset,
showing several prominent molecular bands. Vertical red
dashed lines indicate \ha\ and the Li~I line at 6707.76~\AA.
Cyan and green regions
indicate respectively "peaks" and "valleys" of bands selected
empirically for study here, identified by numbers 1-7 (above bottom axis).
\label{atlas-molbands}}
\end{figure}

Figure~\ref{atlas-molbands} shows the Giraffe HR15n normalized spectrum of
a mid-M star (star 2358 in Jeffries \e 2009, with $V-I_c=3.239$), with
several prominent molecular bands: those studied here are labeled with
numbers from 1 to 7 near the bottom axis.  Wavelengths shown are in the
stellar rest frame.
Vertical red dashed lines indicate \ha\ and the
Li~I line at 6707.76~\AA. The molecular bands cause the spectrum to show
"valleys" and "peaks": we have indicated with a cyan background selected
"peak" regions,
and with green background its corresponding "valley" regions to the right.
For each of the bands $n=1-7$ we define an adimensional band-strength
index $\mu_n$, as
\begin{equation}
\mu_n = <f^{An}>/<f^{Bn}>,
\label{mun-define}
\end{equation}
where $f$ is the (normalized) spectrum intensity, the symbol $<f>$
indicates flux average, and superscripts $An$ and $Bn$ refer to
"peak" and "valley" wavelength ranges, respectively, for index $\mu_n$.
The wavelength ranges used are listed in Table~\ref{band-def}.
Values of $\mu_n$ for our sample stars are listed in Table~\ref{table2}.

\begin{table*}
\caption{Wavelength ranges for definition of spectral indices.}
\label{band-def}
\centering

\begin{tabular}{ccc}
\hline
Index  & $A$ range(s) & $B$ range(s) \\
       & \AA          & \AA          \\
\hline
$\mu_1$ & 6468-6477   & 6479-6487   \\
$\mu_2$ & 6530-6540   & 6544-6552   \\
$\mu_3$ & 6641-6648   & 6651-6658   \\
$\mu_4$ & 6664-6676   & 6681-6695   \\
$\mu_5$ & (6702-6706.26)$+$(6709.26-6713)   & 6718-6726   \\
$\mu_6$ & 6738-6745   & 6747-6755   \\
$\mu_7$ & 6773-6780   & 6782-6790   \\
$\alpha_w$\tablefootmark{a} & (6556.8-6560.8)$+$(6564.8-6568.8) & (6532.8-6542.8)$+$(6582.8-6592.8) \\
$\alpha_c$ & 6560.8-6564.8 & (6532.8-6542.8)$+$(6582.8-6592.8) \\
$\beta_t$ & 6492.6-6502.6 & (6487-6489)$+$(6505-6507) \\
$\beta_c$ & (6495.9-6499.3)$+$(6489-6492)\tablefootmark{b} & (6487-6489)$+$(6505-6507) \\
$\gamma_1$ & (6759-6761)$+$(6763.5-6765.5)$+$ & (6761.5-6763.5)$+$(6765.5-6768.4)$+$ \\
           & (6768.4-6770.0)$+$(6771.7-6774.1) & (6770.0-6771.7)$+$(6774.1-6776.0) \\
$\zeta_1$ & 6624-6626 & 6632.7-6634.7 \\
\hline
\end{tabular}
\tablefoot{
\tablefoottext{a}{The $\alpha_w$ index must include a correction based
on $\mu$, see text.}
\tablefoottext{b}{The 6489-6492~\AA\ range is summed with the main range
with relative weights 1:7.}
}
\end{table*}

In order to obtain the best results from this method, we made a careful
choice of the band boundaries.  First, we have purposely left a
minimum separation of 2\AA\ between the right end
of an interval and the left end of the next, to minimize the effect on
$\mu_n$ of the smearing of bands caused by fast stellar rotation.
In this case, "peak" and "valley" regions increasingly contaminate each
other (the effect of other, narrow absorption lines being much smaller),
with the contaminated wavelength interval being proportional to $v \sin i$.
At these
wavelengths 2\AA\ correspond to about 90~km/s, and very few of our
spectra show $v \sin i$ exceeding this value (Fig.~\ref{vrot-vi}).
Indices $\mu_n$ are therefore expected to be affected by rotation only
for few stars with $v \sin i >90$ km/s.
Next, the $\mu_2$ defining range is such that it avoids contamination by
the neighboring [N II] line, often strongly in emission in CTTS. In the
same way,
the $\mu_4$ range is designed to avoid the He~I $\lambda$6678 emission line,
and the $\mu_5$ range avoids the [SII] $\lambda$6716 emission line.
Finally, since the lithium $\lambda$6707.7 line is
neither a \teff\ nor \logg\ indicator, we excluded a 3\AA\ interval
centered on the Li~line from the "peak" region of the $\mu_5$ range
(as indicated in Fig.\ref{atlas-molbands}). This
choice makes $\mu_5$ totally insensitive to Li EW variations in stars with $v
\sin i$ up to $\sim 70$ km/s, and only a little above it.

The relatively wide wavelength ranges used to define indices $\mu_n$
make these indices to have a generally high S/N:
in the worst case of spectra with S/N=15 (per pixel)
the statistical 1$\sigma$ errors on $\mu_n$ fall in the range 0.0086-0.012;
most stars have however much higher S/N, and errors on $\mu_n$ are thus
significantly lower than these worst-case limits.

\begin{figure*}
\includegraphics[bb=18 20 594 774,width=18cm]{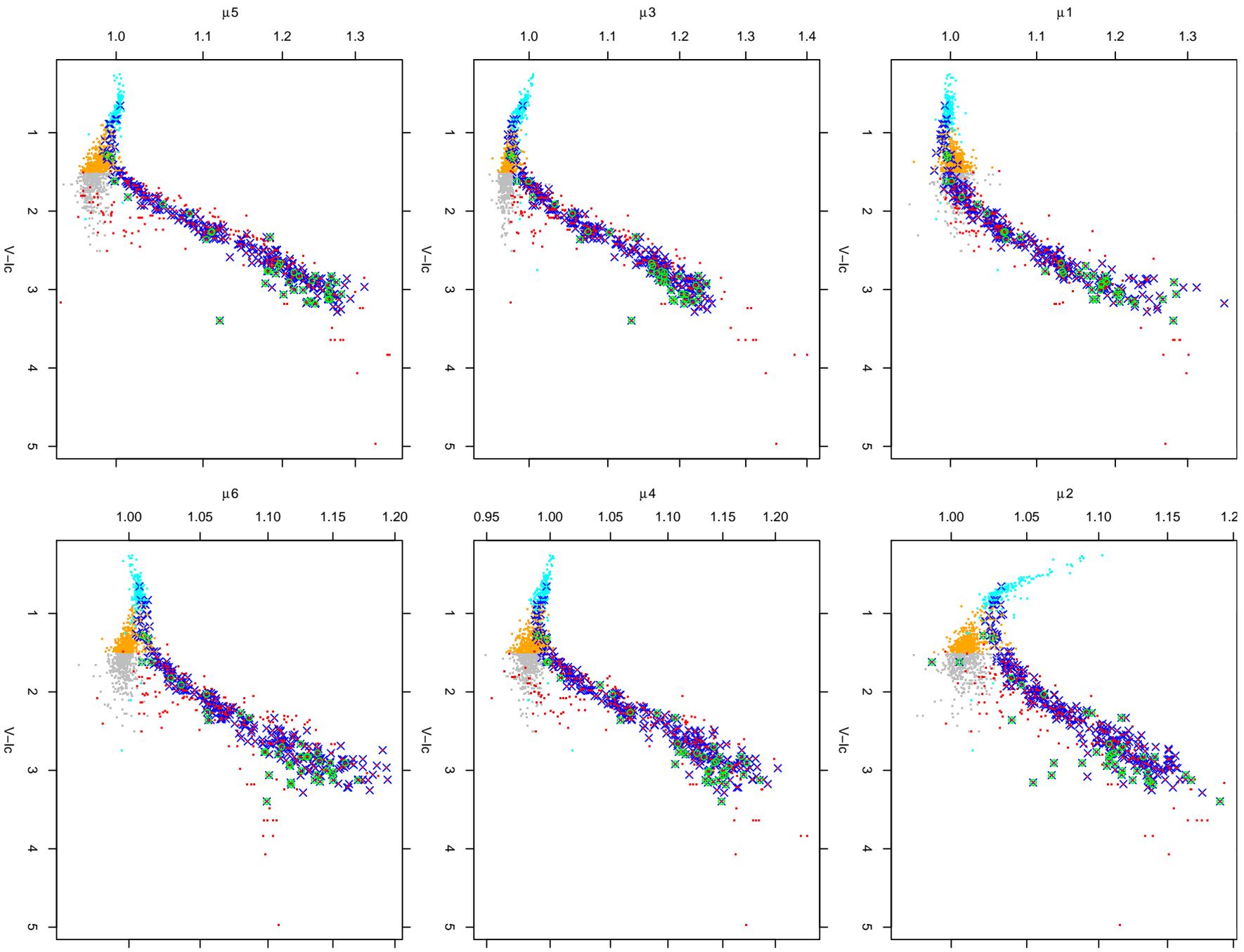}
\caption{Molecular indices $\mu_n$ vs.\ $V-I_c$, for $n=1-6$.
Symbols as in Fig.\ref{v-vi-2}.
\label{mb-vi}}
\end{figure*}

\begin{figure*}
\includegraphics[bb=18 520 594 774,width=18cm]{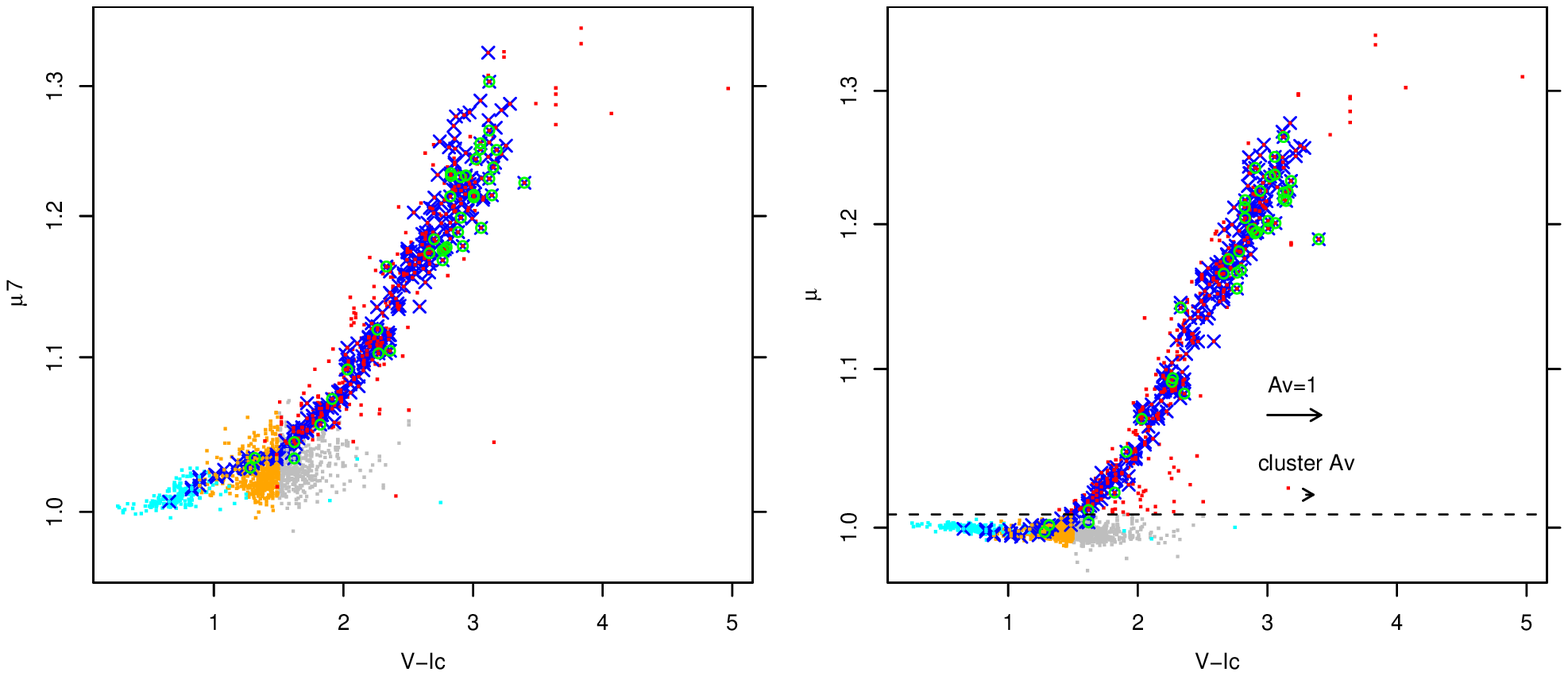}
\caption{$(a, left)$: Molecular index $\mu_7$ vs.\ $V-I_c$.
$(b, right)$: Mean molecular index $\mu$ vs.\ $V-I_c$.
Symbols as in Fig.\ref{v-vi-2}.
In panel $b$, a representative reddening vector is indicated,
as well as the very small cluster reddening vector.
The horizontal dashed line indicates the limiting value $\mu_0$.
\label{mbm-vi}}
\end{figure*}

Figures~\ref{mb-vi} and~\ref{mbm-vi} show the dependence of the seven
indices $\mu_n$ on  $V-I_c$: the general increase in the indices toward
redder, cooler stars is similar but not identical among the different cases.
Only stars with S/N (per pixel) $>15$ are shown.
Index $\mu_2$ is affected by the extreme \ha\ wings in
the earliest-type stars in our sample (cyan dots); index $\mu_6$ appears
to saturate for the reddest (non-member) stars in the sample, but not
for members (blue crosses), which may indicate a dependence on
\logg\ in addition to \teff, especially if compared with the lack of
such saturation in the $\mu_3$ index for the same red stars, which seem
to rule out a reddening effect on the $V-I_c$ color as explanation.
The larger scatter observed among member stars near $V-I_c \sim 3$ may be
due only partially to the lower S/N of their spectra, near the bottom of
the cluster sequence, but especially to their large $v \sin i$ ($>90$
km/s), more frequent in this color range than elsewhere (Fig.~\ref{vrot-vi}).

Indices $\mu_1$, $\mu_3$, $\mu_5$, and $\mu_7$ show the least scatter around
their mean dependence on $V-I_c$, considering only member stars.
Therefore, we define an average $\mu$ index as
\begin{equation}
\mu = 0.25 \; \; (\mu_1 + \mu_3 + \mu_5 + \mu_7),
\label{mu-define}
\end{equation}
tabulated in Table~\ref{table2}, and
whose dependence on $V-I_c$ is shown in Fig.\ref{mbm-vi} (right).
Maximum statistical 1$\sigma$ error on $\mu$ is 0.0057, occurring
mostly for our reddest cluster members.
The strong correlation with $V-I_c$ for the redder cluster stars is very
evident, and interpreted as a \teff\ dependence.
Since $\mu$ is essentially dependent on \teff, with a very small (if
any) dependence on \logg, we consider this index as our primary \teff\
indicator for late-type stars (basically, M stars).
This agrees with the generally accepted adoption of TiO bands as \teff\
diagnostic for M stars.
Stars without TiO bands are located in a narrow strip around $\mu=1$,
which is the reference or "neutral" value for such index. The observed scatter
around this value enables us to define a threshold value $\mu_0=1.008$ above
which the presence of molecular bands, however weak, can be considered
certain (dashed line in the Figure). We make therefore our initial
classification (corresponding to color coding used throughout the paper)
as follows:
\begin{itemize}
\item Stars with $\mu>\mu_0$ are
defined here as {\rm late-type}, and drawn in all figures with {\rm red dots}.
\item Stars with wide \ha\ absorption (see Section~\ref{halpha}) are defined
as {\rm early-type} stars\footnote{Here, early-type must be intended in
a relative sense, and not as indicative of OBA stars; the Sun itself
turns out to be included in this class.} (cyan dots).
\item Stars showing no wide \ha\ absorption, nor TiO bands
are defined as {\rm intermediate-type} stars; if they have $V-I_c \leq 1.4$,
their position in Figure~\ref{mbm-vi} is compatible with no
reddening\footnote{This is not the same as stating that their
reddening is zero, however.}, and they are plotted as orange dots.
\item Stars without wide \ha\ or TiO bands, but having $V-I_c>1.4$, are
still intermediate-type stars, but their optical color is too red to be
compatible with zero extinction (gray dots), since in this latter case TiO
bands would be found in their spectra.
\end{itemize}
The "gray" stars might be not qualitatively different
from the "orange" ones, but their high reddening implies that they are
giants, as we argued in Section~\ref{obs-sample}.
This fact will be very useful
when selecting gravity-dependent spectral features in the
following. In Fig.\ref{mbm-vi} (right) a few of the (non-member) red datapoints
fall at redder $V-I_c$ than the cluster-member locus: they also are
high-reddening stars, which by virtue of their showing
molecular bands are to be classed as M giants.

\begin{figure*}
\includegraphics[bb=18 265 594 774,width=18cm]{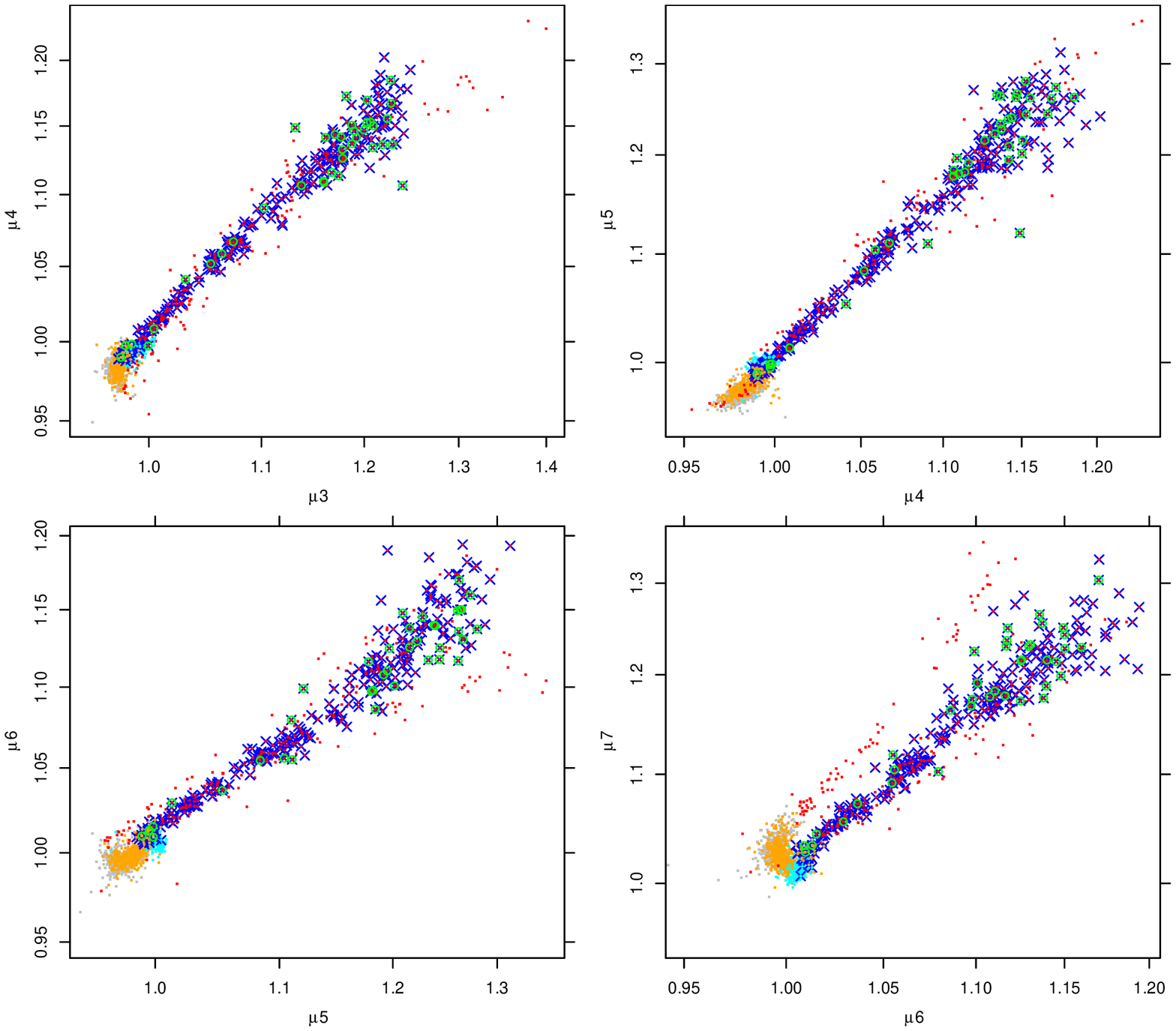}
\caption{Plots of indices $\mu_{n+1}$ vs.\ $\mu_n$, for $n=3-6$.
Symbols as in Fig.\ref{v-vi-2}.
\label{mb-mb}}
\end{figure*}

A better understanding may be gained from the study of the mutual
relationships between pairs of indices $\mu_n$, as shown in
Figure~\ref{mb-mb}. Especially considering the cluster members, these
indices are well correlated with one another, and this further supports
using a single $\mu$ index as representative. The "saturation" affecting
index $\mu_6$ is again visible here, only for very red non-member stars.
The lower right panel of Fig.~\ref{mb-mb} shows however a very interesting
"double"
correlation between $\mu_6$ and $\mu_7$, with a lower branch populated
by members (and other stars), and an upper branch containing only non-members.
Moreover, the upper branch joins smoothly near $\mu_6 \sim 1$ with the
distribution of gray dots (i.e.\ reddened giants), and also by virtue of this
continuity it is highly probable that stars in the upper branch are M giants.
The separation we observe between the two branches,
therefore, is the first example of a gravity-sensitive spectral feature
in the HR15n range. This initial selection of giants enables us to
define suitable template spectra among those in our dataset, which will
in turn enable us to study gravity effects on other spectral features.

The index $\mu$ might be affected by variable metal abundance, or by
veiling, the latter being more likely to occur in young clusters.
In stars where veiling is detected and measured (which is rarely the case
for \gam\ stars, see below), a veiling-corrected $\mu$ may be derived: in
fact, a veiling $r$ (ratio of veiling continuum level to star
continuum level) causes $\mu$ to reduce its distance from its "neutral
value" of 1 by a factor $\sim 1+r$ to a good approximation; this
implies that a veiling-corrected index $\mu^{corr}$ may be computed
from the observed index $\mu^{obs}$ as
\begin{equation}
\mu^{corr} = 1 + (\mu^{obs}-1) \times (1+r).
\label{mu-veil}
\end{equation}
In the following, all indices shown are the observed ones, not the
veiling-corrected ones, even for the few stars with veiling.

\begin{figure*}
\includegraphics[bb=18 520 594 774,width=18cm]{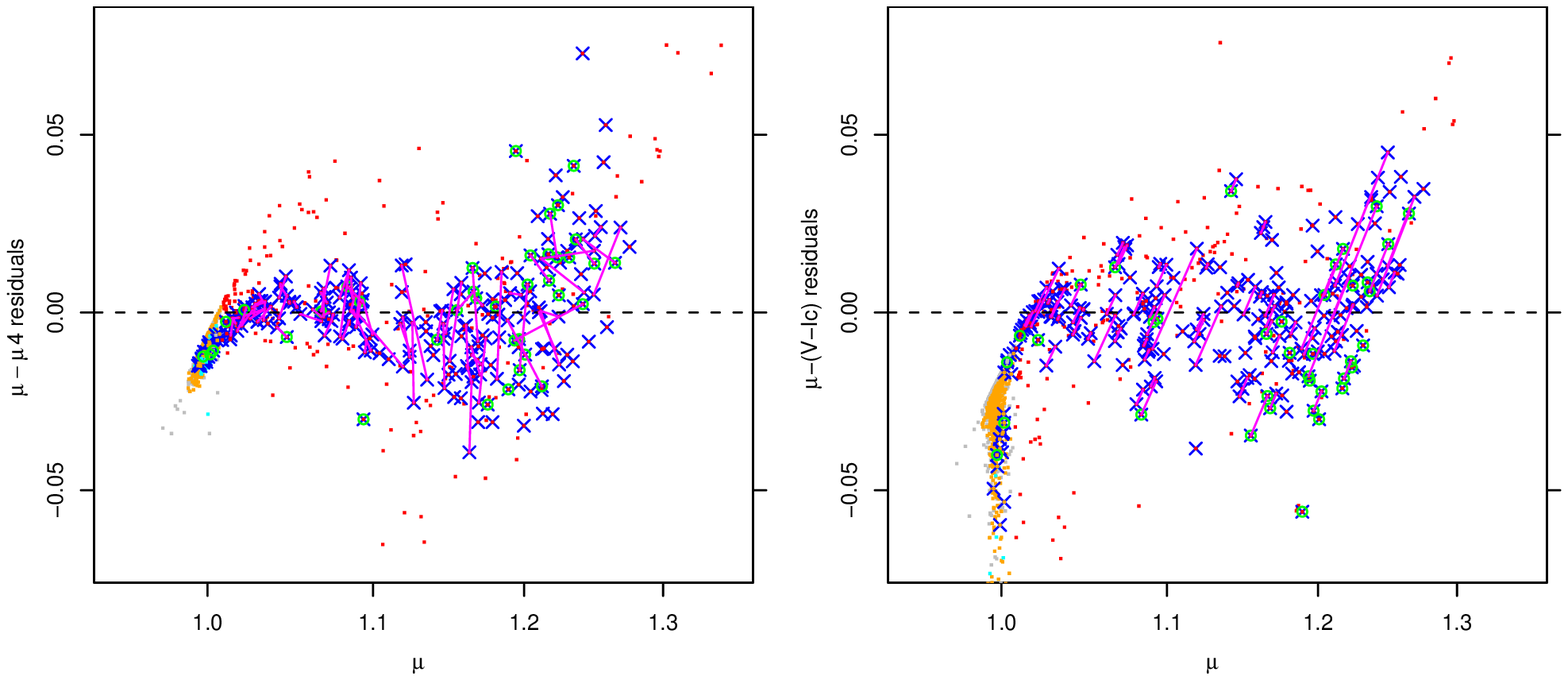}
\caption{$(a, left)$: Best-fit residuals of index $\mu$ vs.\ $\mu_4$.
$(b, right)$: Best-fit residuals of $\mu$ vs.\ $V-I_c$.
Symbols as in Fig.\ref{v-vi-2}.
Magenta segments connect different observations of the same stars.
\label{mb5-mbm-resid}}
\end{figure*}

Since the $(\mu_{n+1},\mu_n)$ relations shown in Fig.~\ref{mb-mb} are
reddening-independent, they may be used to discriminate
true uncertainties on the $\mu$ values from residual
reddening, which may instead affect the $V-I_c$ color in the $(\mu,V-I_c)$
plot, and contribute to the scatter of datapoints in it (together with
photometric color variability).
In order to quantify the contribution by reddening and variability to
$V-I_c$ scatter, we have made a (rough) polynomial fit to both the
$(\mu,V-I_c)$
correlation and the $(\mu,\mu_4)$ correlation (not shown for brevity);
indices $\mu$ and $\mu_4$ are statistically independent since $\mu_4$ does
not enter in the definition of $\mu$, and their correlation is unaffected
by either reddening or variability.
Fitting residuals are shown respectively in
Figures~\ref{mb5-mbm-resid}$a$ and~$b$, as a function of $\mu$:
despite the roughness of the fit and systematic trends in the residuals,
around $1.05<\mu<1.1$ it is clear that the $V-I_c$ vs.\ $\mu$ fit has larger
residuals among members (std.dev.=0.012)
than the $\mu_4$ vs.\ $\mu$ fit (std.dev.=0.0069), a difference which we ascribe
to non-uniform reddening and/or to variability among late-type cluster stars.
Subtracting (in
quadrature) the two standard deviations, and considering the slope of
the $\mu$ vs.\ $V-I_c$ relation and the $A_V/E(V-I)$ ratio,
we may compute a $1\sigma$ range in
extinction of $\Delta A_V = 0.12$ mag, thus comparable to the average
cluster extinction itself. This percentually large variation in $A_V$
should be considered as an upper bound, however, since
Figure~\ref{mb5-mbm-resid} also confirms
that some stellar variability is present, as made clear by the multiple
observations of the same stars in our dataset (magenta
segments). Likely, variable starspots
coverage changes slightly the average \teff\ of the star, causing
a small additional spread in the measured values of $\mu$.
This may explain at least partially why statistical errors on $\mu$
computed above are significantly smaller than the spread of datapoints
in the $(\mu,V-I_c)$ diagram.
It may also be worth remarking that, as discussed by Mohanty
\e (2004), magnetic fields in starspots might also affect
{\rm gravity indicators}, because of the additional contribution of
magnetic pressure (and thus a decrease in gas pressure and {\rm
apparent} gravity) within a spot, with a reduction in effective gravity
by up to 0.25 dex in the cases considered by Mohanty \e (2004).

\subsection{The H$\alpha$ line}
\label{halpha}

Besides molecular bands, a most prominent feature of many spectra in the
sample is the \ha\ line. It occurs in an extreme variety of line shapes
and intensities: from deep absorption to strong emission, from narrow to
very wide, passing through barely noticeable lines. The measured
quantity most commonly used to describe the line is its equivalent width
(EW); however, such a simple scalar quantity does not seem adequate to
parametrize the whole observed variety of \ha, to gain the best insight
this line may offer on the stellar properties.
If we consider, for the moment, only the issue of classification of
normal photospheric spectra, the \ha\ line is known to become strongly
saturated and develop wide absorption wings at types earlier than G0
(\teff\ above $~\sim 6000$~K), with an intensity peaking around type A0-A2,
and then decreasing toward earlier types. For the purpose of this work,
concerned only with stars later than mid-A, the latter decrease will not be
discussed. For stars between mid-A and early-G, instead, we have tried to
define a spectroscopic index measuring the \ha\ line wings.

\begin{figure}
\resizebox{\hsize}{!}{
\includegraphics[bb=20 10 465 475]{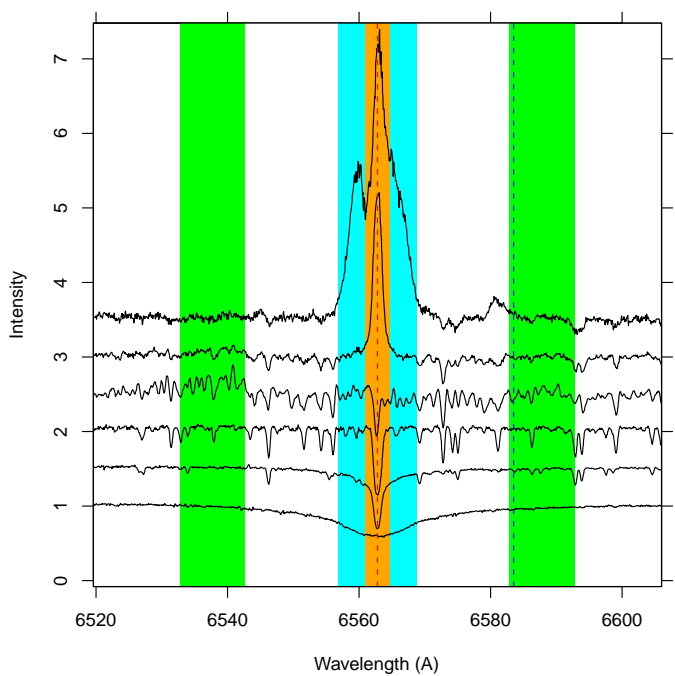}}
\caption{Sample spectra of \gam\ stars in the \ha\ region,
in a sequence of earlier (bottom) to later (middle) spectral types, and
including some narrow- and wide-emission stars (top).
Colored regions indicate wavelength ranges used to define the indices
discussed in the text. Green: continuum region. Orange: \ha\ line core.
Cyan: (inner) line wings.
The two dashed vertical lines indicate the \ha\ line and the [N II]
nebular line.
\label{atlas-halpha}}
\end{figure}

Fig.~\ref{atlas-halpha} shows the spectral region around \ha\ for some
representative
spectra in the \gam\ sample, covering most of the encountered situations
as far as \ha\ is concerned, from earlier (bottom) to later (middle)
spectral types, and including some narrow- and wide-emission stars
(topmost two spectra).
To define \ha\ spectroscopic indices, we have chosen to consider
separately the emission arising from different line parts:
the line core (2~\AA\ from line center, orange in the Figure), and
the inner line wings (beween 2-6~\AA\ from line center, cyan in the Figure).
The line reference continuum is measured between 20-30~\AA\ on both
sides from line center (green in the Figure), in order to minimize the
effect of the wide line wings found in early-type stars.
Weak \ha\ lines, broadened only by rotation, are entirely
contained in the line core region up to $v \sin i \sim 90$~km/s. The same
is true of narrow, chromospheric-like emission lines (e.g., second spectrum
from top in the Figure). The wide absorption wings of early-type stars
(bottom spectrum in the Figure),
and the wide emission profiles of CTTS (topmost
spectrum in the Figure), are well sampled by the inner-wings region selected.
The spectral regions covering the whole wings are only suited for
use in early type stars, since in late-type stars they are
filled by many absorption lines (or bands), and considering them rather
that just the inner wings adds both noise and systematic effects to the
corresponding \ha\ indicator.
The third spectrum from top in Fig.~\ref{atlas-halpha} belongs to a M
star, as evident from the molecular band causing the jump near 6540~\AA\
(sampled by our index $\mu_2$). In these stars, the reference continuum
for \ha\ wings index is not flat, and a correction will be needed.

\begin{figure}
\resizebox{\hsize}{!}{
\includegraphics[bb=20 10 465 475]{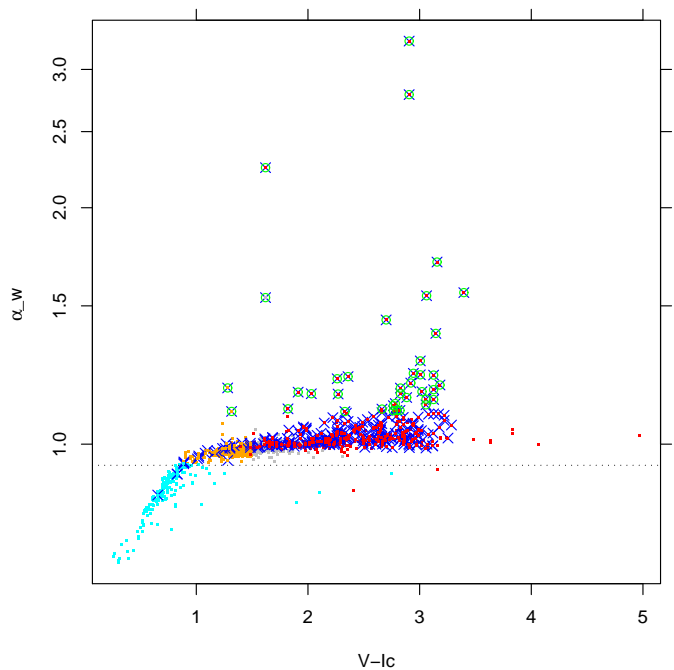}}
\caption{The $\alpha_w$ \ha\ wing index vs.\ $V-I_c$ color.
Symbols as in Fig.\ref{v-vi-2}.
The horizontal dotted line at $\alpha_w=0.94$ shows the adopted limit
for stars with wide \ha\ absorption. Stars (shown with green circles)
with $\alpha_w>1.1$ have instead wide \ha\ {\rm emission}, and are
therefore candidate CTTS.
\label{hawc-vi}}
\end{figure}

\begin{figure}
\resizebox{\hsize}{!}{
\includegraphics[bb=20 10 465 475]{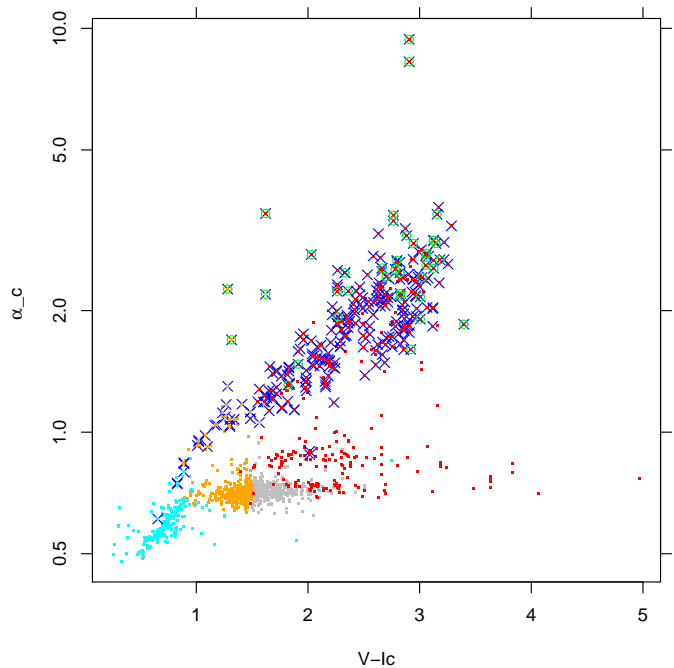}}
\caption{The "chromospheric", narrow \ha\ index $\alpha_c$ vs.\ $V-I_c$
color.  Symbols as in Fig.\ref{v-vi-2}.
\label{hacc-vi}}
\end{figure}

Using these spectral regions, we define two indices, describing the \ha\
core and (inner) wing components respectively, as
\begin{equation}
\alpha_c = <f_{core}>/<f_{continuum}>,
\label{hac-def}
\end{equation}
and
\begin{equation}
\alpha_w = <f_{wings}>/<f_{continuum}> - \; 0.4 \times (1 - \mu).
\label{haw-def}
\end{equation}
Note the corrective term included in the $\alpha_w$ definition: this
compensates the jump in pseudo-continuum emission near 6540~\AA, as
noted above, which would otherwise lead to a downward-bending $\alpha_w$
toward redder stars. This correction ensures instead a flat dependence
on star color for all late-type stars (excluding CTTS) as
can be clearly seen in Figure~\ref{hawc-vi}. Strictly speaking, the correction
is tied to the spectrum jump sampled by index $\mu_2$, but since this index
is also very well correlated to the average molecular index $\mu$, we
use this latter to compute the corrective term.
Fig.~\ref{hawc-vi} also shows that for blue stars $\alpha_w$ is
correlated with color, as expected; deviant points (redward of the main
correlation) are likely reddened stars. We therefore set a threshold
value $\alpha_{w0} = 0.94$, below which we classify stars as {\rm
early-type} (shown in all plots with {\rm cyan dots}), as already
anticipated in Section~\ref{mol-bands}.
The Figure also shows that many stars have $\alpha_w>1.1$, namely far
above the spread typical of most stars, including cluster members. These
stars, like the topmost spectrum of Fig.~\ref{atlas-halpha}, have wide \ha\
{\rm emission} lines, and are therefore candidate CTTS. We indicate them
with green circles throughout all paper figures.
Among the cluster members in the range $1.0<\alpha_w<1.1$, some upwards spread
in $\alpha_w$ values is caused by fast rotation, which broadens an
intrinsically narrow chromospheric \ha\ line so much that the line wings
cause a small increase in $\alpha_w$ above unity;
this cannot however explain datapoints much above the $\alpha_w=1.1$ threshold.
Note also that because of widespread variability in PMS stars' emission
lines, stars may cross the $\alpha_w=1.1$ threshold from one observation
to the next, as found also using different definitions of CTTS status.
Finally, we note the
presence of two stars exhibiting molecular bands (hence shown with red dots)
{\rm below} the threshold $\alpha_{w0} = 0.94$, well separated from all
other late-type stars, and whose location in the diagram cannot be
therefore ascribed to reddening. Such outliers are peculiar stars, which need
to be examined individually; within the Survey, this is the main task of
WG14, to which these objects have been signaled.
A better understanding of our indices ability to find some types of
peculiar stars is gained in Sect.~\ref{uvespop}.

Since we only consider stars with average S/N$>$15 per pixel, {\rm minimum}
S/N in the indices $\alpha_c$, $\alpha_w$ are computed to be
96 and 128, respectively, corresponding to (maximum) 1$\sigma$ expected
errors in the indices of approximately 0.027 and 0.008, respectively,
excluding CTTS.
As for $\mu$, these statistical errors are much smaller than the actual
spread of datapoints in all relevant Figures.
The $\alpha_w$ and $\alpha_c$ values are reported in Table~\ref{table2}.
If the \ha\ line is entirely contained within $\pm 6$~\AA\ from line
rest wavelength, it is also possible to compute the line EW from our two
indices, using the expression (where positive EW values mean emission)
\begin{equation}
EW(H\alpha) (\AA) = 4 \times \alpha_c + 8 \times \alpha_w -12.
\label{ha-ew}
\end{equation}

Figure~\ref{hacc-vi} shows instead the $\alpha_c$ index vs.\ $V-I_c$.
This Figure shows very characteristic patterns for different
classes of stars: the strong \ha\ absorption of early-type stars makes
their $\alpha_c$ index to attain the lowest values in the sample.
Most of the cluster
members, instead, are found in a characteristic band in this diagram.
The very few members stars above the main cluster locus are strong
emission-line stars, and the single one below it might be a spurious
candidate member. For most cluster stars, this strong \ha\ core
emission is not accompanied by emission in the \ha\ wings (see
Fig.~\ref{hawc-vi}): \ha\ lines in these stars are therefore
narrow, as characteristics of chromospheric emission, not of
CTTS stars. The cluster locus in Fig.~\ref{hacc-vi} describes thus
the dependence of chromospheric emission on spectral type (for which
$V-I_c$ is a good proxy {\rm in the particular case of \gam\ stars}).
Since chromospheric emission is known to decay with stellar age $t$ as
$\propto t^{-1/2}$ (Skumanich 1972), the group of non-cluster
late-type stars (red dots)
at lower $\alpha_c$ values ($\alpha_c <1$) is likely to be identified
with much older low-mass MS field stars. Moreover, we
identify the small "tail" of red points attached to the cloud of
(background) gray stars as M~giants.
Most non-cluster, intermediate-type stars (MS stars
in particular) have only weak, very uniform \ha\ absorption.

\begin{figure}
\resizebox{\hsize}{!}{
\includegraphics[bb=20 10 465 475]{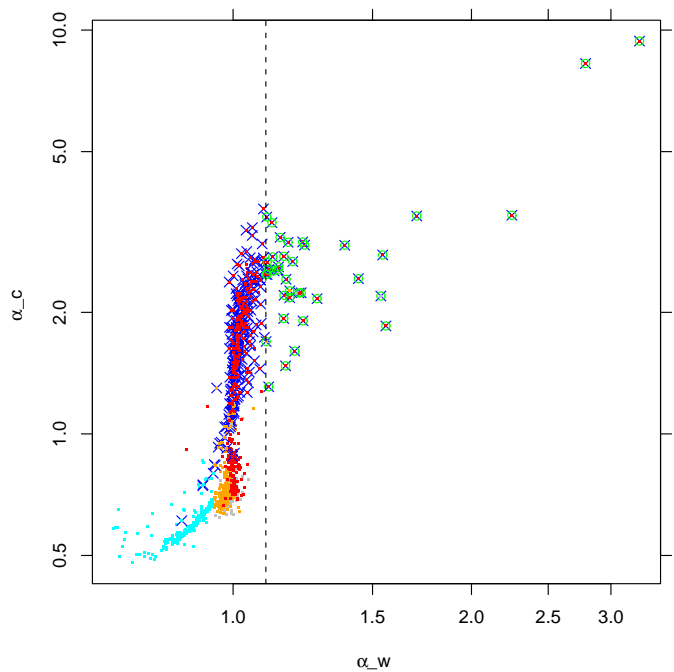}}
\caption{Index-index diagram $\alpha_c$ vs.\ $\alpha_w$.
Symbols as in Fig.\ref{v-vi-2}.
The vertical dashed line indicates the limiting value $\alpha_{w}=1.1$ for
selection of stars with wide \ha\ emission, i.e.\ likely CTTS.
\label{hacc-hawc}}
\end{figure}

A reddening-free $\alpha_c$ vs.\ $\alpha_w$ diagram (shown in
Figure~\ref{hacc-hawc}) is the most useful diagnostics of
the nature of \ha\ emission.
Stars in the lower left part are characterized by {\rm wide}
\ha\ {\rm absorption} lines; {\rm narrow} \ha\ {\rm emission}
accounts for the
vertical strip of datapoints around $\alpha_w=1$; stars with {\rm wide}
\ha\ {\rm emission} are candidate CTTS (green circles rightward of the dotted
vertical line).
Most cluster stars have a strong narrow \ha\ emission, but comparatively
few \gam\ cluster stars have CTTS-like \ha\ emission.  Note also the very
tight (reddening-independent) correlation between $\alpha_c$ and $\alpha_w$
for the early-type stars (cyan dots), with the notable exception of a
few outlier datapoints. Explanations for these latter may include
chromospheric \ha\ emission cores (as, e.g., for the candidate early-type
members), or fast rotation, present in several of the earliest-type stars
shown. Again, a detailed study of these outliers is deferred to a future
work.

\ha\ wings for stars colder than early-A (namely, all stars in our
dataset) are reported to be a gravity-insensitive feature (e.g., Gray and
Corbally 2009); they are also metallicity-independent, since the
hydrogen abundance is essentially the same for varying metallicity.
Therefore, we consider our $\alpha_w$ index a primary \teff\ indicator
for early-type stars in the studied sample.

\subsection{The 6490-6500 \AA\ region}
\label{quintet}

\begin{figure}
\resizebox{\hsize}{!}{
\includegraphics[bb=20 10 465 475]{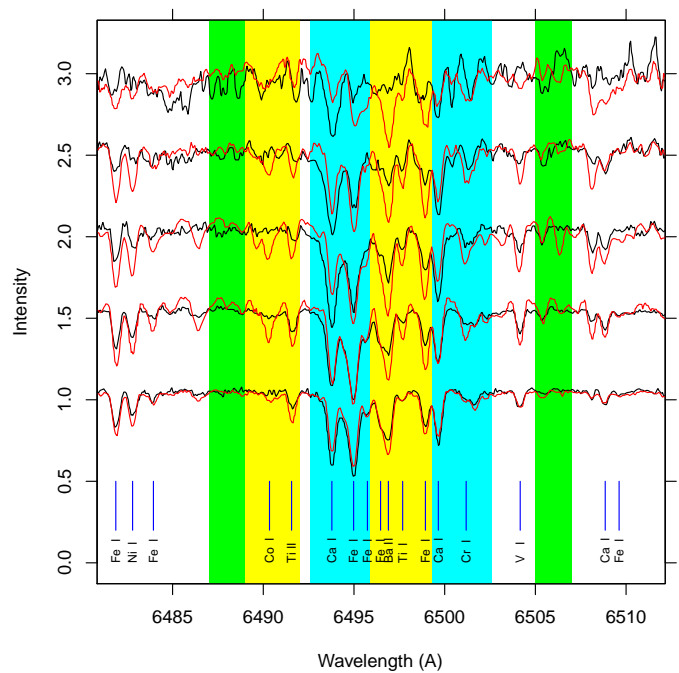}}
\caption{Atlas of spectra in the 6485-6510\AA\ region, containing the
"quintet" of strong Ca~I, Fe~I, and Ba~II lines. These and other
prominent lines are indicated near the bottom axis. Normalized,
telluric-line corrected spectra for MS stars are shown in black, in a
temperature sequence (hotter to the bottom), and offset by 0.5 units for
clarity. Red spectra are for giant stars at similar temperatures as the
respective MS spectra. Gravity-sensitive lines stand out clearly from
the comparison.  Spectral ranges used to compute the $\beta_t$
index are indicated with cyan color (and the enclosed yellow range),
while green ranges indicate the index reference continuum. Yellow ranges
are instead used to compute the $\beta_c$ index, with the same reference
continuum as for $\beta_t$.
\label{atlas-quartet}}
\end{figure}

After \ha\ and molecular bands, the most conspicuous feature in the
HR15n spectral range is the close group of lines between 6490-6500 \AA.
The strongest lines in this group, observed in our spectra are those of
Ca~I $\lambda$6493.78, Fe~I $\lambda$6494.98, Ba~II $\lambda$6496.897,
Fe~I $\lambda$6498.94, and Ca~I $\lambda$6499.65
(plus weaker lines of Fe~I and Ti~I, see Figure~\ref{atlas-quartet}).
The bulk absorption of all lines in this "quintet" is a strong, characteristic
feature in the spectra, which represents well the overall strength of
metallic lines in the spectrum.
This closely-spaced group of lines
becomes easily blended with even moderately fast rotation, in which case EWs
are not easily computed for the individual lines: this is again an instance
where the use of a spectral index becomes of advantage.

We first define an index which measures the {\rm total} intensity of the
"quintet" lines as
\begin{equation}
\beta_t = <f^A>/<f^B>,
\label{qrt-define}
\end{equation}
where the $A$ and $B$ wavelength ranges are defined in Table~\ref{band-def}.
The $B$ range has the role of a reference (pseudo-)continuum. With
reference to Fig.~\ref{atlas-quartet}, the $A$ range is comprised of the
cyan-colored bands (and the yellow region bracketed by them), while the
reference continuum is extracted from both green-colored regions on the
two sides. The black spectra shown in the Figure are ordered in a
temperature sequence (increasing from top to bottom, from early-M to
late-G types); prominent spectral lines are also indicated.
Here and in following analogous plots, a temperature sequence for MS/PMS
stars is defined on the basis of the $\mu$ index for late-type (M)
stars, and on the basis of $V-I_c$ color for hotter stars, extinction
toward these stars being low and uniform enough for a good
correspondence between optical color and temperature (Sect.~\ref{obs-sample}).
The meaning of the yellow-colored bands, and
of the red-colored spectra will be explained below. Stars of the
earliest spectral types are not included among the spectra shown here:
they exhibit a gradual increase in the intensity of the "quintet" lines,
from the earliest types toward intermediate-type stars.

\begin{figure}
\resizebox{\hsize}{!}{
\includegraphics[bb=20 10 465 475]{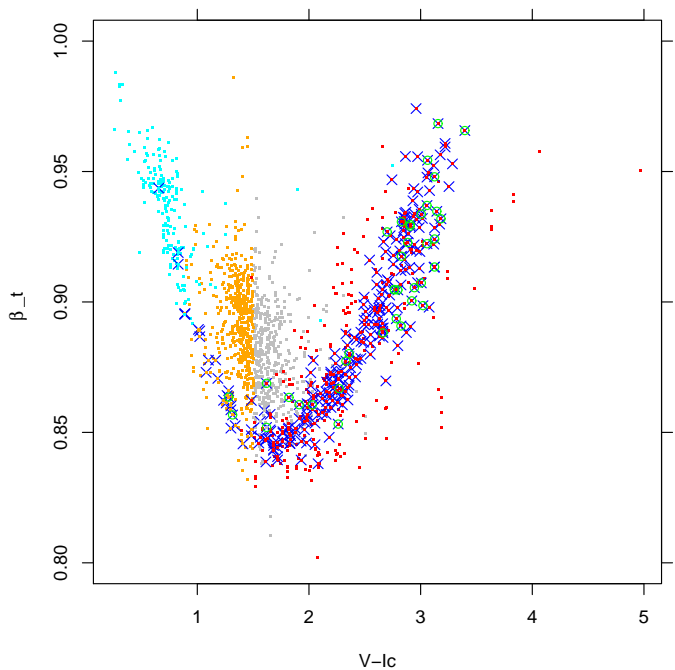}}
\caption{The total "quintet" intensity index $\beta_t$ vs.\ $V-I_c$ color,
with the usual meaning of symbols.
\label{qrt3-vi}}
\end{figure}

The usefulness of the $\beta_t$ index just defined becomes clear when
plotting it vs.\ the $V-I_c$ color (Figure~\ref{qrt3-vi}). Considering for
the moment only cluster members, which especially near solar
temperatures have near-ZAMS gravities, we observe a very marked
monotonic decrease in the $\beta_t$ index, starting from its "neutral"
value $\beta_t=1$ at early-A types toward a minimum value of $\beta_t
\sim 0.85$ right close to the transition between intermediate-type and late-type
(M-type) stars, where molecular bands start to appear. In particular,
$\beta_t$ shows a well-behaved decrease in the "gap" not covered by
either $\alpha_w$ nor $\mu$ indices: $\beta_t$ provides
therefore a good, {\rm highly significant \teff\ indicator} for {\rm
intermediate-type}
stars. The ensuing increase toward unity for colder stars does not introduce
ambiguity, since the $\mu$ index is able to discriminate effectively
stars on the two sides of the $\beta_t$ minimum. We cannot make any
statement at the moment about the large cloud of (orange/gray)
datapoints around $V-I_c \sim 1.5$ above the $\beta_t$ minimum, since they
are (background) giants, for which their reddened $V-I_c$ color is not an
useful quantity.

\begin{figure}
\resizebox{\hsize}{!}{
\includegraphics[bb=20 10 465 475]{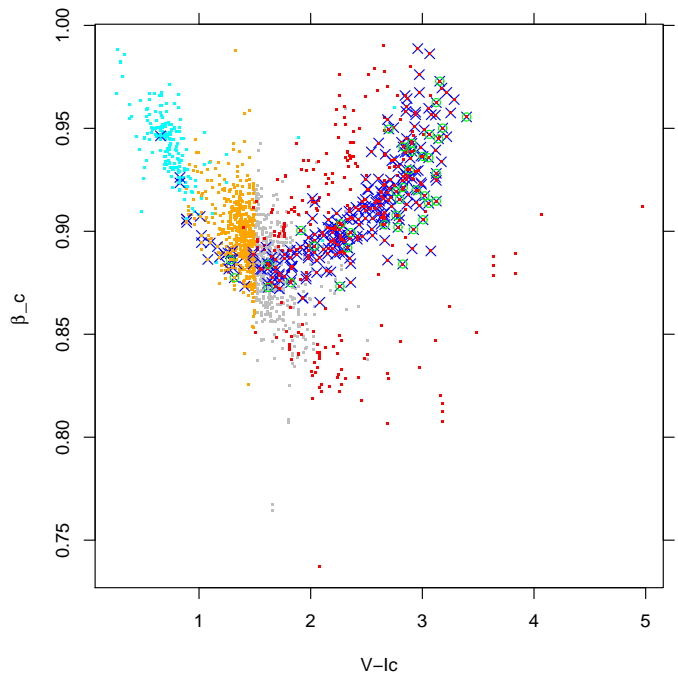}}
\caption{The "quintet core" index $\beta_c$ vs.\ $V-I_c$, showing the
strong gravity-related differentiation of late-type stars for $V-I_c>1.5$.
\label{qrtc3b-vi}}
\end{figure}

We come now to consider the red-colored spectra shown in
Fig.~\ref{atlas-quartet}: in general, they exhibit overall similarities
to the respective overplotted black-colored spectra,
chosen of similar effective temperatures, as estimated from the $\mu$
index for late-type stars and overall spectrum appearance for hotter
stars. The "black" and "red" series
of spectra belong respectively to the lower and upper branches
in Figure~\ref{mb-mb} (lower-right panel), which were argued in
Sect.~\ref{mol-bands} to correspond to cluster (and MS)
and giant stars, respectively. Therefore, the comparison between
overplotted black and red spectra in Fig.~\ref{atlas-quartet} enables us
to study {\rm gravity effects} in the "quintet" lines. It is immediately
clear that most of the quintet lines are highly sensitive to gravity:
in particular, the two Ca~I lines become {\rm weaker} in giants than in
MS stars (high-gravity lines), while on the contrary the Ba~II and Fe~I
$\lambda$6498.94 are {\rm enhanced} in giants (low-gravity lines).
The Fe~I $\lambda$6494.98 behaves in an intermediate way, as
a high-gravity (henceforth high-g) line in G~stars
(bottom spectrum in Fig.~\ref{atlas-quartet})
while as a low-gravity (low-g) one in M stars (top two spectra in Figure). Since
high-g lines in G~stars are relatively rare in our wavelength
range, we put this line into the high-g line group. Comparing the
MS-star spectra (black) with overplotted giant-star spectra (red) shown
in the Figure (especially the topmost ones), it is evident that in MS
stars the quintet is dominated by the outer (Ca~I) lines, while in
giants it is dominated by the inner, "core" lines. This simple
observation permits to use this line group as a quick and easy mean to
visually discriminate between dwarfs and giants in the low-mass range.
Accordingly, we have defined a new "quintet core" index $\beta_c$, this
time using only the yellow-colored region in Fig.~\ref{atlas-quartet} (see also
Table~\ref{band-def}). Since the high-g Ca~I $\lambda$6499.65
lines is very close to the low-g Fe~I $\lambda$6498.94 line, the
boundaries of the "core" region have been set with great care.
To compute the $\beta_c$ index we have added to
low-g lines in the quintet "core" region also the Co~I
$\lambda$6490.34 and Ti~II $\lambda$6491.56 lines falling in the
yellow-colored region, blueward of the quintet proper, whose dependence
on gravity is also evident from Fig.~\ref{atlas-quartet}.
In order to obtain the best
results for the G-star range, a weighted-mean index was found to be most
useful, of the form
\begin{equation}
\beta_c = (7/8 \; \times <f^A> + 1/8 \; \times <f^{A^1}>)\; /<f^B>,
\label{qrtc-define}
\end{equation}
where the $A$ and $A^1$ wavelength ranges refer respectively to the
quintet "core" and to the blueward, additional region, respectively.
The $\beta_t$ and $\beta_c$ values are reported in Table~\ref{table2}.
The observed dependence of the $\beta_c$ index on $V-I_c$ color is shown
in Figure~\ref{qrtc3b-vi}. An extremely clear separation can be seen between
cluster stars and giants, especially in the M-star range, with the
former stars having always $\beta_c>0.87$, while giants having $\beta_c$
values lower by 0.05-0.1 or more. But this diagram shows also a most
interesting separation between $\beta_c$ values of cluster members and
other M stars {\rm on the opposite side with respect to giants}: the
most plausible interpretation of this separation is that the $\beta_c$
index is able to {\rm tell apart the slight gravity difference between MS and
PMS stars}, in the M-star range.
To our knowledge, gravity differences between MS and PMS late-type stars
have never been measured before {\rm using this wavelength range}. This
has been done, instead, using better-known gravity diagnostics such as
the Na~I lines at 8183-8195~\AA\ (e.g., Mohanty \e 2004, Lyo \e 2004,
2008, Riddick \e 2007, Lawson \e 2009).
A thorough investigation of the diagnostic capabilities which may be
offered by the "quintet" lines is therefore most interesting.

\begin{figure}
\resizebox{\hsize}{!}{
\includegraphics[bb=20 10 465 475]{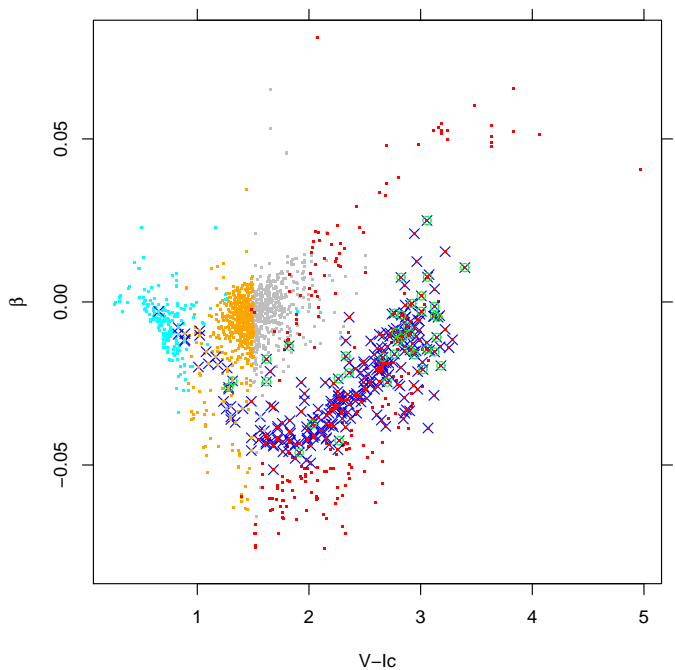}}
\caption{The quintet "core/edge" ratio index $\beta$ vs.\ $V-I_c$, which
best defines gravity differentiation at all spectral types.
\label{beta-vi}}
\end{figure}

The contrasting properties of low- and high-g lines in the quintet
are best put into evidence by using a "relative difference" index, which
we now define as
\begin{equation}
\beta = (\beta_t-\beta_c)/\beta_t,
\label{beta-define}
\end{equation}
and whose dependence on $V-I_c$ is shown in Figure~\ref{beta-vi}.
This Figure, where giants are now placed above the MS/PMS stars, shows again
clearly the MS/PMS separation for M stars with $V-I_c<2.6$, but also that the
index $\beta$ is able to discern MS from PMS stars even through the
intermediate-type
range (orange dots), with higher-gravity MS GKM stars describing a locus
in the diagram below that occupied by cluster members of same $V-I_c$ colors.
The $\beta$ values are reported in Table~\ref{table2}.
Maximum 1$\sigma$ errors (at minimum S/N=15) for $\beta_t$, $\beta_c$,
and $\beta$ are
0.01, 0.013, and 0.012, respectively. It is worth remembering that
such relatively large errors are only appropriate for the lowest-mass
cluster members with lowest S/N values, while for most
other stars in our dataset, especially those of early types, S/N values
are much higher than 15, being the median S/N $\sim 65$.
Note also that the quintet spectral region may be affected by many
telluric lines (Fig.~\ref{telluric}), and their subtraction is therefore
particularly important to compute $\beta_t$ and $\beta_c$ most accurately.

Some warnings about the $\beta_c$ (and thus $\beta$) index are however
necessary: first, while the total quintet index $\beta_t$ is not very
sensitive to fast rotation, the closeness in wavelength between high-
and low-g lines used to compute $\beta_c$ makes that index rather sensitive
to fast rotation (above $\sim 30-50$ km/s),
since in this case adjacent rotation-broadened lines tend
to merge, and cross-contamination between the low-g and high-g bands
make the $\beta_c$ index approach its "neutral" value of unity (the
neutral value of $\beta$ is instead zero). This rotation effect may be partially
responsible for the large scatter of $\beta_c$ near $V-I_c \sim 3$ for
cluster members (Fig.~\ref{qrtc3b-vi}), since these stars are mostly
fast rotators (Fig.~\ref{vrot-vi}).
Moreover, the Ba~II $\lambda$6496.897 line was found to be enhanced in
young star clusters, and might be affected by NLTE effects,
especially in young stars with strong chromospheric activity (D'Orazi \e
2009b, 2012, and references therein), and this may also contribute
to the scatter seen in Figs.~\ref{qrtc3b-vi} and~\ref{beta-vi}.
Second, the Ba~II $\lambda$6496.897 line is one of the most gravity sensitive
lines in the quintet "core" region,
but it may be strongly enhanced also in the so-called Barium stars, leading to
completely spurious values for the $\beta_c$ index. These stars are
however rare, and none appears to be present in our dataset. Since they
are known to have high abundances not only of barium but also of other
s-process elements such as strontium and yttrium, examination of lines
of these elements (a few of which are found in the HR15n range) is
needed to determine whether a star with a strong Ba~II line is a
Barium star or a low-gravity star.
Finally, indices $\beta_t$, $\beta_c$, and $\beta$ are expected to be highly
sensitive to metallicity, as will be discussed in Sect.~\ref{elodie}.

\begin{figure}
\resizebox{\hsize}{!}{
\includegraphics[bb=20 10 465 475]{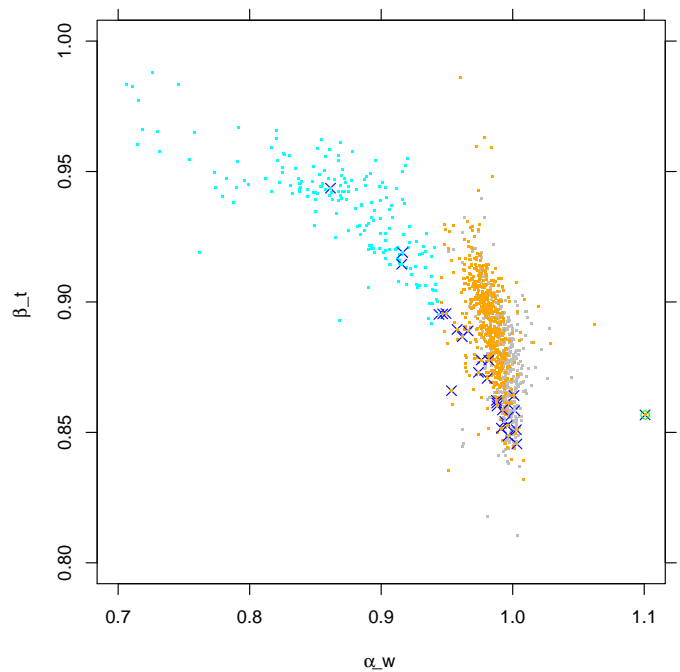}}
\caption{Index $\beta_t$ vs.\ the temperature-sensitive \ha\ wings index
$\alpha_w$. Late-type stars (comprising most emission-line CTTS)
are not plotted.
\label{qrt3-hawc}}
\end{figure}

Having already defined temperature indicators such as the $\alpha_w$ and
$\mu$ indices, it is now possible to compare the "quintet" indices with
them, in completely reddening-free representations. A
$(\beta_t,\alpha_w)$ diagram is shown in Figure~\ref{qrt3-hawc},
excluding late-type (M-type) stars for clarity.
This Figure shows that the decrease in $\beta_t$ with $\alpha_w$ (thus,
with temperature) is not uniform, but shows a sort of plateau, with a
non-negligible spread of datapoints; this might be due to metal
abundance effects, and/or to gravity. Candidate cluster members fall,
with one exception, well within the main band comprising most stars in
the sample, and fit comfortably the picture where deviation from this
main locus are only due to gravity or metallicity, both of which are
expected to be (nearly) constant for cluster stars in this mass range.
The locus of orange datapoints (mostly giant stars) above the position
of cluster stars might be ascribed to both low-gravity effects, or to
reduced metallicity. This latter effect is almost surely responsible for
the few orange datapoints with $0.95<\alpha_w<1$ and $\beta_t>0.94$. A
study of these interesting outliers, and of similar ones below the main
band in the Figure, is however beyond the scope of this work, and will
be made in a future work; some insight on their nature will be
nevertheless gained from suitable reference spectra in Sect.\ref{uvespop}.

\begin{figure}
\resizebox{\hsize}{!}{
\includegraphics[bb=20 10 465 475]{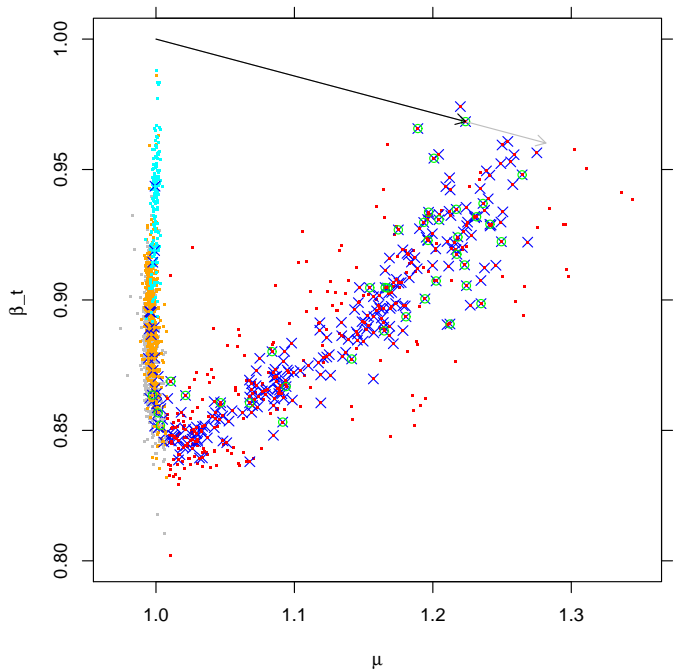}}
\caption{Index $\beta_t$ vs.\ the temperature-sensitive TiO index $\mu$.
Heavily veiled CTTS, if present, would fill the empty triangular region,
converging toward the point $(1,1)$.
The procedure for determination of veiling $r$ is exemplified by the comparison
between the black arrow and its gray prolongement; see text for details.
At most two-three moderate-veiling stars may be present in this dataset,
near $\mu \sim 1.2$.
\label{qrt3-mbm}}
\end{figure}

Next, Figure~\ref{qrt3-mbm} shows a $(\beta_t,\mu)$ diagram describing
the behavior of late-type stars in a reddening-independent way. While
low S/N is a limiting factor for the reddest stars, the Figure shows
nevertheless a practically empty triangular region, with only a few
non-member stars in it. We recall that the neutral value of both
$\beta_t$ and $\mu$ is unity, i.e.\ the value approached only by the
most massive stars in the dataset. This value would also be approached
by heavily veiled PMS stars, which are to be found (exclusively) among
CTTS (green circles): the fact that CTTS datapoints in this cluster stay
away from the central part of this diagram, but on the contrary follow
the same locus as other, non-CTTS, cluster members, implies that no
significant veiling is present in their spectra. The only possible
exception are the 2-3 CTTS with $\beta_t>0.95$, for which at most a
small veiling factor $r \sim 0.2$ may be derived.
The procedure for quantitative determination of veiling $r$ is
exemplified in the Figure, by the comparison
between the black arrow (joining the neutral point with the measured
position in the diagram for a CTTS) and its gray prolongement
(extrapolated up to the cluster locus):
the length ratio between them is 1:$r$ (see equation~\ref{mu-veil}).
It should be noted that the procedure just outlined becomes unreliable
(and overestimates $r$) for
very fast-rotating stars, whose indices are compromised as already explained.

\begin{figure}
\resizebox{\hsize}{!}{
\includegraphics[bb=20 10 465 475]{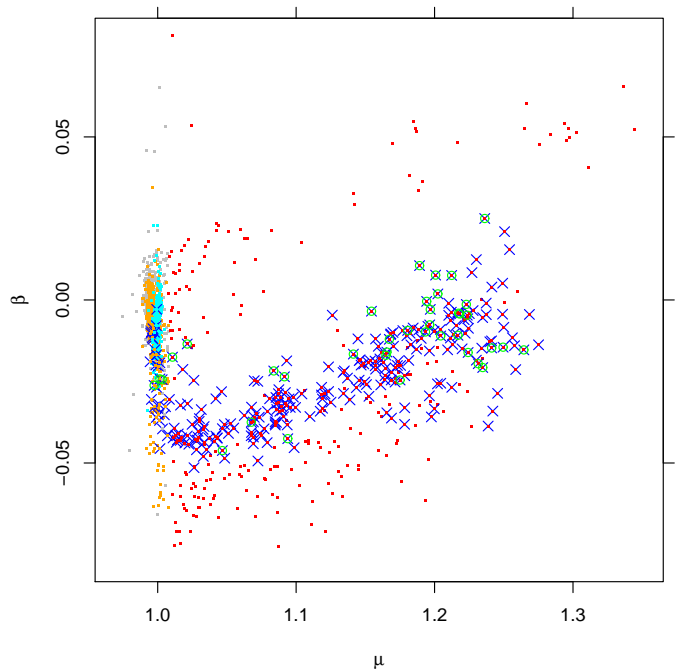}}
\caption{Gravity index $\beta$ vs.\ $\mu$. This reddening-free diagram
shows clearly the difference between gravities of MS and PMS late-type stars,
and even more between these and giant stars.
\label{beta-mbm}}
\end{figure}

\begin{figure}
\resizebox{\hsize}{!}{
\includegraphics[bb=20 10 465 475]{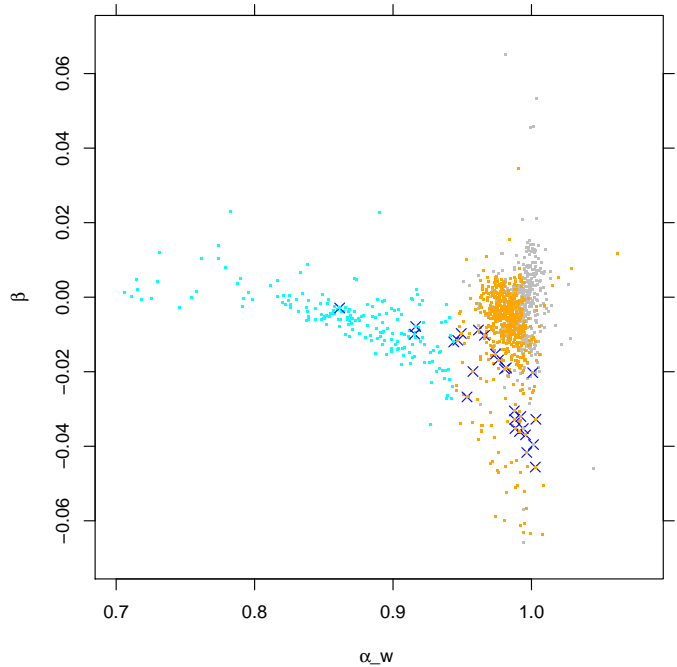}}
\caption{Gravity index $\beta$ vs.\ $\alpha_w$, not showing late-type
stars. Difference in gravities between intermediate-type MS, PMS, and giant
stars are also clear from this diagram.
\label{beta-haiwc3}}
\end{figure}

\begin{figure}
\resizebox{\hsize}{!}{
\includegraphics[bb=20 10 465 475]{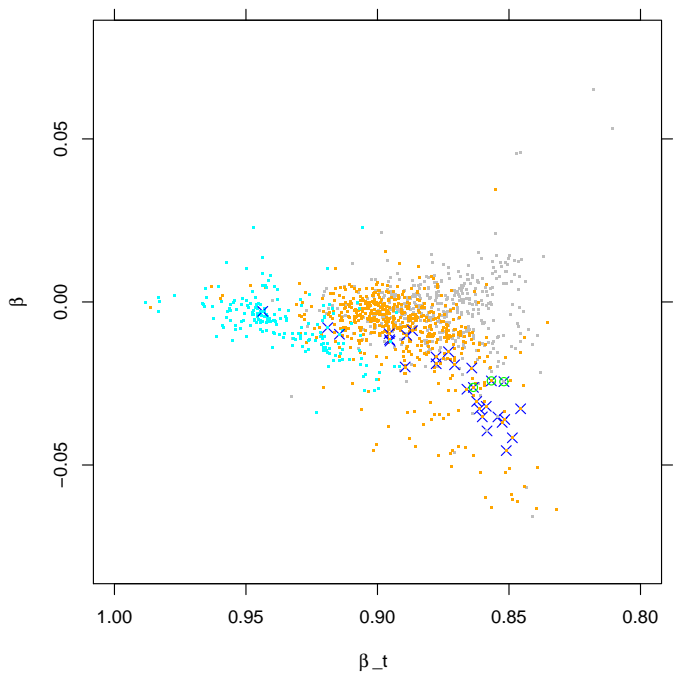}}
\caption{Index $\beta$ vs.\ total quintet index $\beta_t$, the latter
measuring temperature in intermediate-type stars.
\label{beta-qrt3}}
\end{figure}

We now turn our attention to the gravity index $\beta$, as a function of
different temperature indices. First, a $(\beta,\mu)$ diagram is shown
in Figure~\ref{beta-mbm}, which again shows the good separation between MS,
PMS, and (especially) giant stars already seen in Fig.~\ref{beta-vi}, but here
in a reddening-independent way. While reddening is not of great concern for
the "clean" case of \gam\ stars, it is of great importance and nuisance
for many other (younger) clusters such as Cha~I, and therefore we devote some
attention to build reddening-independent diagnostics, with a more general
purpose in mind than the study of \gam\ alone.
That the power of the $\beta$ index as a gravity diagnostics extends to
intermediate-type stars as well as to M stars is demonstrated by the
$(\beta,\alpha_w)$ diagram of Figure~\ref{beta-haiwc3} (excluding M stars).
The MS locus is clearly seen, starting at $(\beta,\alpha_w)=(0,0.7)$ and
bending downwards (orange datapoints) in a regular way;
the bulk of giants (orange/gray)
remain instead near $\beta \sim 0$, well above MS stars with the same
$\alpha_w$ index. Cluster stars of intermediate type fall in between, where they
are expected to lie, just below the envelope of giant stars (i.e.,
with \logg\ just above them, see Fig.~\ref{teff-g-isochr}).
The few orange and gray outliers with $\beta>0.03$ are again interesting
peculiar stars, with strongly enhanced low-g lines, which will be studied
elsewhere.

Since the $\alpha_w$ index may not be a good \teff\ indicator for
intermediate-type
stars, we use instead the $\beta_t$ index in this mass range.
Figure~\ref{beta-qrt3} is a $(\beta,\beta_t)$ diagram, again excluding M
stars. Again a separation between MS, PMS, and giant stars is observed,
with PMS cluster stars distinctly above higher-gravity MS stars, which
confirms the usefulness of such representation. Some
mixing can be seen between giants and earlier-type MS stars (as well
as PMS stars) and might be due to a dependence of $\beta_t$ on
metallicity (or \logg) as well as on \teff.

\subsection{Temperature-sensitive features near 6630 \AA}
\label{tg}

\begin{figure}
\resizebox{\hsize}{!}{
\includegraphics[bb=20 10 465 475]{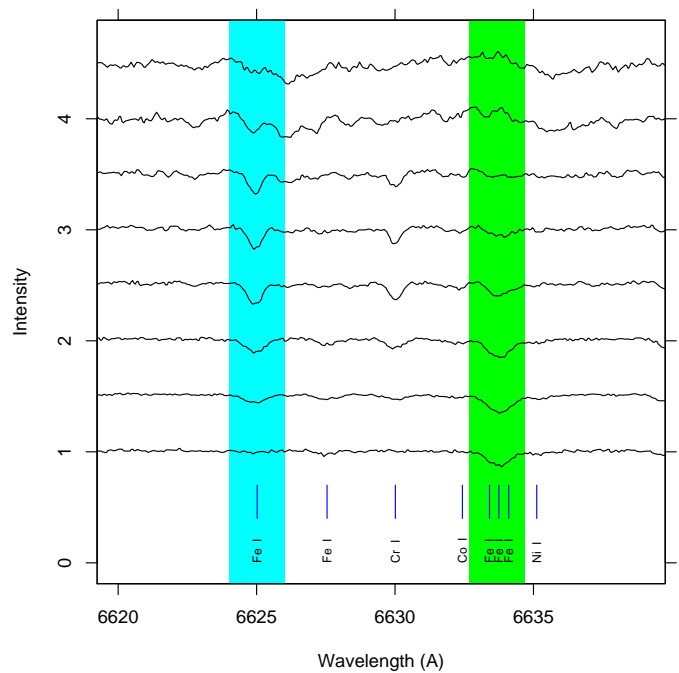}}
\caption{Atlas of spectra of intermediate-type and late-type cluster members
in the range 6620-6638\AA,
showing the wavelength ranges used to define our index $\zeta_1$,
containing Fe~I lines and also sampling a TiO molecular jump, as
evident from the two topmost spectra. Spectra form a temperature
sequence, hotter to the bottom.
The green (cyan) band indicates the range used for high- (low-)
temperature lines.
\label{atlas-tg}}
\end{figure}

\begin{figure}
\resizebox{\hsize}{!}{
\includegraphics[bb=20 10 465 475]{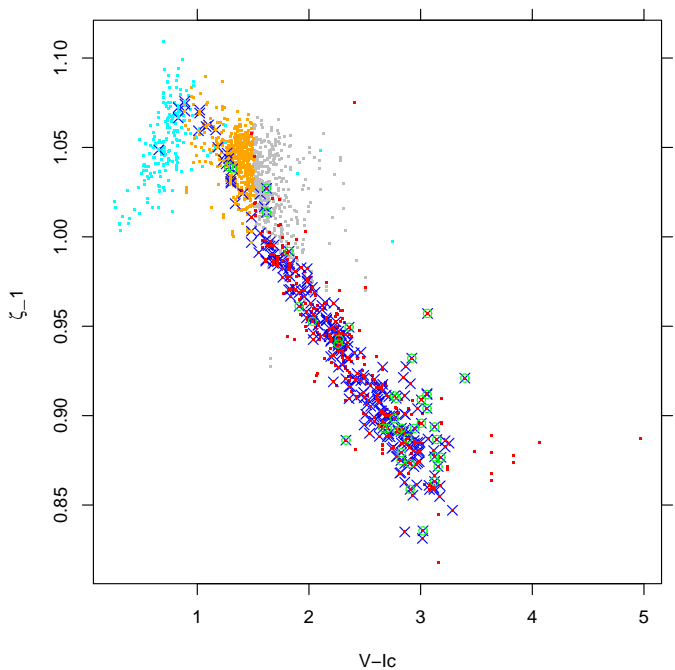}}
\caption{Behavior of the $\zeta_1$ index vs.\ $V-I_c$. Note the strong,
monotonic temperature dependence from early-G types down to the
latest-type sample stars.
\label{tg-vi}}
\end{figure}

We have then considered a set of additional lines, useful to compute a
further index, sensitive to temperature in intermediate-type stars. Not many
lines are found in the HR15n range to this aim, since most lines
exhibit an intensity peak in this \teff\ range and are no good temperature
indicators. The lines chosen here are all Fe~I lines, $\lambda\lambda$6625.022,
6633.412, 6633.749, and 6634.106. Figure~\ref{atlas-tg} shows the
relevant spectral region for a set of cluster member stars, with
temperatures increasing from top to bottom according to $V-I_c$ color (the three
lowest spectra are GK stars, the topmost five are M stars).
We compute an index $\zeta_1$ as
\begin{equation}
\zeta_1 = <f^A>/<f^B>,
\label{tg-define}
\end{equation}
with $A$ and $B$ ranges defined in Table~\ref{band-def}. We see that the
$\lambda$6625.022 line has a strongly different temperature sensitivity
than the other three (unresolved) Fe~I lines near 6634\AA. In the
coolest stars (top two spectra in Figure) all these lines have largely
disappeared, but in their place the two extremes of a molecular TiO band
are found: this chance coincidence has the favorable effect that the
trend exhibited by the $\zeta_1$ index with $V-I_c$ color continues
monotonically beyond the intermediate-type stars, until the latest spectral
types considered. This is
shown in Figure~\ref{tg-vi}: after a short rise in the early-type range,
this index attains its maximum values for early-G stars (cyan-orange
dots transition), to decrease sharply and monotonically throughout all
lower-temperature stars. No pronounced gravity effects are apparent: for
instance, no separate MS and giant branches are identifiable in the Figure.
Metallicity effects can be expected, but are not obvious from the Figure.
The temperature dependence of the $\zeta_1$ index for intermediate-type stars
is therefore useful to reinforce temperature indications obtained from
the $\beta_t$ index. It is also worth noting that the TiO band giving
the $\zeta_1$ index its extended cool-star temperature sensitivity was
{\rm not} included among those sampled by our $\mu_n$ indices, and
therefore this new index provides also a new, independent \teff\
diagnostics for M stars as well.
Maximum 1$\sigma$ errors on $\zeta_1$ are estimated to be 0.018, for the
latest-type cluster members.
Being defined using narrow wavelength intervals, the $\zeta_1$ index
is affected by fast rotation above $v \sin i \sim 50$ km/s.
The $\zeta_1$ values are reported in Table~\ref{table2}.

\begin{figure}
\resizebox{\hsize}{!}{
\includegraphics[bb=20 10 465 475]{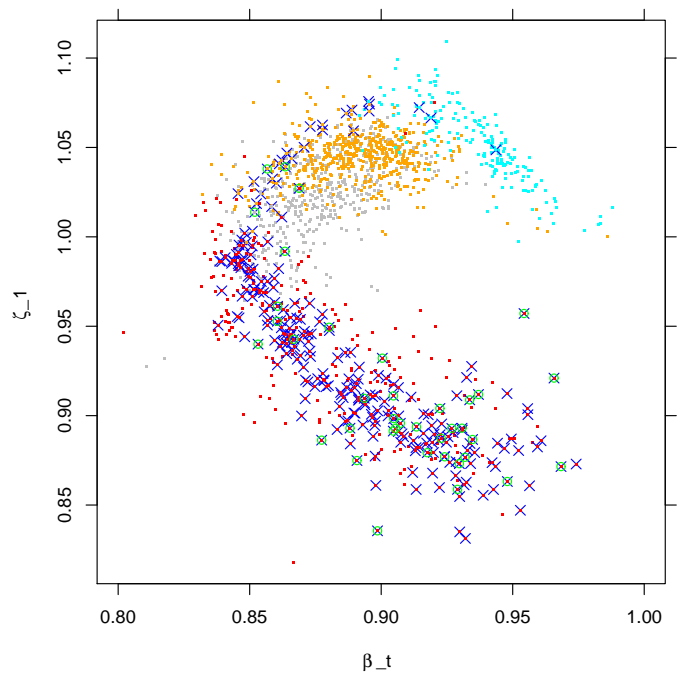}}
\caption{A $\zeta_1$ vs.\ $\beta_t$ diagram. MS and PMS stars define
a characteristics C-shaped pattern. Veiled CTTS can be easily recognized
since they would depart from that locus toward point $(1,1)$, and their
veiling $r$ would thus be measured. Again, only two such stars are
present in the sample.
\label{tg-qrt3}}
\end{figure}

The fact that the $\beta_t$ index {\rm departs} increasingly from
unity, going from early-G to early-M stars, while $\zeta_1$ {\rm
approaches} unity across the same spectral-type range, has the
interesting practical consequence that the comparison between these two
indices may be used as an effective {\rm veiling diagnostic}.
This is demostrated in Figure~\ref{tg-qrt3}: in this $(\zeta_1,\beta_t)$
diagram, cluster datapoints follow a very characteristic C-shaped
pattern, while the locus of giants is slightly offset. Apart from
early-type stars (cyan dots), or likely metal-poor giants (orange dots
mixed with cyan dots), no other stars fall near the neutral point
$(\zeta_1,\beta_t)=(1,1)$. Veiled stars would depart at almost
right-angles from the PMS locus,
outlined by non-CTTS cluster stars, toward this neutral point, by an
amount proportional to veiling $r$ (see equation~\ref{mu-veil}). The
C-shaped pattern means that veiling-related effects are non-degenerate
with respect to temperature effects, for intermediate- and late-type stars in a
cluster. In \gam, this plot shows again that veiling may be present at
most in two cluster CTTS, with values between $r \sim 0.2-0.5$.
The independent veiling measurements obtainable from the
$(\zeta_1,\beta_t)$ and $(\beta_t,\mu)$ diagrams (see
Fig.~\ref{qrt3-mbm}) can also be of great help in estimating errors on
veiling $r$.

\subsection{Gravity-sensitive features in the 6750-6780 \AA\ region}
\label{lowg}

\begin{figure}
\resizebox{\hsize}{!}{
\includegraphics[bb=20 10 465 475]{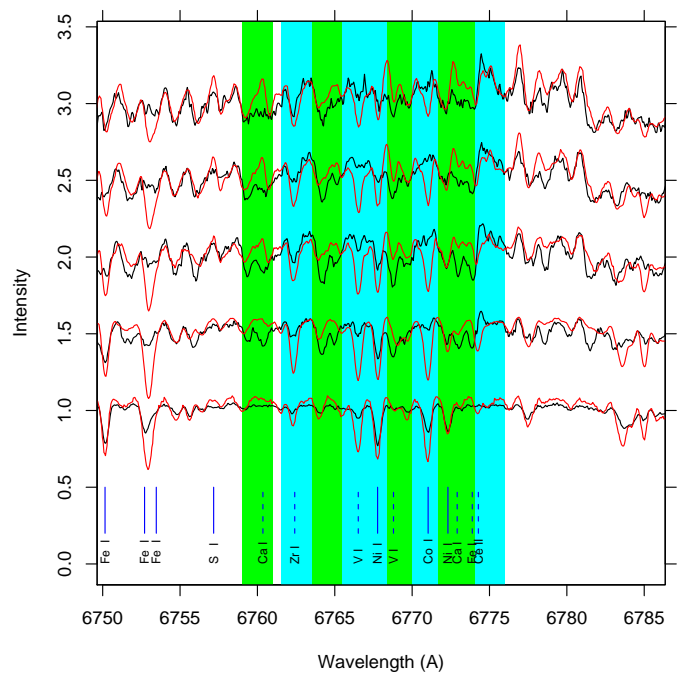}}
\caption{Atlas of MS (black) and giants (red) spectra in the wavelength
range 6750-6785\AA, used to define the $\gamma_1$ gravity-sensitive
index. Lines with better-known parameters are indicated at the bottom
with solid blue segments, while some lines with less certain parameters
are indicated with dashed segments. Strongly gravity-dependent (low-g)
lines are evident. Cyan- and green-colored wavelength ranges are used to
sample respectively low-g and high-g lines in the $\gamma_1$ index definition.
\label{atlas-lowg}}
\end{figure}

\begin{figure}
\resizebox{\hsize}{!}{
\includegraphics[bb=20 10 465 475]{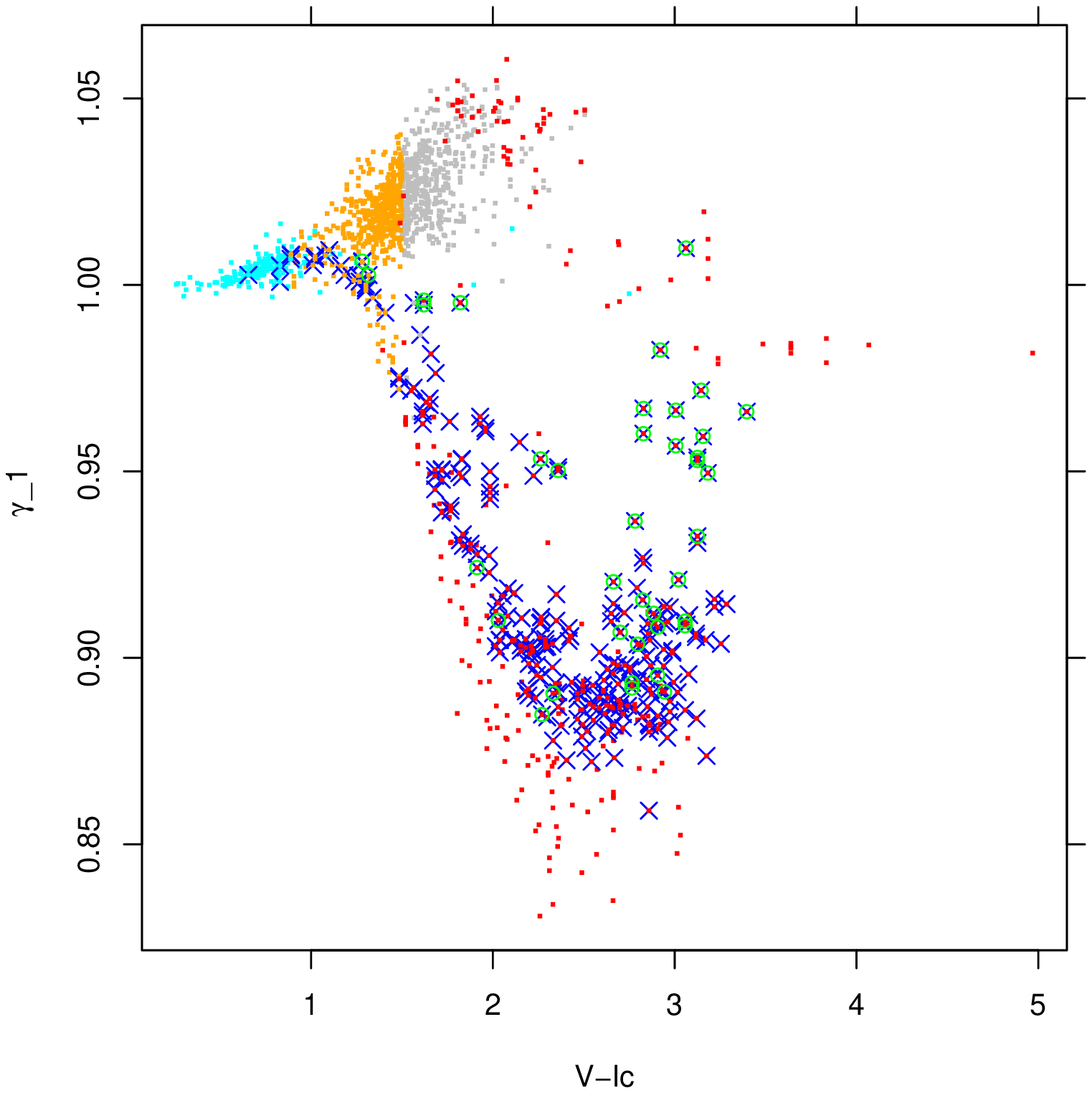}}
\caption{The $\gamma_1$ gravity-sensitive index vs.\ $V-I_c$ color,
showing the extreme difference in $\gamma_1$ values between giants and MS
stars, and also the MS/PMS separation.
\label{lowg-vi}}
\end{figure}

Now we consider a wavelength region in the reddest part of the HR15n
range, where many very interesting gravity-sensitive lines are found.
Figure~\ref{atlas-lowg} shows a number of spectra for intermediate- and
late-type MS stars (black) and giants (red), in a temperature sequence
(increasing from top to bottom). We see that there are extreme differences in
individual lines between MS stars and giants, despite the same
temperatures as measured from the TiO bands (and the $\mu$ index).
Most striking is the almost complete suppression of the
$\lambda\lambda$6766,6771 lines in MS stars, while the same lines are
quite strong in giants. Low-g lines are also found at
$\lambda\lambda$6762,6767.5,
although not as strongly affected as the former two.
The identification of some of these lines is not certain: the Co~I
$\lambda$6771.033 and Ni~I $\lambda$6767.768 lines are identified with
better confidence, while the identification with Zr~I $\lambda$6762.413
and V~I $\lambda$6766.519 is much less certain (as are the atomic parameters
of the two latter lines); other lines may also contribute to the
observed absorption features.

We define a new index which takes advantage of these strongly
gravity-sensitive lines, from the flux ratio between the low-g line
ranges (cyan-colored wavelength ranges in Fig.~\ref{atlas-lowg}),
and contrasting high-g features nearby (green-colored ranges in the
Figure). The new $\gamma_1$ index is then defined as usual
\begin{equation}
\gamma_1 = <f^A>/<f^B>,
\label{gg-define}
\end{equation}
with the $A$ (high-g lines) and $B$ (low-g lines) wavelength regions
defined in the Figure and in Table~\ref{band-def}. Because of the
closeness between the high-g and low-g ranges (in fact, they alternate
in a comb-like pattern) and their relative narrowness, the $\gamma_1$
index becomes sensitive to rotation (like the $\beta$ index):
high-g and low-g contributions tend to mix together, and the
index tends toward its neutral unity value for fast rotators.
Maximum statistical 1$\sigma$ errors on $\gamma_1$ are on the order of 0.015.
The $\gamma_1$ values are reported in Table~\ref{table2}.

The behavior of $\gamma_1$ vs.\ $V-I_c$ color is shown in
Figure~\ref{lowg-vi}: the extreme MS vs.\ giants differences visible in
Fig.~\ref{atlas-lowg} are here mirrored by the huge gap between
late-type giants and MS stars. Again, PMS stars form a band above that
of MS stars, up to $V-I_c \sim 3$, although the distinction is not as clear as
when using $\beta$ in the late-K range (transition between orange and red dots).
The rotation effect on $\gamma_1$ may be seen for some fast-rotating CTTS
stars near $V-I_c \sim 3$, which tend to mix with giants as their $\gamma_1$
index is "diluted" toward the neutral value of unity.

\begin{figure}
\resizebox{\hsize}{!}{
\includegraphics[bb=20 10 465 475]{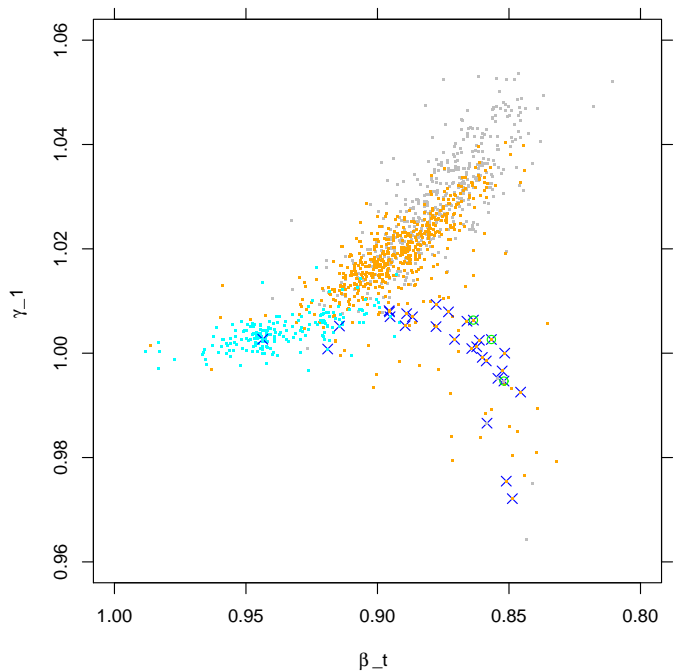}}
\caption{The $\gamma_1$ vs.\ $\beta_t$ diagram, showing the
effectiveness of $\gamma_1$ to discriminate gravities between MS, PMS,
and giants also in the case of intermediate-type stars, with $\beta_t$ acting as
a temperature index.  Late-type stars are not shown.
\label{lowg-qrt3}}
\end{figure}

A reddening-free diagram of $\gamma_1$ vs.\ $\beta_t$ is useful to
better inspect the index behavior for intermediate-type stars, and is shown in
Figure~\ref{lowg-qrt3}: the distinction between giants (with $\gamma_1$
increasing with lower temperatures), PMS stars ($\gamma_1$ first flat, then
mildly decreasing), and MS stars ($\gamma_1$ decreasing with
lower temperatures) is very noticeable, and confirms the usefulness of this
index.

\subsection{Global temperature and gravity indices}
\label{tau-gam}

\begin{figure}
\resizebox{\hsize}{!}{
\includegraphics[bb=20 10 465 475]{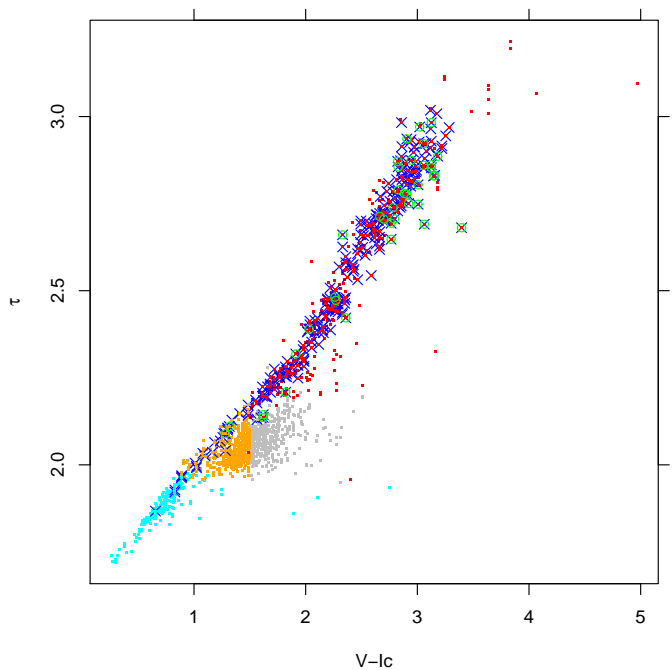}}
\caption{Composite temperature index $\tau$ vs. $V-I_c$, for all sample
stars.
\label{tau1-vi}}
\end{figure}

In the previous subsections we have introduced four temperature-sensitive
indices: the \ha\ wings index $\alpha_w$ for early-type stars, the total
"quintet" intensity index $\beta_t$ and the Fe~I ratio index $\zeta_1$
for intermediate-type stars, and the TiO band index $\mu$ for late-type
stars. Moreover, the quintet "core/edge" index $\beta$ and the
$\gamma_1$ index have been shown to be very good gravity indicators.
Here we attempt to build more global temperature and gravity indices,
starting from those already introduced.

We define a new composite index $\tau$ as
\begin{equation}
\tau = min(\alpha_w,1) - \beta_t - \zeta_1 + 3 \times \mu.
\label{tau-define}
\end{equation}
The $\alpha_w$ index is used here only for its sensitivity to the
absorption wings of early-type stars, while in cases where $\alpha_w>1$
(fast-rotating WTTS and especially CTTS) it is "reset" to unity,
since in those cases it is no temperature
indicator. In the three broad spectral-type ranges we have studied,
individual contributions to index $\tau$ behave as follows:
\begin{itemize}
\item For early-type stars, $\alpha_w$ is the main contributor to
$\tau$; the contrasting contributions of $\beta_t$ and $\zeta_1$ tend to
cancel each other out, and $\mu$ is flat and irrelevant.
\item For intermediate-type stars, $\alpha_w$ and $\mu$ are flat, and
contributions from $\beta_t$ and $\zeta_1$ reinforce each other.
\item For late-type stars, $\alpha_w$ is flat, and the
strong contributions by (three times) $\mu$ and (minus) $\zeta_1$ more
than overcompensate the opposite contribution by $\beta_t$.
\end{itemize}
Maximum 1$\sigma$ errors on $\tau$ are on the order of 0.026.  The net
result is shown in Figure~\ref{tau1-vi}: the strong, direct, almost
linear correlation with $V-I_c$ across the whole sampled color range
leaves little doubt about the real usefulness of the $\tau$ index just
defined. It also argues in favor of a nearly uniform reddening for \gam\
cluster stars (except perhaps for the latest-type cluster members, as already
noted), without which no such narrow correlation would have been obtained.
The two CTTS falling below the main band near $V-I_c \sim 3$ are those
stars for which a non-zero veiling was suspected from previous diagrams;
Fig.~\ref{tau1-vi} helps to confirm this fact once again, although in
this latter diagram also a high reddening may have a similar effect.

We have noted in the relevant sections that all indices used in the
definition of $\tau$ are almost entirely independent of gravity, with
some dependence expected only for $\beta_t$. They are also relatively
unaffected by rotation, at least if it does not exceed $\sim
50-90$~km/s. We expect therefore $\tau$ to be a quite robust temperature
indicator, across the spectral type range sampled by our dataset (mid-A
to mid-M), at least for MS and PMS stars whose gravities are not too
dissimilar.

\begin{figure}
\resizebox{\hsize}{!}{
\includegraphics[bb=20 10 465 475]{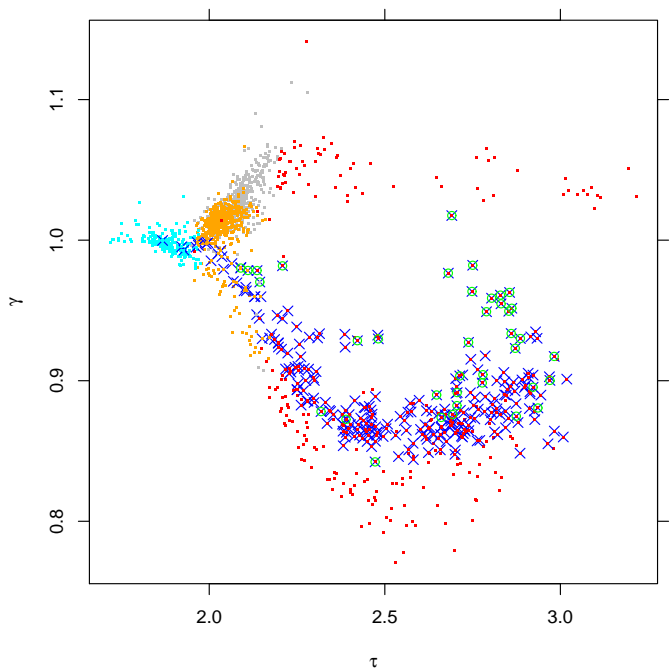}}
\caption{Composite gravity index $\gamma$ vs.\ temperature index $\tau$,
for all sample stars. The gravity difference between MS and PMS stars is
best studied by using this reddening-free diagram encompassing all
spectral types.
\label{gam1-tau1}}
\end{figure}

\begin{figure}
\resizebox{\hsize}{!}{
\includegraphics[bb=20 10 465 475]{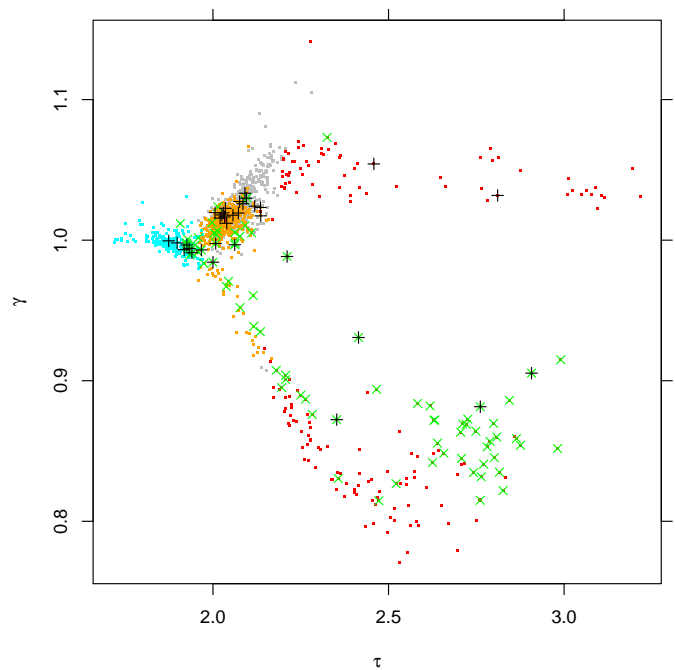}}
\caption{Same as Fig.~\ref{gam1-tau1}, but omitting cluster candidate
members, to better appreciate the locus of MS stars alone.
Stars with lithium EW $>150$~m\AA\ are indicated with black plus
signs, while those with narrow, chromospheric \ha\ emission with green
crosses.
\label{gam1-tau1-field}}
\end{figure}

We also define a composite index $\gamma$ using previous
gravity-sensitive indices as
\begin{equation}
\gamma = \beta + \gamma_1,
\label{gam-define}
\end{equation}
where the two indices $\beta$ and $\gamma_1$ reinforce each other as
gravity indicators.
Maximum 1$\sigma$ errors on $\gamma$ are on the order of 0.019.
The $\tau$ and $\gamma$ values are reported in Table~\ref{table2}.
Using now directly $\tau$ as a temperature proxy,
rather than the $V-I_c$ color, we obtain finally a most interesting
{\rm $(\gamma,\tau)$ diagram}, shown in Figure~\ref{gam1-tau1}.
We regard this (reddening-free) diagram as one of the most important
results of this work. The giants are very well separated from
higher-gravity stars, at all temperatures for which a separation was
expected, starting at early-G types (cyan-orange transition in the
Figure) and even more for colder stars.  Most importantly, the
cluster PMS stars (at 7-10~Myr) are quite clearly separated from
higher-gravity (by only 0.5-1.0 dex) main-sequence stars, also starting
at early-G types. As expected, very fast-rotating PMS stars (green
circles near diagram center) have a $\gamma$ index slightly compromised,
and approaching its neutral value of unity: they appear
as lower-gravity than they probably are.

To better appreciate the degree of separation between MS and PMS stars,
we plot in Figure~\ref{gam1-tau1-field} the same diagram, but without cluster
members. Moreover, we have added symbols for lithium-rich stars (Li
EW$>150$~m\AA, black plus signs), and for chromospherically-active stars
(from the \ha\ core index $\alpha_c$, green crosses). Five stars have
both lithium and chromospheric emission, and might be additional members
missed by our (preliminary) member-selection criteria; for instance, they
might be binary members that happen to lie outside the X-ray field of view.
Their position in the $(\gamma,\tau)$ diagram above the MS locus also suggests
their being PMS stars.  Many giants have
also moderately strong lithium lines. Except for the few stars just
mentioned, comparison between the diagrams of
Figs.~\ref{gam1-tau1} and~\ref{gam1-tau1-field} shows that the locus
occupied by PMS stars is practically devoid of MS stars up until $\tau
\sim 2.6$, and that some overlap between the MS and PMS samples at
redder colors (higher $\tau$) is limited to just a few stars, {\rm all
with strong chromospheric emission}. We cannot even rule out that these
latter are PMS stars, perhaps older than \gam\ stars and belonging to a
dispersed population.

Metal abundance variations among sample stars may be responsible for
some scatter of datapoints, in both the MS and giant-stars regions.
We also admit that this representation is purposely
aimed at characterizing low-mass near-MS stars, and does not do justice
to the wide variety of spectral features shown by GK giant stars, which
are found in a rather compressed region of the $(\gamma,\tau)$
diagram. Moreover, this latter giant-star region contains the diagram
neutral point at $(\gamma,\tau)=(1,2)$, where extremely metal-poor stars
are probably to be found. This emphasizes that this diagram is not
suitable to test for chemical abundances (which will be studied in
Sect.\ref{elodie}). Veiling also would have the
effect of making a star converge toward the neutral point: to correct
for this, veiling corrections should be made {\rm before} placing stars
in this diagram, with the techniques discussed in the previous sections.

Outliers in the $(\gamma,\tau)$ diagram above the giant-star locus, or
above the early-type star locus are the same stars which were outliers
in, e.g., the $(\beta,\beta_t)$ diagram of Fig.~\ref{beta-qrt3}. Due to
their lack in the studied sample, we could not test the performance of
our indices in the case of early-type PMS stars, such as the Herbig
Ae/Be stars: in these cases, \ha\ is mostly in emission, in which case
it cannot be used as a \teff\ indicator via the $\alpha_w$ index. This
latter would be probably high enough to signal the star as an accreting
star as for CTTS. Assuming that all other indices are unaffected with
respect to, e.g., normal A stars, we would find such a star as a (green)
datapoint near $(\gamma,\tau)=(1,2)$, and confirmation of these stars
nature will require a careful individual analysis of their spectrum.
Again, no such case is found in
the \gam\ sample in Fig.~\ref{gam1-tau1}. Testing these expectations can
only be done by using actual spectra of Herbig stars, however.

\begin{figure}
\resizebox{\hsize}{!}{
\includegraphics[bb=20 10 465 475]{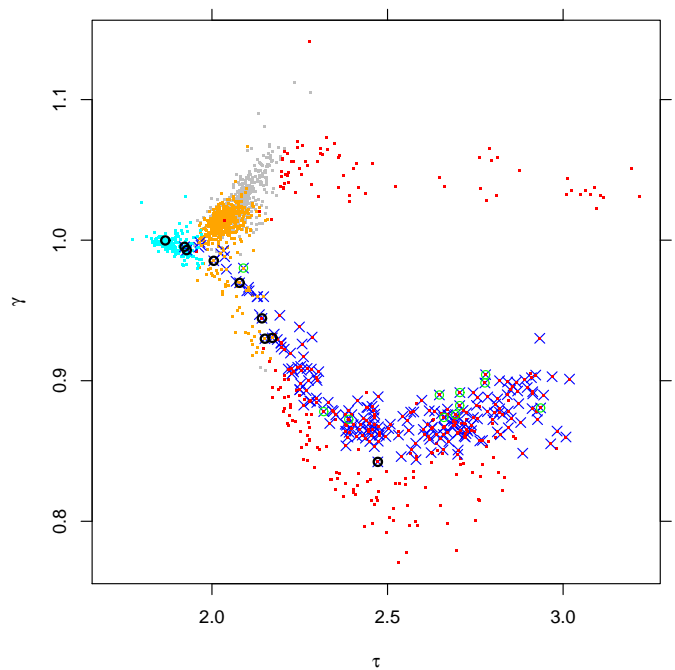}}
\caption{Same as Fig.~\ref{gam1-tau1}, but omitting all stars with $v
\sin i >30$~km/s. The spread of datapoints is significantly reduced with
respect to Fig.~\ref{gam1-tau1}, and the MS and PMS loci better defined.
Thick black circles indicate the position of six stars (nine individual
spectra), all candidate members, falling near the ZAMS in the CMD of
Fig.~\ref{v-vi-2}. They here appear to have intermediate gravities between
the MS and other cluster members.
\label{gam1-tau1-slow}}
\end{figure}

The usefulness of the $(\gamma,\tau)$ diagram is emphasized when we
consider only slowly-rotating stars ($v \sin i <30$ km/s), as we do in
Figure~\ref{gam1-tau1-slow}: here the PMS band becomes significantly
narrower than in the complete diagram; the average trend of $\gamma$
vs.\ $\tau$ is very well defined, distinctly for MS and \gam\ cluster stars, so
that a mapping between this plot and the $(T_{eff},\log g)$ plane
becomes a realistic possibility. In order to quantitatively do that, we need to
calibrate our $\gamma$ and $\tau$ indices, a problem which will be
examined in the next Sections. From an empirical point of view, however,
the $(\gamma,\tau)$ diagram may retain its usefulness: given the regular
progression, at a given temperature (or $\tau$), between the gravities
of MS, PMS, and giant stars, we may reasonably expect
that the sensitivity of $\gamma$ to gravity extends to
lower-gravity stars than the \gam\ PMS stars. Therefore, we expect that
stars in younger clusters than \gam, characterized by lower gravities,
will form a band at higher $\gamma$ values: in this way, the
$(\gamma,\tau)$ diagram may become an empirical {\rm age indicator} for
PMS stars. It is worth remarking that such a (relative) age estimate
would be unaffected by highly nonuniform reddening, as typical of very
young clusters, or distance uncertainties, being determined using purely
spectroscopic indicators. It may also be helpful as an additional
membership criterion, where other membership criteria may fail (e.g., absent
lithium for stars in the so-called lithium chasm;
inconsistent RVs for individual SB1 binaries; missing X-ray data, etc.).

\begin{figure}
\resizebox{\hsize}{!}{
\includegraphics[bb=20 10 465 475]{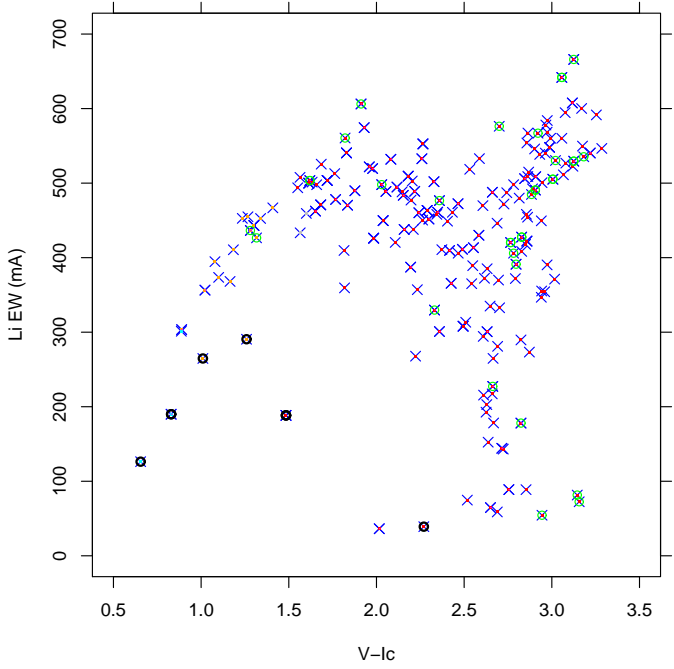}}
\caption{A lithium EW vs.\ $V-I_c$ diagram, for members only. Thick
black circles indicate the same spectra as in Fig~\ref{gam1-tau1-slow}.
\label{li-vi}}
\end{figure}

As a first, tentative test of the $(\gamma,\tau)$ diagram usefulness in
this latter respect, we remark that in Fig.~\ref{v-vi-2} six stars,
flagged as members, can be seen to lie below the main cluster locus,
which means nearer to the ZAMS if they lie at about the same distance as \gam.
Their placement in the
$(\gamma,\tau)$ diagram is shown in Fig.~\ref{gam1-tau1-slow} (bold
black circles): they {\rm all} lie at the interface between the MS and
the PMS bands, suggesting that their gravities (and ages) are consistently
intermediate between \gam\ members and MS stars. This is well consistent with
these stars having a young age, but nevertheless older than \gam\
members, and with their belonging to a more dispersed population of PMS
stars. To test this further, we show in
Figure~\ref{li-vi} a diagram of lithium EW\footnote{Lithium EWs from
release {\rm GESiDR1Final}, uncorrected for blends.} vs.\ $V-I_c$ color,
for PMS stars only
(Li-rich giants are therefore not included): while a clear Li-depletion pattern
can be seen (to
be studied in detail by Franciosini \e, in preparation), the six "older" stars
(bold circles) again appear to follow very consistently a different pattern,
which according to models is typical of significantly older stars, of
about 30-50~Myr (see, e.g., Jeffries \e 2003).
The information which can be obtained from our $(\gamma,\tau)$ diagram
on gravity and age of PMS stars is therefore highly coherent with other
evidence, of completely independent origin.

\section{Photometric calibration of index $\tau$}
\label{calibr}

\begin{figure*}
\includegraphics[bb=18 520 594 774,width=18cm]{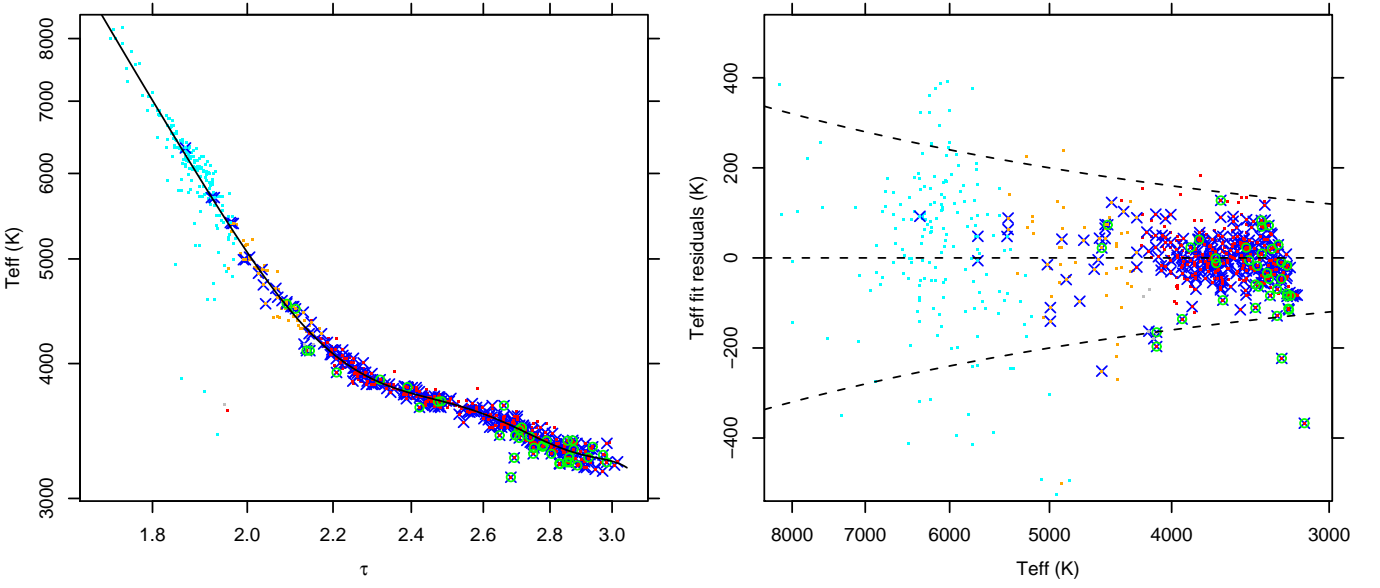}
\caption{Calibration between $\tau$ and \teff, as derived from observed
$V-I_c$ colors, average cluster reddening, and KH95 calibrations.
Left: \teff\ vs.\ $\tau$; the polynomial best-fit calibration is shown
as a solid line. Right: fitting residuals; datapoints outside the dashed
curves were not used in the fit.
\label{tau1-vi-calibr}}
\end{figure*}

\begin{figure}
\resizebox{\hsize}{!}{
\includegraphics[bb=20 10 465 475]{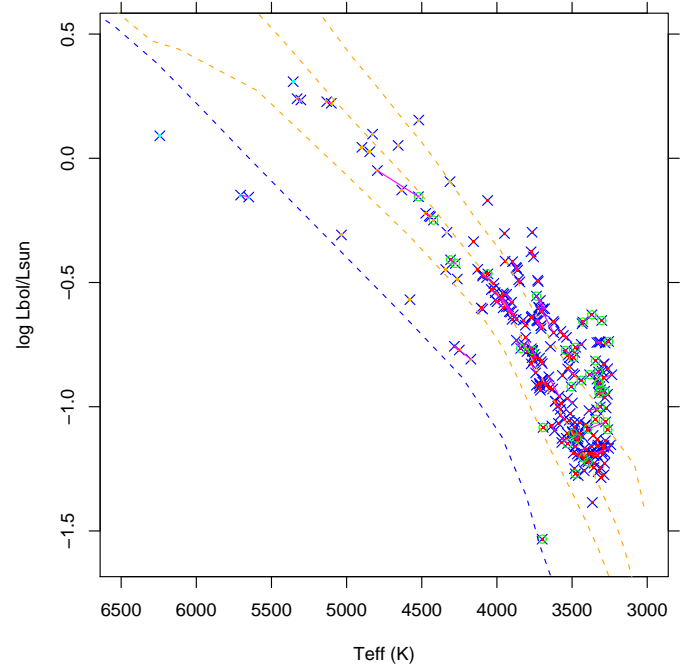}}
\caption{HR diagram for \gam\ members. The ZAMS is indicated with a blue
dashed line. Orange dashed lines are Siess isochrones for ages 5, 10,
and 20~Myr. Magenta segments connect different observations of the same
stars.
\label{hrd}}
\end{figure}

While we will devote Sect.~\ref{results} to calibration of our indices based
among others on the Gaia benchmark stars, a set of bright stars
included in the Gaia-ESO Survey whose stellar parameters are accurately known
(Jofre et al.\ 2013a,b, Heiter et al., Pancino et al., both in preparation),
we have first attempted
a calibration of our temperature index $\tau$ on the basis of
the widely used Kenyon and Hartmann (1995; KH95) color-\teff\ calibrations.
This approach relies on the fact that for \gam\ cluster stars reddening is
low and uniform, a fact supported by our data, and on its value as measured
by Jeffries \e (2009). We consider together MS and PMS stars (which have
similar reddenings), while we do not use in the calibration giant stars,
which may have much higher reddening, and which are easily selected from
the $(\gamma,\tau)$ diagram. A few early-type stars may also have large
distance and high reddening, but these latter are easily found as outliers in
the fit, and discarded. We use de-reddened $V-I_c$ colors\footnote{We
adopt for foreground MS stars the same reddening as for cluster members,
since as discussed in Section~\ref{obs-sample} the bulk of the former
stars lie at distances close to the cluster itself.},
and the $V-I_c$-\teff\ calibration of KH95
for $V-I_c<1.8$, and that of Stauffer \e (1998) for $V-I_c>1.8$, which
better matches the Pleiades data, to assign a \teff\ value to
MS/PMS stars. The resulting diagram of \teff\ vs.\ $\tau$ is
shown in Figure~\ref{tau1-vi-calibr} (left), together with a
seventh-degree polynomial fit.
The fact that no separation is seen in this Figure between cluster
members and MS foreground stars reassures us about the applicability of
this calibration to both subsamples.
After a first best-fit, datapoints
deviant from the fitted values by more than $|\Delta T_{eff}| = 0.04
\; T_{eff}$ were discarded, and the fit repeated, with the result being
shown in the Figure. The final best-fit formula is:
\begin{eqnarray*}
T_{eff} = & 5606071.88 -17422058.54\; \tau +23206500.2\; \tau^2 \\
 & -17090080.44\; \tau^3 +7496636.79\; \tau^4 -1956240.22\; \tau^5 \\
 & +281022.26\; \tau^6 -17141.93\; \tau^7.
\end{eqnarray*}
High-reddening early-type stars are easily seen, as
well as two likely veiled cluster stars near $\tau \sim 2.7$. Best-fit
residuals are shown in the same Figure (right panel); the
$|\Delta T_{eff}| = 0.04\; T_{eff}$ limits are shown by the two dashed
curves (points outside of the region enclosed by these curves were
not included in the fit). For stars inside these limits, the residual
standard deviations are 109~K for early-type stars, 73~K for
intermediate-type
stars, and 50~K for late-type stars. For intermediate- and late-type stars,
standard deviations are essentially unchanged when considering only
cluster members, while too few early-type members are present to draw
solid conclusions about them.

Eventually, we have tested our ability to recover a HR diagram for \gam\
members, correcting the photometric data using \teff\ as derived
from the $\tau$ index.  Because of the spread of
datapoints in the fit above, and the already narrow locus occupied by
members in the CMD of Fig.~\ref{v-vi-2}, we do not expect this to yield
a big improvement in a region with low and uniform reddening like the
\gam\ cluster, but there are large potential gains in younger or
more distant regions with patchy or higher reddening.
We use the best-fit calibration for $\tau$ to re-assign a \teff\ value
to each cluster star. Then, using again the KH95 (plus Stauffer \e
corrections) calibration, we compute the expected intrinsic colors
$(V-I_c)_0$; these, together with the observed $V-I_c$, permit
to compute individual extinction values $A_V$ for all cluster stars.
Using these $A_V$ values, the observed $V$ magnitudes, and bolometric
corrections from KH95, we compute bolometric luminosities $L_{bol}$,
and are able to construct the HR diagram shown in Figure~\ref{hrd}.
Adopting as typical \teff\ error for the bulk of late-type cluster
stars the standard deviation computed above (not the much smaller
statistical errors on computed \teff), this propagates into median errors
of 0.084~mag on $E(V-I)$, and of 0.09~dex on $\log L_{bol}$, comparable to the
symbols size in the Figure.
In the Figure we also show the Siess \e (2000) ZAMS (blue dashed line)
and isochrones for ages 5, 10, and 20~Myrs (orange dashed lines). A hint
of a binary sequence (or perhaps of an age range)
is seen at \teff\ below 4000~K (despite we left out
of our study recognized SB2 stars), where also a narrow
cluster sequence is found, well described by the 10~Myr isochrone.
Datapoints near the ZAMS (9 individual spectra for 6 stars) are again
found, for those stars discussed in Sect.~\ref{tau-gam} to be older PMS
stars. Overall, the result is good, and gives us confidence on the reliability
of the procedure followed to calibrate \teff\ in the \gam\ cluster.

%============================ calibration ===================================

\section{Reference data sets for general calibration}
\label{refdata}

Next, we address the problem of calibrating our temperature and
gravity indices in the general case, and also of studying the effect of
non-solar metallicities on our indices.
Most publically available sets of spectra are of much lower spectral
resolution than that of HR15n spectra; this fact renders the use of our
indices impossible, since some of them are defined over narrow
wavelength intervals, or even adjacent intervals containing lines with
strongly contrasting sensitivity to, e.g., gravity. We therefore must
consider exclusively intermediate- to high-resolution spectra, of a
large number of stars covering a region of the parameter space
(\teff, \logg, \fe) as large as possible.
First, we use the spectra from the UVES Paranal Observatory
Project\footnote{ESO DDT Program ID 266.D-5655.}
(UVES-POP, see Bagnulo \e 2003).  This set of spectra comprises
relatively bright stars (both field and cluster stars), observed with the
UVES spectrograph at very high S/N.
Only the field stars are
considered here, as they cover a wider parameter range (O-M, plus
some peculiar stars, Wolf-Rayet, and Carbon stars). The spectral
resolution of UVES being much higher that that of Giraffe/HR15n spectra,
spectral degradation is needed by convolution with a suitable
line-spread function (assumed Gaussian). One big shortcoming of the
composition of the UVES-POP sample is the lack of main-sequence
stars later than about K5. Another issue is the non-homogeneous nature
of the stellar parameters, collected from literature without a detailed check,
as explained in Bagnulo \e (2003): a visual inspection of spectra shows
indeed that in a few cases the spectral types appear rather inaccurate.
Nevertheless, the wide coverage of the (\teff, \logg) plane makes this
set an interesting one for a study of the properties of our indices.
Overall, the number of UVES-POP stars with valid data in the HR15n
wavelength range is about 300.
It is important to realize that the composition of the UVES-POP sample
is {\rm not} representative of the average spatial density of the
different spectral types, but instead contains an exceedingly large
proportion of peculiar stars, rare Wolf-Rayet (WR) stars, and supergiants; this
offers us the opportunity to test the performances of our indices for
these rare stellar types.

A much larger set of spectra is available from the ELODIE~3.1
stellar library (Prugniel and Soubiran 2001), containing 1962 spectra
of 1388 stars, with a wide coverage in \teff, \logg, and \fe,
spectral resolution $R \sim 42000$ and high S/N,
and covering the wavelength range 3850-6800~\AA.
These spectra are a subset of the complete ELODIE archive, corresponding
to stars with best-studied fundamental parameters. Accompanying the
library, a table of literature parameters including quality flags helps
to select the most reliable measurements, unlike stars in the UVES-POP sample.
One shortcoming of the ELODIE spectra, instead, is the way in which
telluric-line contamination was treated, namely blanking completely the
spectrum in correspondence to known telluric lines: these latter are
especially dense in the wavelength region 6490-6500~\AA, where the
line "quintet" is found (Sect.~\ref{quintet}), from which the
temperature- and gravity-sensitive indices $\beta_t$ and $\beta$ are
computed. These indices, computed from the ELODIE spectra, are therefore
slightly altered by such "holes" in their extraction regions, in a
non-systematic way because of the lack of correlation between the star
radial velocity (RV) and its \teff\ and \logg. In practice, this has a
similar effect as a reduced S/N in these particular
indices, increasing the scatter of datapoints in the relevant diagrams.
We find, nevertheless, that this is not too critical in practice, and we
will be able to derive useful calibration information from the ELODIE
spectra despite this feature.
The ELODIE set of spectra is slightly richer in cold stars than the
UVES-POP set, although the number of MS stars cooler than mid-K remains
only a handful; moreover, the fraction of peculiar stars appears to be
negligible, unlike the UVES-POP set. Overall, the two sets are
complementary in several aspects, each one providing useful information
on our indices.

To them, we have added about two dozen Gaia-ESO Survey benchmark stars
(Jofre et al.\ 2013a,b, Heiter et al., in preparation, Pancino et al.,
in preparation), which for the time being provide only a
limited coverage of parameter space, and are therefore used only as a
final check of results obtained using the two larger reference datasets
from UVES-POP and ELODIE.
Furthermore, recent results from Gaia-ESO Survey UVES observations of
hundreds stars
are used as a check, including a comparison between common UVES-Giraffe
targets from the Survey \gam\ dataset.
Finally, results from Gaia-ESO Survey HR15n observations of
young clusters are also examined, since the member population
of each cluster constitutes an age- and metallicity-homogeneous sample, and
is predicted to lie along definite sequences in the stellar parameter
space.

\section{Calibration results}
\label{results}

\subsection{UVES-POP dataset}
\label{uvespop}

\begin{figure}
\resizebox{\hsize}{!}{
\includegraphics[bb=20 10 465 475]{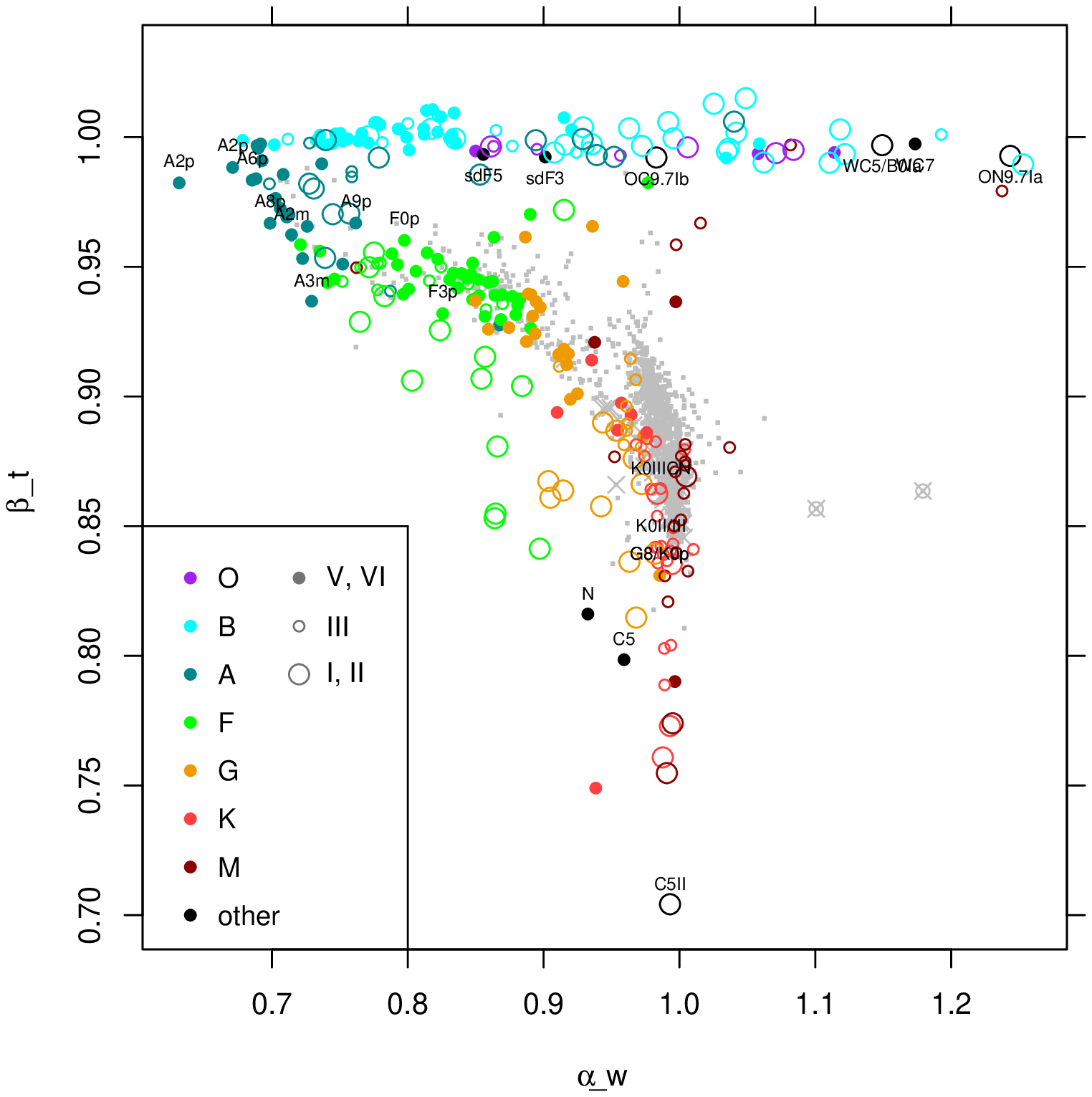}}
\caption{Index $\beta_t$ vs.\ \ha\ wings index $\alpha_w$, for UVES-POP
stars (colored symbols). Datapoints from the \gam\ dataset
are shown in gray (excluding stars of types later than about K7).
\gam\ cluster members are indicated by gray
crosses, and cluster CTTS in particular by empty gray circles.
As explained in the legend, filled colored dots are UVES-POP stars of
luminosity classes IV-V (or undefined), smaller empty circles are of class
III (including intermediate classes such as II/III, or III/IV), and
bigger empty circles are of luminosity classes I-II. Spectral classes
are color-coded, as in the legend. Wolf-Rayet stars, OC/ON stars,
subdwarfs, and carbon stars are plotted with black symbols.
These and stars with peculiar spectra are labeled.
\label{pop-qrt3-haiwc3}}
\end{figure}

We present here the analysis of the UVES-POP spectra. For each of them,
all spectral indices were computed, following
exactly the same procedure (including normalization in exactly the same
HR15n wavelength range). RVs needed to
place spectra in the stellar rest frames were taken from SIMBAD
database, where they are found for only 264 UVES-POP stars.
Since the UVES-POP sample lacks entirely MS stars later than mid-K, this
dataset is not useful to calibrate indices for \gam\ cluster stars, most
of which are in fact later than mid-K; however, we find it useful to
understand some other interesting properties of our indices. We start
from a diagram of indices $\beta_t$ vs.\ $\alpha_w$, as shown in
Figure~\ref{pop-qrt3-haiwc3}.
This "reversed question mark" diagram is a very important starting point
in our analysis. We mentioned in Sect.\ref{halpha} that the system of indices
we developed is suitable for stars with $T_{eff} \leq 8000$~K, and this
diagram makes it clear why: stars hotter than 8000-10000~K (i.e., OB and
WR stars) all belong to the upper strip in the diagram, while cooler
stars (such as those in the \gam\ dataset) populate the lower, S-shaped
part of the diagram. This pattern is easily understood, by recalling
that $\alpha_w=1$ means a \ha\ line without broad wings, such as that in
K-M stars, while $\beta_t=1$ means essentially a lack of metallic lines
(Ca~I, Fe~I, Ti~I, and Ba~II all contribute to $\beta_t$), which is
found in OB-WR stars, where these elements are found in higher ionization
stages, or in very metal-poor subdwarfs, two of which shown in the
Figure. As temperature decreases from the very highest values found in
O-WR stars (which often also have \ha\ emission, and thus $\alpha_w>1$),
the position of stars in the diagram moves along the $\beta_t=1$ strip
from right to left, until class A0-A2 is reached, where the \ha\ line
attains its maximum width, and the index $\alpha_w$ its minimum value.
At the same spectral type, metallic lines start to be relevant in the
spectrum, and $\beta_t$ starts to depart from its neutral value of
unity, toward lower values, decreasing monotonically until mid-K types,
as explained in Sect.~\ref{quintet}.
With the $(\beta_t,\alpha_w)$ diagram, we are
therefore able to discriminate between stars respectively hotter and
colder than type early-A, selecting only the latter ones for analysis
using our indices. The diagram should instead {\rm not} be used to infer
a temperature for OB stars, since for these stars the width of the \ha\
wings, as measured by the $\alpha_w$ index, is affected by gravity as
well as by temperature (e.g., B\"ohm-Vitense 1989, Gray and Corbally
2009), and also because of the
additional complication of \ha\ going into emission in many OB-WR stars.

There are other interesting lessons to be learned from the
$(\beta_t,\alpha_w)$ diagram of Fig.~\ref{pop-qrt3-haiwc3}. The S-shaped
strip going from early-A to late-G stars is populated (in the UVES-POP
sample) by stars belonging to all luminosity classes, from V (filled
dots), through III (small empty dots) to I (bigger empty circles); yet,
only these latter supergiant stars segregate from class III-V stars,
lying lower than the main strip in the diagram. Therefore, the position
of stars in this part of the diagram is {\rm insensitive to gravity}
except for supergiant stars; since, as mentioned in Sect.~\ref{halpha} and
further discussed below, the $\alpha_w$ index is independent of gravity in this
spectral-type range, this implies that also $\beta_t$ is independent of
gravity for early-A to (at least) late-K, and luminosity classes V to III.
Besides FG supergiants, also KM supergiants are found at lower $\beta_t$
values than KM dwarfs/giants: $\beta_t$ values for these latter are
usually above 0.83 (Sect.~\ref{quintet}), while in Fig.~\ref{pop-qrt3-haiwc3} a
number of supergiant stars are found below that value.
The diagram is also useful to select metallic-line Am stars, since they
have stronger metal lines (i.e., lower $\beta_t$) than expected on the
basis of their Balmer lines (i.e., the $\alpha_w$ index). One such
example can be seen in the Figure. One peculiar A star (type A2p)
is also found to deviate from the main strip in the diagram. Carbon stars are
also found in a peculiar position in this diagram ($\beta_t<0.85$ and
$0.9<\alpha_w<1.0$), near the KM-supergiant locus. Three such stars are
labeled 'N', 'C5', and 'C5II' in the Figure; one more star in the same
part of the diagram (HD~115236),
for which a K5 spectral type is listed in the UVES-POP
page\footnote{http://www.eso.org/sci/observing/tools/uvespop/ field\_stars\_uptonow.html},
is instead reported in SIMBAD to have a type of S6/8~B. This latter type
explains why the position of this star in the diagram is so different from
other K-type stars in the sample, and strengthens our argument about
this location in the diagram being typical of Carbon stars and similar
objects. Because a very large number of absorption lines is present in
the HR15n spectra of Carbon stars, in excess of those found in "normal"
stars, we are unable to state with certainty whether their
$\beta_t$ enhancement is caused by stronger absorption from the same lines as
in "normal" stars or by other lines.

Finally, there are a few stars in the gap between the OB-star strip and
the solar-star strip: while this is the expected position of mid-late M
stars (for which both $\beta_t$ and $\alpha_w$ tend to unity), for FGK
stars this is indicative of low metallicity, as we will discuss in
more detail below. At very low metallicities, FGK stars may even mix
with OB stars in this diagram (as is the case for the two sdF stars
labeled in the Figure, namely HD~140283 and HD~84937):
should this happen, a visual
inspection of the spectra themselves would solve the ambiguity, since
metal-poor stars will have only very weak lines, while OB stars in the
same region of the $(\beta_t,\alpha_w)$ diagram will show strong lines
of He~I and ionized metals at the very least near 6678~\AA.
The ambiguity between late-M and OB stars is instead solved by looking at
molecular index $\mu$, which has a value of unity for OB stars, but is much
higher for M stars (Sect.~\ref{mol-bands}).

\begin{figure}
\resizebox{\hsize}{!}{
\includegraphics[bb=20 10 465 475]{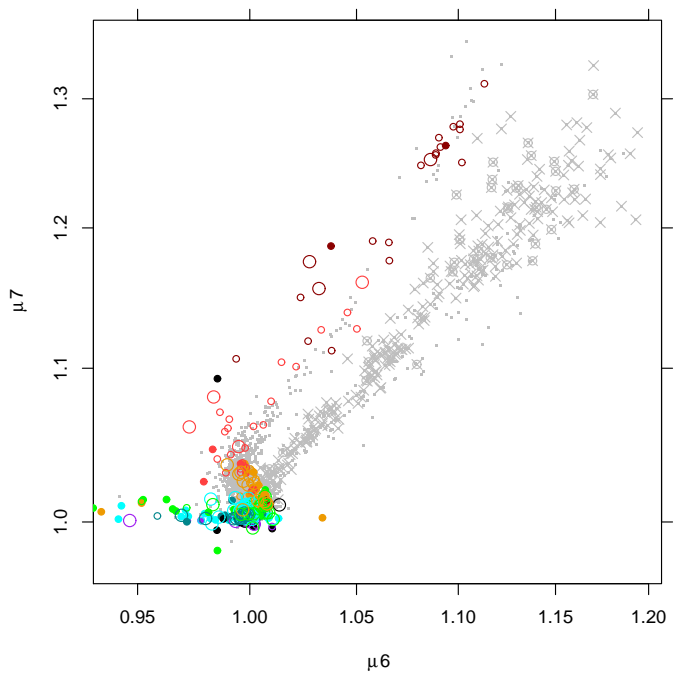}}
\caption{Plot of $\mu_7$ vs.\ $\mu_6$ for UVES-POP stars.
Note the lack of stars in the region occupied by MS/PMS M stars.
Symbols as in Fig.~\ref{pop-qrt3-haiwc3}, with the addition of stars
later than K7 from the \gam\ dataset.
\label{pop-mu6-mu7}}
\end{figure}

Next, we examine the diagram $(\mu_7,\mu_6)$ of Figure~\ref{pop-mu6-mu7},
which describes the behavior of two distinct molecular bands in the HR15n
spectral range. In Sect.~\ref{mol-bands}, this diagram was the first instance
of a gravity segregation between M giants and dwarfs. The UVES-POP
sample does not contain any M dwarf, and accordingly no datapoint is
found in the Figure along the M-dwarf strip (gray dots are \gam\
datapoints; gray crosses are \gam\ cluster members), confirming our
findings from Sect.~\ref{mol-bands}. Interesting in this diagram is also the
location of the few KM supergiant stars, along a strip placed still
above that occupied by KM giants (which in turn was above the dwarf
strip): this shows that the gravity dependence of the combination of
$(\mu_7,\mu_6)$ indices continues monotonically across the whole range of
M-star gravities, from dwarfs through giants up to supergiants. The
extension to supergiant stars is a new result, which could not have
been obtained in Sect.~\ref{mol-bands} because of the
lack of M supergiants in the \gam\ dataset. This lack is also suggested
by comparing the ranges of $\beta_t$ values found respectively in the \gam\ and
UVES-POP datasets, as discussed above.

\begin{figure}
\resizebox{\hsize}{!}{
\includegraphics[bb=20 10 465 475]{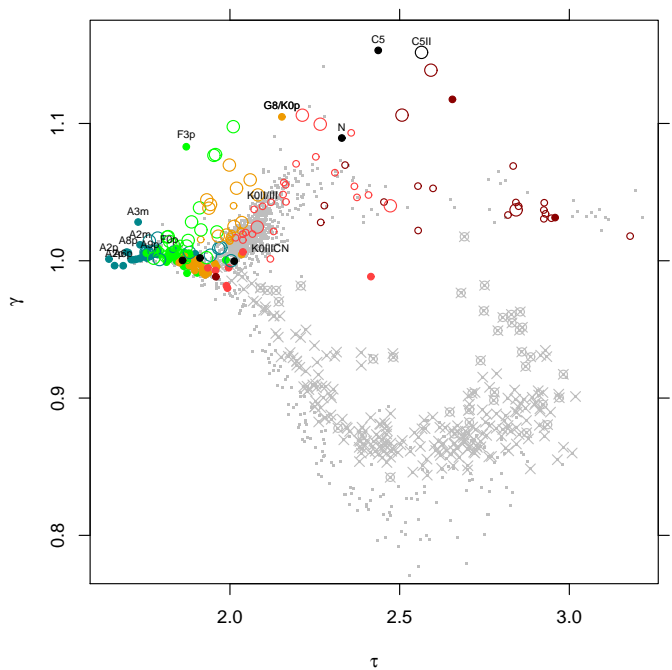}}
\caption{$(\gamma,\tau)$ diagram for UVES-POP stars, excluding stars
earlier than A2.
Carbon stars and stars with peculiar spectra are labeled.
Symbols as in Fig.~\ref{pop-mu6-mu7}.
\label{pop-gamma-tau}}
\end{figure}

We eventually show in Figure~\ref{pop-gamma-tau} the $(\gamma,\tau)$ diagram
for UVES-POP stars, with spectral types later than A2 (gray symbols
refer to the \gam\ dataset as above). The lack of UVES-POP dwarf stars later
than mid-K is evidently confirmed by the lack of non-gray datapoints in
the lower part of the diagram. Noteworthy is the position occupied by
some stars with peculiar spectra, with values for $\gamma$ index mimicking
low-gravity stars. FGKM supergiants are found above the locus of dwarfs
and giants of the same spectral types, confirming the effectiveness of
$\gamma$ as a gravity index for these stars as well. We note that this does
{\rm not} happen for A-type supergiants (blue circles, near
$(\gamma,\tau)=(1,2)$): the reason for this may be twofold: at these
temperatures, the $\alpha_w$ index which is an essential ingredient of
temperature index $\tau$ starts to depend to gravity, and the metallic
lines which contribute to $\gamma$ become too weak to be an useful
gravity indicator.
The position of Carbon stars (labeled in the Figure), near that of
KM supergiants is also remarkable, and both these types of stars may account for
many of the "outlier" stars mentioned in Sect.~\ref{all-diag}. The star found
in isolation near the center of the diagram (red filled dot in the Figure)
is the already mentioned S6/8 star HD~115236, in a location not populated
by normal dwarfs or giants.
In such "suspicious" cases, an accurate examination of the spectrum
is needed to ascertain the peculiar nature of the star, while our indices may
only provide an initial hint in this direction.

\subsection{ELODIE dataset}
\label{elodie}

\begin{figure}
\resizebox{\hsize}{!}{
\includegraphics[bb=20 10 465 475]{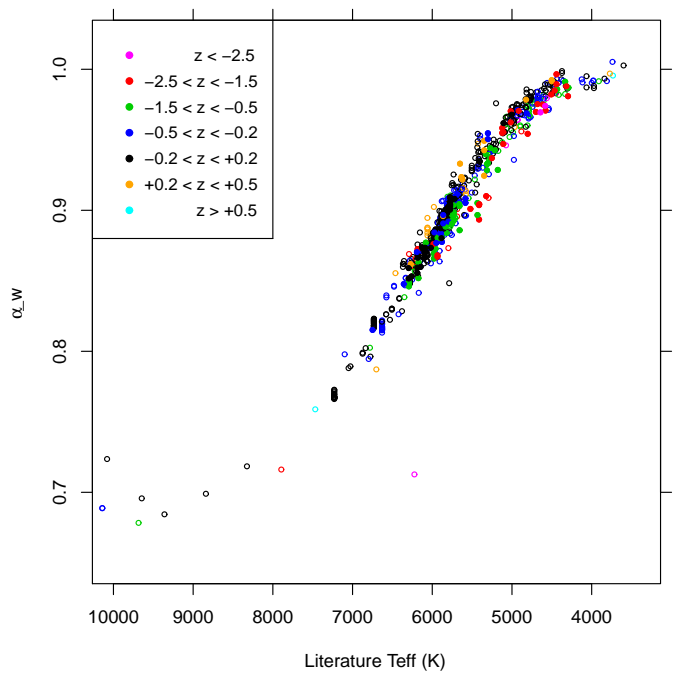}}
\caption{Index $\alpha_w$ vs.\ literature \teff, for the ELODIE spectra.
Datapoints are color-coded according to their metallicity (from
literature), as in the legend. Filled dots indicate stars with most
reliable determinations of \teff\ and \fe, while empty dots are
stars with less reliable parameters.
\label{elo-teff-haiwc3}}
\end{figure}

\begin{figure}
\resizebox{\hsize}{!}{
\includegraphics[bb=20 10 465 475]{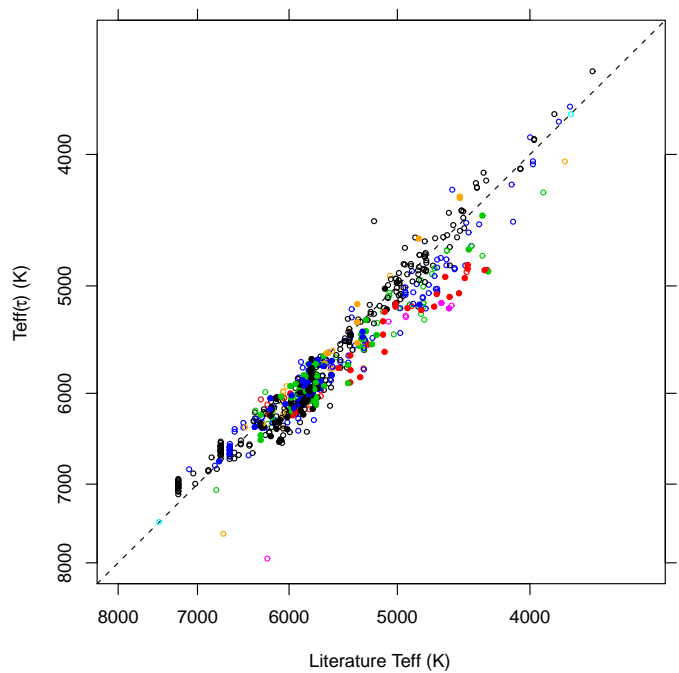}}
\caption{Effective temperature as derived from index $\tau$ (in
paper~I), vs.\ literature \teff, for the ELODIE spectra.
The dashed line represents identity.
Symbols as in Fig.~\ref{elo-teff-haiwc3}.
\label{elo-teff-tefftau}}
\end{figure}

We examine now results from the ELODIE dataset. 
We start from a diagram of our index $\alpha_w$ vs.\ the literature
\teff\ values for ELODIE sample stars (Figure~\ref{elo-teff-haiwc3}).
Here and in the following, we have considered only ELODIE stars with no
indication of SB2 status, no "uncertain RV" flag, and without close visual
companions possibly
contaminating the stellar spectrum. We have plotted color-coded symbols
according to (literature) metallicity $z=[Fe/H]$, choosing seven $z$ ranges
(with finer spacing near $z=0$), as explained in the Figure legend;
moreover, we use filled dots only for
stars having the best-quality flag ($q=4$) set for both
\teff\ and \fe\ in the stellar parameter table accompanying the ELODIE spectra
themselves\footnote{http://www.obs.u-bordeaux1.fr/m2a/soubiran/elodie\_library.html}.
Empty dots are instead used for all remaining stars, with lower-quality
parameter determinations. The $(\alpha_w,T_{eff})$ diagram of
Fig.~\ref{elo-teff-haiwc3} shows a very good correlation between our
$\alpha_w$ index and \teff, at least on the \teff\ range 4500-7500~K.
This is a confirmation, if ever needed, that the width of \ha\ line
wings is a very good measure of \teff\ in this range, for all gravities
and metallicities. This gives further support to our use of the $\alpha_w$
index as a primary temperature indicator, in the above \teff\ range.
\teff\ standard deviations around a quadratic best-fit line are
105~K for $6000 K < T_{eff} <7000 K$ and 116~K for $5000 K < T_{eff} <6000 K$.

Next, in Figure~\ref{elo-teff-tefftau} we plot the effective temperature
$T_{eff}(\tau)$ we have derived from the measured values of the $\tau$
index on ELODIE spectra and its calibration following KH95 (Sect.~\ref{calibr}),
vs.\ literature \teff. The agreement is not bad, but inferior with
respect to that shown using the $\alpha_w$ alone: in this case,
standard deviations of the
differences between $T_{eff}(\tau)$ and literature \teff\ are
203~K for $6000 K < T_{eff} <7000 K$, 165~K for $5000 K < T_{eff} <6000 K$,
and 213~K for $4000 K < T_{eff} <5000 K$.
We recall however,
that $\tau$ is useful as a \teff\ indicator over a much wider \teff\
range than $\alpha_w$, so the question whether or not the spread of
datapoints in Fig.~\ref{elo-teff-tefftau} can be minimized is an
important one. We note that the spread around the identity relation
(dashed line) seems to be mostly related to differences in metallicity
$z$: the lowest-$z$ stars (red and magenta datapoints) fall mostly below
the dashed line, while high-$z$ stars (orange) fall above it.
In order to minimize the spread in the diagram, and obtain a more
accurate \teff\ indicator, it is therefore important to be able to
estimate $z$ from our HR15n spectra themselves.
We remark that the weak $z$ dependence of $T_{eff}(\tau)$ implies that
also index $\tau$ (of which $T_{eff}(\tau)$ is a monotonic function) is
only slightly depending on $z$.

\begin{figure}
\resizebox{\hsize}{!}{
\includegraphics[bb=20 10 465 475]{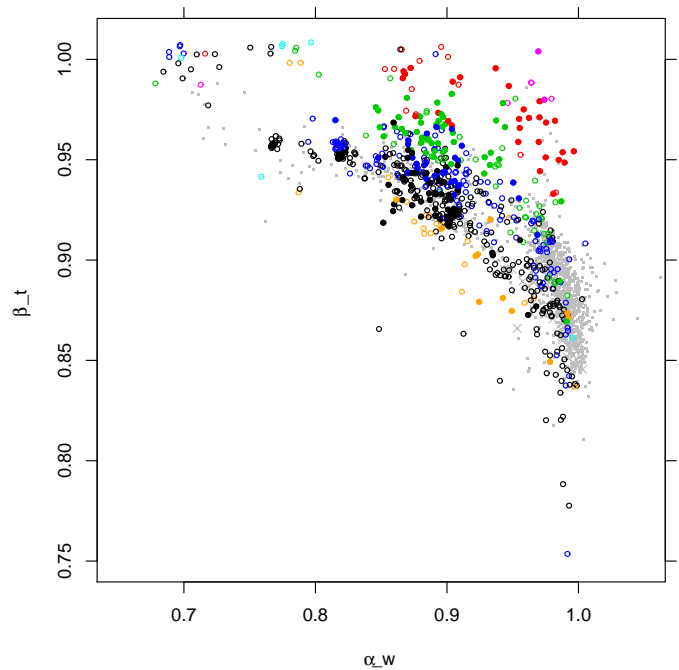}}
\caption{A $\beta_t$ vs.\ $\alpha_w$ diagram, analogous to
Fig.~\ref{pop-qrt3-haiwc3}, for ELODIE spectra.
Colored symbols are as in Fig.\ref{elo-teff-haiwc3}, while gray symbols
indicate stars from the \gam\ dataset, as in Fig.\ref{pop-qrt3-haiwc3}.
\label{elo-qrt3-haiwc3}}
\end{figure}

\begin{figure}
\resizebox{\hsize}{!}{
\includegraphics[bb=20 10 465 475]{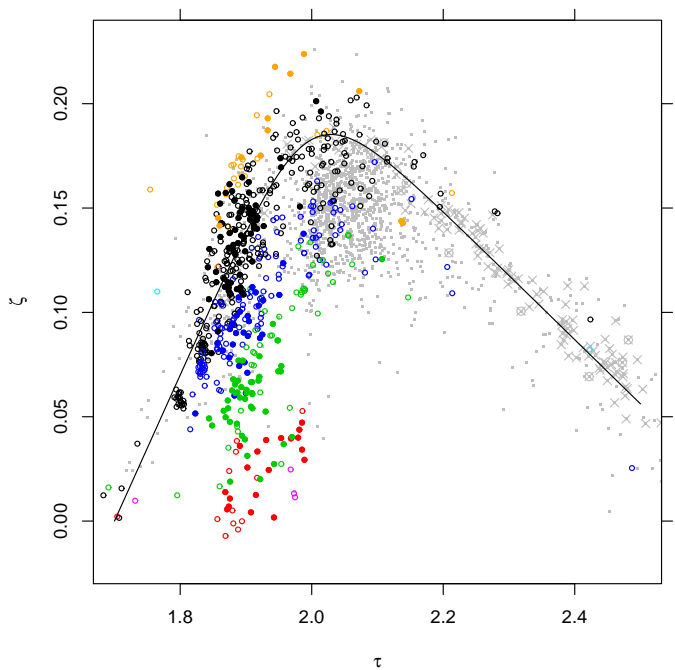}}
\caption{Index $\zeta$ vs.\ $\tau$ for
ELODIE spectra with $T_{eff}<10000$~K, and for \gam\ dataset stars.
The solid line is an approximate analytic description of the behavior
of solar-metallicity ELODIE stars (mostly for $\tau<2.1$) and of \gam\ cluster
members (for $\tau>2.1$).
Symbols as in Fig.~\ref{elo-qrt3-haiwc3}.
\label{elo-zeta-tau}}
\end{figure}

\begin{figure}
\resizebox{\hsize}{!}{
\includegraphics[bb=20 10 465 475]{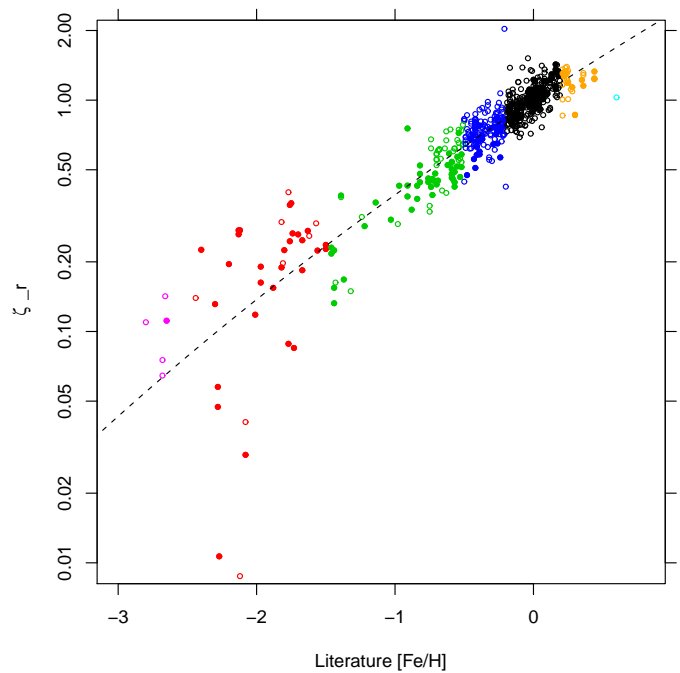}}
\caption{Ratio $\zeta_r$ between measured $\zeta$ and the
function $Z(\tau)$, vs.\ literature metallicity $z=[Fe/H]$, for
ELODIE spectra with $\tau>1.8$ ($T_{eff}<7000$~K).
The dashed line is an analytical description of the relation found (see
text).  Symbols as in Fig.~\ref{elo-teff-haiwc3}.
\label{elo-feh-zratio}}
\end{figure}

We have already assessed that $\alpha_w$ is a metallicity-independent index,
while it is likely that $\beta_t$ is highly $z$-dependent. The
$(\beta_t,\alpha_w)$ diagram of Figure~\ref{elo-qrt3-haiwc3} for ELODIE
spectra shows this latter fact rather clearly: datapoints of different colors
(i.e., $z$) are distributed in layers, with lower-$z$ stars found at
gradually higher values of $\beta_t$ than solar-metallicity stars.
We discussed above that, at least for luminosity classes V to III,
$\beta_t$ is practically independent of gravity, so that only the $z$
dependence shifts a star position up- or downwards, for a fixed \teff\
(and $\alpha_w$), except for supergiants.
In the specific case of the ELODIE spectra, however,
we remarked about the masking of telluric lines, adopted by the authors
of the spectral library, having the effect of increasing scatter in the
computed values of $\beta_t$ (and $\beta$) index, which accounts at
least partially for the scatter found in Fig.~\ref{elo-qrt3-haiwc3}.
The question is therefore whether another indicator of $z$ can be
obtained in the HR15n wavelength range, from a less telluric-contaminated
spectrum region.

We therefore consider a new combination of indices,
which we define as
\begin{equation}
\zeta = \zeta_1 - \beta_t.
\label{zeta}
\end{equation}
In this index, the telluric-line effect on $\beta_t$ is mitigated by the
addition of the telluric-free $\zeta_1$ index.
To understand its usefulness as a metallicity indicator, we plot $\zeta$
vs.\ $\tau$ in Figure~\ref{elo-zeta-tau}. Index $\tau$ was shown above to
be only little affected by metallicity; on the contrary, $\zeta$
is strongly dependent on $z$ (hence the choice of its name). This
statement is of course valid only within the limitations of the ELODIE
dataset, since for $\tau \geq 2.1$ ($T_{eff}\leq 4500$~K) very few
datapoints are available with known stellar parameter (even
less with {\rm most reliable} parameter determinations - filled dots). In the
Figure we show also datapoints from the \gam\ dataset (gray) and an analytical
approximation $Z(\tau)$ to the locus of solar-metallicity stars
(black dots), continued by \gam\ cluster members (gray crosses) at lower
temperatures (higher $\tau$). The analytical form of this function is
\begin{equation}
Z(\tau) = \frac{1}{(1/(0.7\tau-1.19)^5+1/(0.825-0.308\tau)^5)^{1/5}}
\label{Zeta}
\end{equation}
and we define a "$\zeta$ ratio" index $\zeta_r$ as
\begin{equation}
\zeta_r = \zeta / Z(\tau),
\label{zeta-r}
\end{equation}
which describes well the weakening of metal lines at lower $z$ values,
at least for stars in the \teff\ range 4500-7500~K, and {\rm irrespective of
gravity}.
We explicitly remark that our formulation of function $Z(\tau)$ is a
zero-order approximation to the true dependence of $\zeta$ on $\tau$ for
solar-metallicity stars, mostly because of the heterogeneous nature of
the ELODIE literature parameters: a much better knowledge of this function (and
in turn of the derived index $\zeta_r$) would be obtained from a study
of a single-metallicity stellar sample, such as that of
a cluster with a richer population of solar-type stars than \gam.

The relation between the newly defined index $\zeta_r$ and literature
metallicity $z$ is shown in Figure~\ref{elo-feh-zratio}, for $\tau>1.8$
($T_{eff}<7000$~K).
The dashed line is an analytical approximation of the
correlation found, of the form
\begin{equation}
Z_r = (1+0.09 \; z)^{10}.
\label{zeta-r-fit}
\end{equation}
At the lowest metallicity values ($z<-2$), Figure~\ref{elo-feh-zratio} shows a
considerable scatter with respect to our analytical approximation, which
should be therefore regarded as tentative; however, the still limited
tests provided by both the Survey UVES results and the benchmark stars in
Section~\ref{bench} below show a smaller scatter, and reassure us about
the validity of our approximation.
Inverting equation~\ref{zeta-r-fit} we have computed new metallicity values
$z(\zeta,\tau)$ from $\zeta$ and $\tau$, for ELODIE stars with
$\tau>1.8$; standard deviations for the
differences $z(\zeta,\tau)-z$ (where $z$ is the literature metallicity)
are then found to be 0.68 for $z<-1.5$, 0.22 for $-1.5<z<-0.5$, and 0.15
for $z>-0.5$. We conclude that our new index $\zeta_r$ does indeed provide a
good metallicity diagnostics, at least for stars with $T_{eff}<7000$~K,
and all gravities. We basically lack a suitable reference
sample of lower-temperature stars with measured metallicities
to properly assess the effectiveness of this index for M stars.

\begin{figure}
\resizebox{\hsize}{!}{
\includegraphics[bb=20 10 465 475]{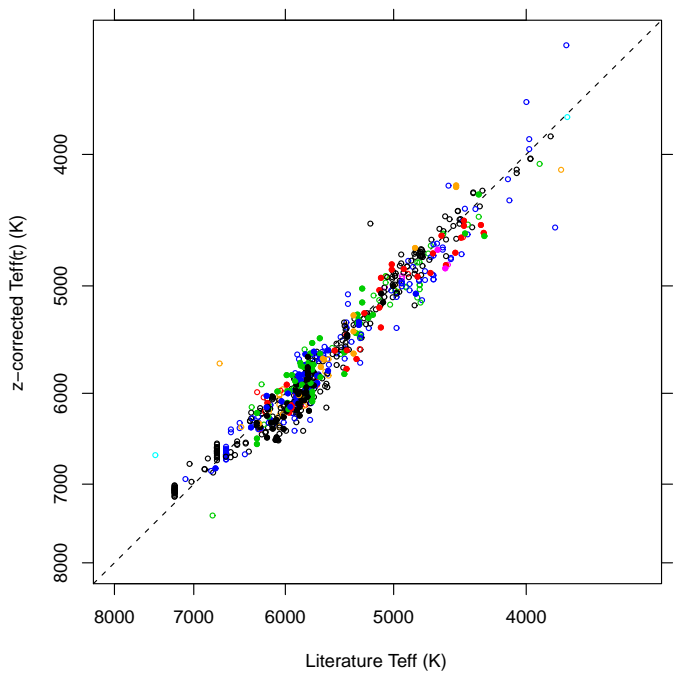}}
\caption{A diagram analogous to Fig.~\ref{elo-teff-tefftau}, but showing
the metallicity-corrected temperature $T_{eff}^z(\tau)$.
Symbols as in Fig.~\ref{elo-teff-haiwc3}.
\label{elo-teff-teffz}}
\end{figure}

Having established a valid metallicity index, we take advantage of it to
attempt a metallicity correction to the $T_{eff}(\tau)$ function
obtained in Sect.~\ref{calibr}. The needed correction term must be such as to
lower the estimated temperature for low-metallicity stars between 4000-5000~K,
while leaving unaltered temperatures near 6000~K, for any metallicity
(see Fig.~\ref{elo-teff-tefftau}). An useful empirical expression for the
metallicity-corrected temperature $T_{eff}^z(\tau)$ (again, not
from a best fit) is found to be
\begin{equation}
T_{eff}^z(\tau) = T_{eff}(\tau) \times
10^{0.00003 (1-\zeta_r) (T_{eff}(\tau)-6300)}
\label{teffz}
\end{equation}
with $T_{eff}(\tau)$ and $T_{eff}^z(\tau)$ in units of~K.
This newly determined temperature is compared in
Figure~\ref{elo-teff-teffz} with literature \teff\ values: apart from
very few outliers (most of which stars without best-quality literature
parameters) a good correlation is obtained, with no appreciable
systematics related to metallicity.
Standard deviations of the
differences between $T_{eff}^z(\tau)$ and literature \teff\ are
230~K for $6000 K < T_{eff} <7000 K$, 175~K for $5000 K < T_{eff} <6000 K$,
and 135~K for $4000 K < T_{eff} <5000 K$.

We note that these standard deviations are almost a factor of 2 {\rm
larger} than those found using photometry in Sect.~\ref{calibr}.
This is surprising, since aside from
systematic errors possibly introduced by the adopted KH95 calibration,
photometric colors may be affected by reddening (anyway by a small
amount, as discussed in paper~I). Doubtlessly, the telluric contamination
to $\beta_t$ (and thus to $\tau$ and $T_{eff}(\tau)$) increases the
spread in our derived \teff\ for ELODIE stars. Moreover, the heterogeneous
assembly of literature data on stellar \teff, \logg, and $z$ also
contribute to the spread in (the abscissa of) Fig.~\ref{elo-teff-teffz},
while the photometric $V-I$ colors used in Sect.~\ref{calibr} were all taken
from the same source (Jeffries \e 2009). For all these reasons, it is
probably safe to consider, for solar-metallicity stars, the calibration
found in Sect.~\ref{calibr} as the most reliable one; for metal-poor (or
metal-rich) stars, the
metallicity correction introduced above becomes necessary, although it
probably results in a less reliable \teff\ determination compared to
solar-metallicity stars.

\begin{figure}
\resizebox{\hsize}{!}{
\includegraphics[bb=20 10 465 475]{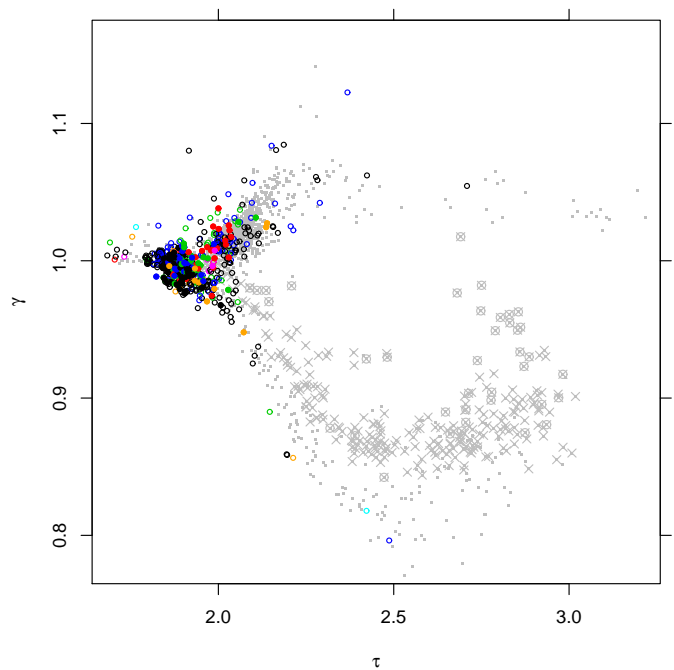}}
\caption{$(\gamma,\tau)$ diagram for ELODIE stars, excluding stars
with literature $T_{eff}>10000$~K, and for \gam\ dataset stars.
Symbols as in Fig.~\ref{elo-qrt3-haiwc3}.
\label{elo-gamma-tau}}
\end{figure}

\begin{figure}
\resizebox{\hsize}{!}{
\includegraphics[bb=20 10 465 475]{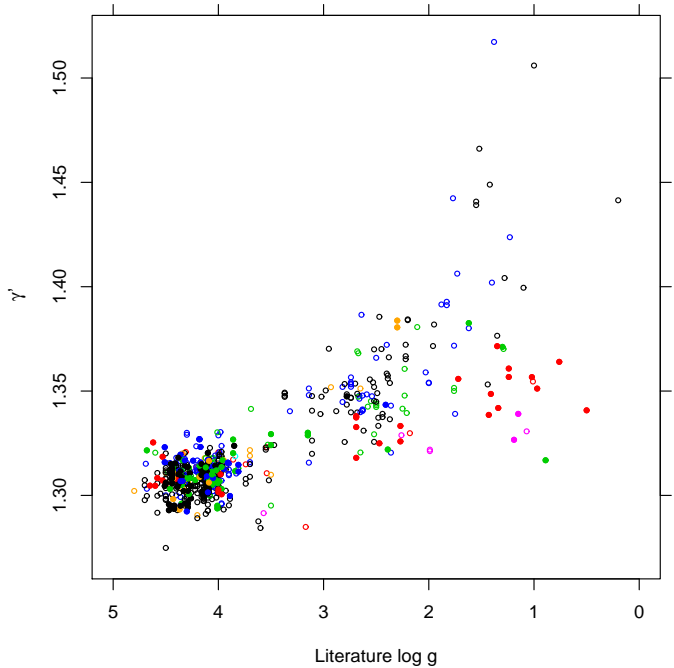}}
\caption{Composite index $\gamma'$ vs.\ literature \logg\ for
ELODIE spectra with $T_{eff}<10000$~K.
Symbols as in Fig.~\ref{elo-teff-haiwc3}.
\label{elo-logg-gammatau}}
\end{figure}

We now try to calibrate the gravity-sensitive index $\gamma$.
We remind that $\gamma = \gamma_1 + \beta$, and
$\beta = 1-\beta_c/\beta_t$. Since {\rm both} $\beta_c$ and $\beta_t$ are
affected by the telluric-line masking present in ELODIE data, index $\gamma$
is expected to be affected by more significant scatter with respect to
other indices discussed above. Nevertheless, useful information can
still be gained from this index for ELODIE spectra.
First, in Figure~\ref{elo-gamma-tau} we show a $(\gamma,\tau)$
diagram for ELODIE stars (with gray symbols referring to the \gam\
dataset). As expected, MS stars later than mid-K ($\tau>2$ and
$\gamma<0.95$) are only a handful; they fall all near the lower envelope
of \gam\ MS stars, i.e.\ well below the locus populated by cluster PMS stars,
in perfect agreement with our results in Sect.~\ref{tau-gam}. The number of
M giants is not large either: in fact, it is less than the UVES-POP giants,
so that again the two datasets are complementary. The diagram is in
agreement with the bulk of ELODIE stars being FGK stars of classes V to
III.

To calibrate the dependence of $\gamma$ on \logg, we first observe that
MS stars with $1.8<\tau<2.0$, with nearly constant \logg, do not fall on
a line at constant $\gamma$, but show a slope in their $\gamma$ vs.\
$\tau$ locus, indicating that $\gamma$ depends slightly on temperature, in
addition to gravity. Limiting ourselves to F,G, and early-K stars
(because of the ELODIE sample composition), we find that a
better \teff-independent gravity index may be taken as
\begin{equation}
\gamma' = \gamma+\tau/6
\label{gamma-prime}
\end{equation}
whose dependence on (literature) \logg\ is shown in
Figure~\ref{elo-logg-gammatau}. From this latter Figure we observe that
the scatter of datapoints is fairly large, compared to analogous diagram
involving \teff\ or $z$, considered above; this was in part expected, as
discussed above, because of telluric contamination. But there is also an
evident systematic trend with stellar metallicity, as seen from some
layering of datapoints with different colors. We therefore attempt again
to apply a metallicity correction to $\gamma'$, based on the metallicity
index $\zeta_r$ derived above.

\begin{table*}
\caption{Properties of spectral indices.}
\label{index-tab}
\centering
\begin{tabular}{ccc}
\hline
Index  & Description & Indicator type \\
\hline
$\mu_1 - \mu_7$ & TiO and CaH bands & \teff\ late-type ($\mu7$ also \logg) \\
$\alpha_w$      & \ha\ wings & \teff\ early-type, CTTS if $\alpha_w>1.1$ \\
$\alpha_c$      & \ha\ core & chromospheres \\
$\beta_t$       & metal line "quintet" $\lambda\lambda\; 6493-6500$ & \teff\ intermediate type \\
$\beta_c$       & low-gravity "quintet" lines & \logg\ \\
$\zeta_1$       & Fe~I lines $\lambda\lambda\; 6625-6635 $ & \teff\ intermediate type \\
$\gamma_1$      & low-gravity lines $\lambda\lambda\; 6758-6776$ & \logg\ \\
\hline
$\mu$    & $0.25\; (\mu_1+\mu_3+\mu_5+\mu_7)$ & \teff\ late-type \\
$\beta$  & $1-\beta_c/\beta_t$ & \logg \\
$\tau$   & $min(\alpha_w,1)-\beta_t-\zeta_1+3\times\mu$ & \teff\ all types \\
$\gamma$ & $\beta+\gamma_1$ & \logg\ all types \\
$\zeta$  & $\zeta_1 - \beta_t$ & metallicity \\
$\zeta_r$ & $\zeta/Z(\tau)$  &  \teff-corrected metallicity \\
$\gamma'^z$ & $1.305 + (\gamma + \tau/6 -1.305)/\sqrt{\zeta_r}$ & \teff-
and $z$-corrected gravity \\
\hline
\end{tabular}
\end{table*}

\begin{figure}
\resizebox{\hsize}{!}{
\includegraphics[bb=20 10 465 475]{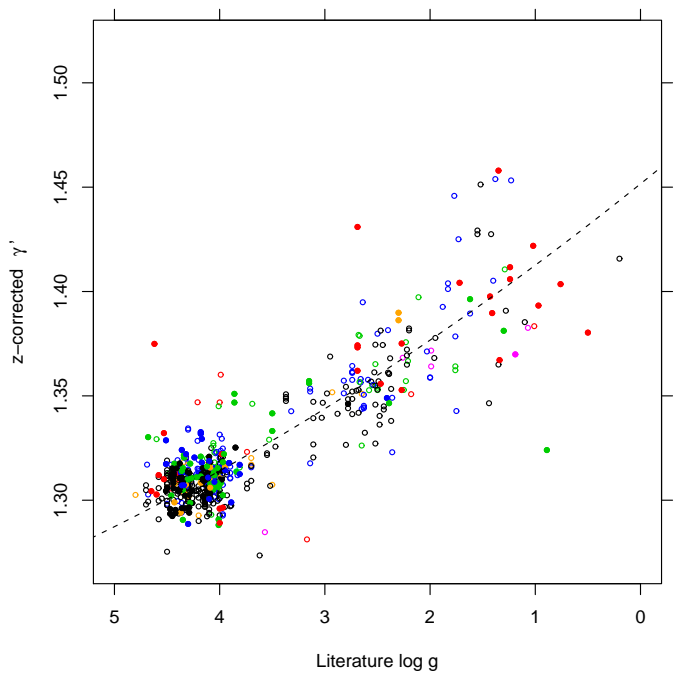}}
\caption{A diagram analogous to Fig.~\ref{elo-logg-gammatau}, but using
the metallicity-corrected index $\gamma'^z$.
Symbols as in Fig.~\ref{elo-teff-haiwc3}.
The dashed line is a quadratic best fit to the data.
\label{elo-logg-gammatauz}}
\end{figure}

\begin{figure}
\resizebox{\hsize}{!}{
\includegraphics[bb=20 10 465 475]{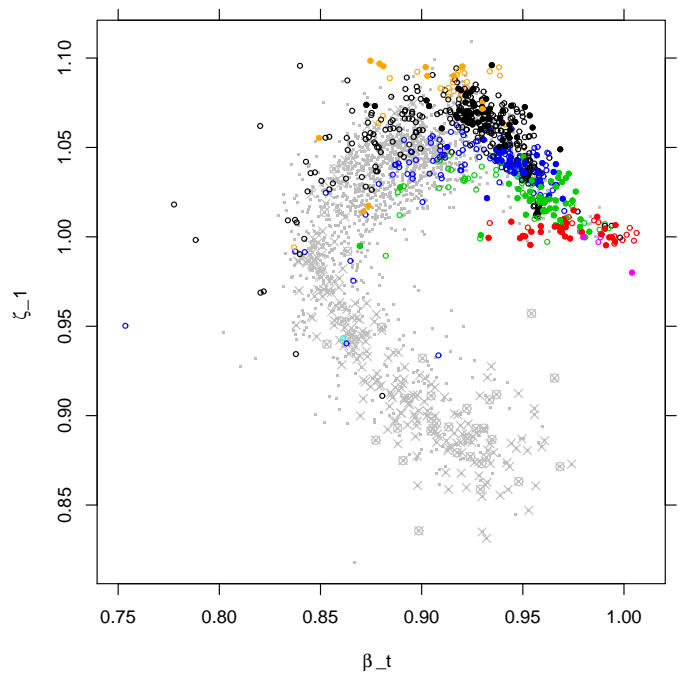}}
\caption{Diagram of $\zeta_1$ vs.\ $\beta_t$, for ELODIE and \gam\
spectra with $T_{eff}<10000$~K.
Symbols as in Fig.~\ref{elo-qrt3-haiwc3}.
\label{elo-tg-qrt3}}
\end{figure}

We found useful this empirical correction to obtain a
metallicity-corrected $\gamma'^z$ index:
\begin{equation}
\gamma'^z = 1.305 + (\gamma'-1.305)/\sqrt{\zeta_r}.
\label{gamma-primez}
\end{equation}
The $\gamma'^z$ index is shown in Figure~\ref{elo-logg-gammatauz} vs.\
literature \logg.
The main properties of all indices are summarized in Table~\ref{index-tab}.
In the Figure, the significant metallicity-related segregation seen in
Figure~\ref{elo-logg-gammatau} is no longer visible.
Despite the scatter still present, using this diagram
it is possible to discriminate effectively between gravities of MS stars,
giants, and supergiants, with very few outlier datapoints, irrespective
of metallicity.
In the Figure, a quadratic best fit\footnote{The best-fit
formula is $\gamma'^z = 1.4516-0.0405 \log g +0.00153 (\log g)^2$.}
to the data is also shown.
Residuals in \logg\ around the best-fit curve have standard
deviations of 0.437~dex for dwarfs ($\gamma'^z < 1.33$) and 0.88~dex for
giants/supergiants ($\gamma'^z > 1.33$). We remark again that these
results do not apply to dwarfs later than mid-K, and that some of the spread
seen is attributable to telluric-line masking in ELODIE data.

We conclude this Section by examining the $(\zeta_1,\beta_t)$ diagram
introduced in Sect.~\ref{tg}, where its usefulness as a PMS veiling diagnostics
was discussed. Such a diagram for ELODIE stars is shown in
Figure~\ref{elo-tg-qrt3}. From the UVES-POP dataset analysis discussed
above, we already know that the few stars to the left of the main locus
(especially those with $\beta_t<0.83$) are probably supergiants, a few
of which are also found among ELODIE stars. The layering of stars
with different metallicities in the upper part of the diagram is also not a
surprise, given the discussion on index $\zeta$ above. What is more
interesting is that metallicity appears to be the dominant cause for
vertical spread of G-K star datapoints near $\beta_t \sim 0.9$, with
gravity being only a secondary factor.
This same region in the diagram was found in Sect.~\ref{tg} to be populated
by giants, with no MS/PMS stars, but there was ambiguity as to whether
such a displacement with respect to the MS star locus was a gravity
or metallicity effect; from Fig.~\ref{elo-tg-qrt3} the dominant role of
metallicity is instead evident. Interestingly, when applied to a
constant-metallicity population such as a star cluster, this finding
implies that G-K cluster stars are expected to be found only in a narrow
strip in this diagram (again, this could not be well tested with \gam\
because of its little population of G-type stars). In a very young cluster,
G-K stars are still in their PMS stage, with gravities below those of MS
G-K stars, but this {\rm will not change their position in this diagram}.
At the same time, PMS veiling might be present, and it {\rm certainly
changes the star position in the same diagram}. Therefore, the above
results strengthen our arguments from Sect.~\ref{tg} that this diagram provides
a very useful veiling diagnostics for PMS stars, from G-type to M-type stars.

\subsection{Gravity calibration for PMS stars}
\label{pms}

We mentioned in Sect.~\ref{tau-gam} that late-type PMS stars occupy a region of
the ($\log g, T_{eff}$) plane not populated by other classes of normal stars.
Therefore, our gravity index could not be calibrated in the range
relevant to PMS stars using either ELODIE or UVES-POP stars. We devise
here an alternative method to obtain \logg\ values for the PMS stars, as
well as for lower-MS stars.
The PMS temperature calibration is assumed not to differ from that valid for
MS or giant stars, since the $\tau$ index was shown above not to depend
on gravity.

First, an empirical polynomial fit to the lower MS was derived, for all
non-member stars in the \gam\ dataset. The best-fit formula, for all
stars with $\gamma < min\; (1, 0.565+0.2 \times \tau)$, is
\begin{eqnarray*}
\gamma_{MS}(\tau)= & -116.2025396\; +229.7914669\; \tau\;
                     -176.8744508\; \tau^2 \\
       & +66.8400769\; \tau^3\; -12.4184477\; \tau^4\; +0.9085821\; \tau^5.
\label{gamma-ms}
\end{eqnarray*}
Then, the following formula gives a good approximation to the expected
\logg\ of members stars in the \gam\ cluster, based on its age and Siess
\e (2000) isochrones (see Fig.~\ref{teff-g-isochr})
\begin{equation}
\log g_{PMS} = 2.31066 + 2\; (\gamma -1.03) / (\gamma_{MS}(\tau) -1.03)
\label{logg-pms}
\end{equation}
and also its extrapolation to younger PMS stages appears satisfactory.
There is however an 'interface' region, between these lower-MS and PMS
stars, and more massive MS and giant stars, where the two formulations for
\logg, i.e.\ that from Equation~\ref{logg-pms} and that from the $\gamma'^z$
best fit, must agree seamlessly. This is achieved by a further correction
to the above formula for $\log g_{PMS}$, needed only for stars with
$\tau<2.07$, such that the final expression for $\log g_{PMS}$ for these
latter stars becomes
\begin{eqnarray*}
\log g_{PMS} = & 2\; (\gamma -1.03) / (\gamma_{MS}(\tau) -1.03)\; -361.637 \\
 & +350.02806\; \tau\; -84.15922\; \tau^2\; \; \; \; \; \; \; \; \; \; \; \;
(\tau<2.07).
\label{logg-pms-final}
\end{eqnarray*}
This completes the quantitative calibration of our indices in terms of
physical quantities, at least for the range of properties typical of
low-mass clusters.

\begin{figure}
\resizebox{\hsize}{!}{
\includegraphics[bb=20 10 465 475]{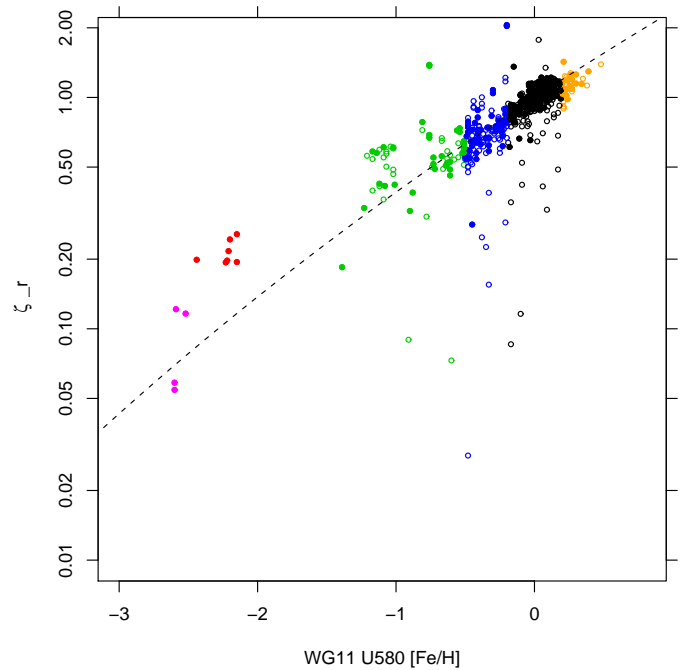}}
\caption{Index $\zeta_r$ for Gaia-ESO Survey UVES spectra in the HR15n range,
vs.\ metallicity $z=[Fe/H]$ as derived by WG11 from the complete
UVES spectra.  Only stars with $\tau>1.8$ are shown.
Dot colors indicate different metallicity ranges,
as in Fig.~\ref{elo-feh-zratio}. Filled dots correspond to UVES spectra with
S/N$>70$, while empty dots are spectra with $20<$S/N$<70$.
The dashed line shows the same relation as in Fig.~\ref{elo-feh-zratio},
and is not a best-fit to the UVES data.
\label{uges-feh-zratio}}
\end{figure}

\begin{figure}
\resizebox{\hsize}{!}{
\includegraphics[bb=20 10 465 475]{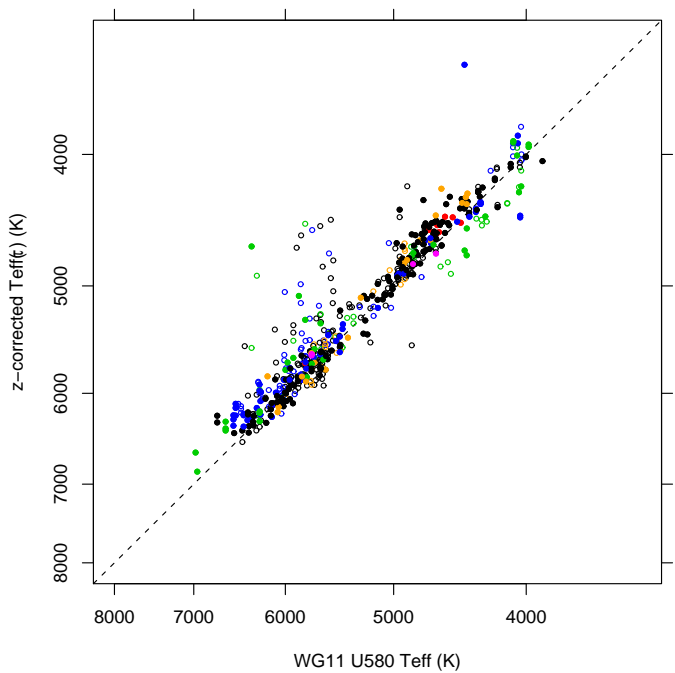}}
\caption{Metallicity-corrected temperature $T_{eff}^z(\tau)$ derived by
us for Gaia-ESO Survey UVES spectra in the HR15n range, vs.\ $T_{eff}$ as
derived by Survey WG11.
Symbols as in Fig.~\ref{uges-feh-zratio}.
The dashed line shows the same relation as in Fig.~\ref{elo-teff-teffz}.
\label{uges-teff-teffz}}
\end{figure}

\begin{figure}[b]
\resizebox{\hsize}{!}{
\includegraphics[bb=20 10 465 475]{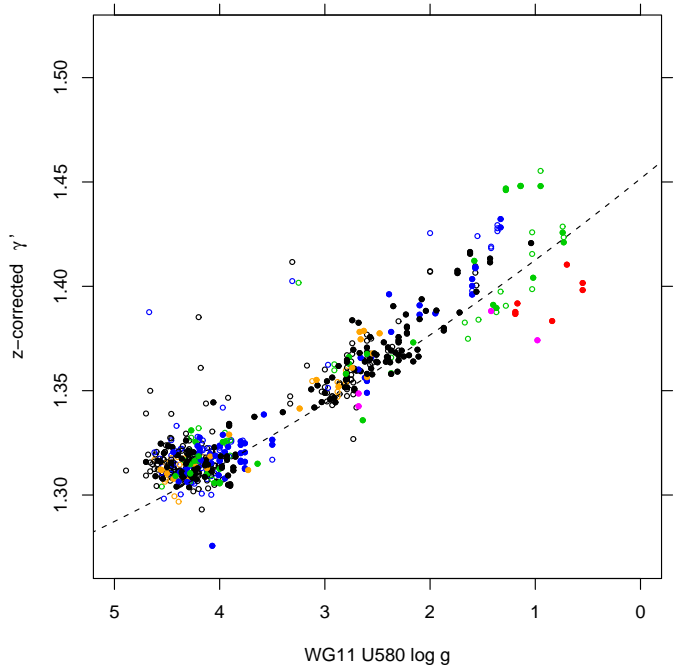}}
\caption{Metallicity-corrected index $\gamma'^z$ derived by
us for Gaia-ESO Survey UVES spectra in the HR15n range, vs.\ \logg\ as
derived by WG11.
Symbols as in Fig.~\ref{uges-feh-zratio}.
The dashed line shows the same relation as in Fig.~\ref{elo-logg-gammatauz}.
\label{uges-logg-gammatauz}}
\end{figure}

\subsection{Other Gaia-ESO Survey data: UVES results, and benchmark stars}
\label{bench}

In order to test how consistent are the calibrations made with other
results being obtained within the Survey, more homogeneously than the
literature data used thus far, we consider now the sample of all UVES
observations (with the 580~nm grating) of Gaia-ESO Survey stars,
with stellar parameters from internal data release {\rm
GESiDR1Final}, derived from
contributions by a large number of consortium members (Working Group
WG11, Smiljanic et al., in preparation).
As for ELODIE spectra, we have degraded the UVES spectra to the HR15n
resolution and wavelength range.
Results are shown in Figures~\ref{uges-feh-zratio},
\ref{uges-teff-teffz}, and~\ref{uges-logg-gammatauz}.
Dashed lines in these Figures are not best fits to the data, but are the
same analytical curves shown in Figures~\ref{elo-feh-zratio},
\ref{elo-teff-teffz}, and~\ref{elo-logg-gammatauz}. The agreement
between the calibration based on ELODIE data and the results from UVES
is good, especially considering the spectra with highest S/N ($>70$,
filled dots in the Figures), and reassures us on the validity of our results.

\begin{figure}
\resizebox{\hsize}{!}{
\includegraphics[bb=20 10 465 475]{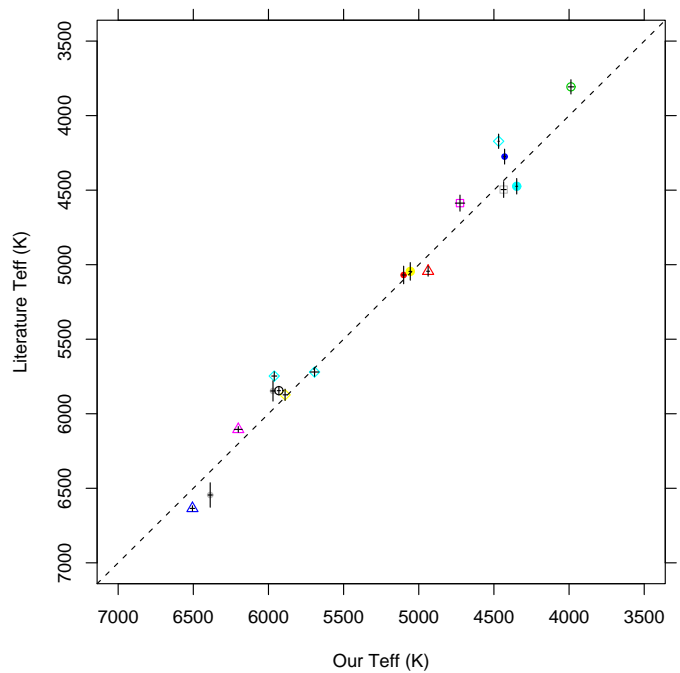}}
\caption{Comparison of \teff\ as derived by our method, and literature
\teff\ for Gaia-ESO Survey benchmark stars. Dotted line represents identity.
\label{bench-teff}}
\end{figure}

\begin{figure}
\resizebox{\hsize}{!}{
\includegraphics[bb=20 10 465 475]{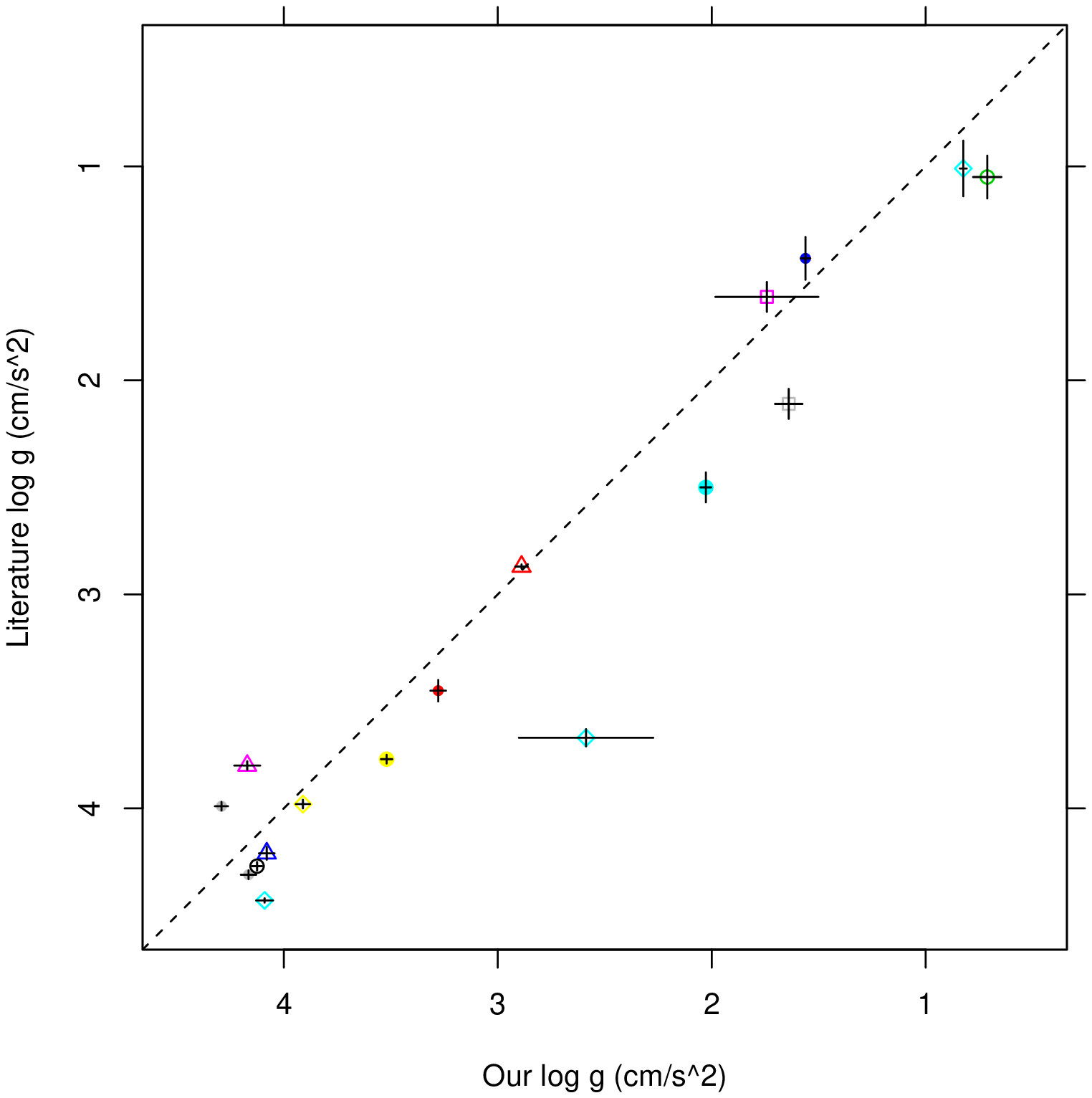}}
\caption{Comparison of \logg\ as derived by our method, and literature
\logg\ for benchmark stars. Dotted line represents identity.
\label{bench-logg}}
\end{figure}

\begin{figure}
\resizebox{\hsize}{!}{
\includegraphics[bb=20 10 465 475]{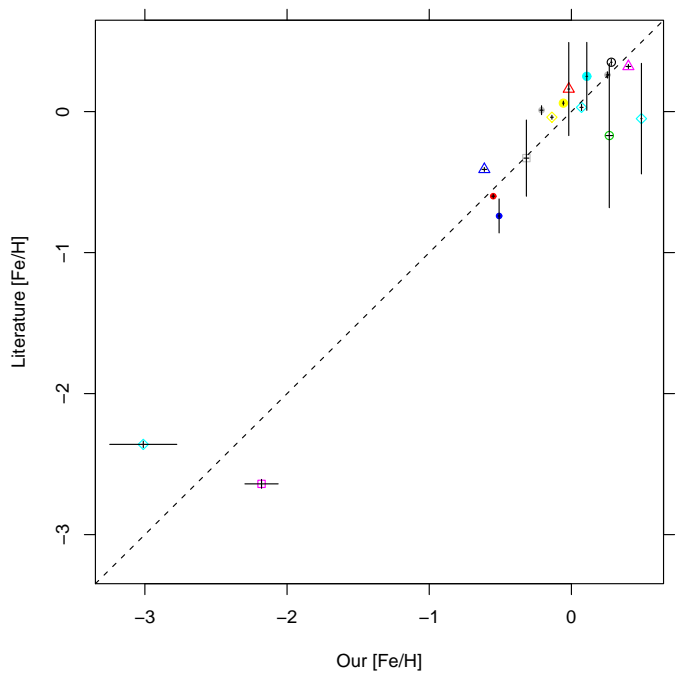}}
\caption{Comparison of \fe\ as derived by our method, and literature
\fe\ for benchmark stars. Dotted line represents identity.
\label{bench-feh}}
\end{figure}

Eventually, we made a comparison between our derived parameters for
Gaia-ESO Survey
benchmark stars, and their accurate literature values (Heiter \e, in
preparation, Jofr\'e \e 2013a,b). In the Survey, each benchmark star
was observed
with all instrument configurations used in the Survey, including HR15n.
For each star, we have selected the HR15n spectrum with the highest
S/N. Figures~\ref{bench-teff}, \ref{bench-logg}, and~\ref{bench-feh}
show the results obtained.
The standard deviations of the residuals are 133~K for \teff, 0.34~dex for
\logg\ (0.24~dex considering only stars with $z>-1$),
and 0.29~dex for \fe\ (0.22~dex considering only stars with $z>-1$).
We note that some scatter in these values
comes from uncertainties in the parameters of the benchmark stars themselves
(e.g., Jofr\'e \e 2013a,b), and that the benchmark sample lacks for the moment
M dwarfs (Pancino et al., in preparation), for which a smaller \teff\
uncertainty was estimated in Sect.~\ref{calibr}.

\subsection{Comparison between UVES and Giraffe results for \gam\ stars}
\label{gir-uves}

\begin{figure}
\resizebox{\hsize}{!}{
\includegraphics[bb=20 10 465 475]{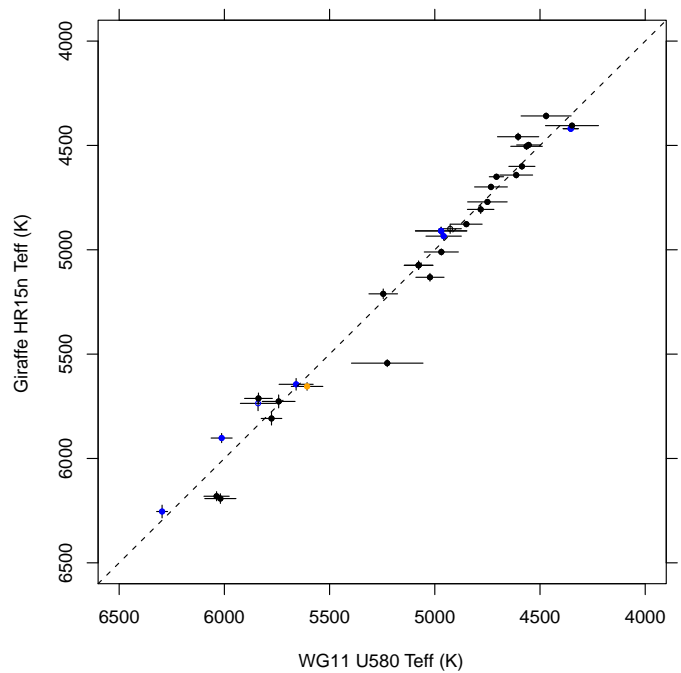}}
\caption{Comparison between \teff\ values derived for common UVES and Giraffe
targets, respectively by WG11 from UVES data and by our method from
Giraffe data.  Symbols as in Fig.~\ref{uges-feh-zratio}.
The dashed line represents identity.
\label{uves-giraffe-teff}}
\end{figure}

\begin{figure}
\resizebox{\hsize}{!}{
\includegraphics[bb=20 10 465 475]{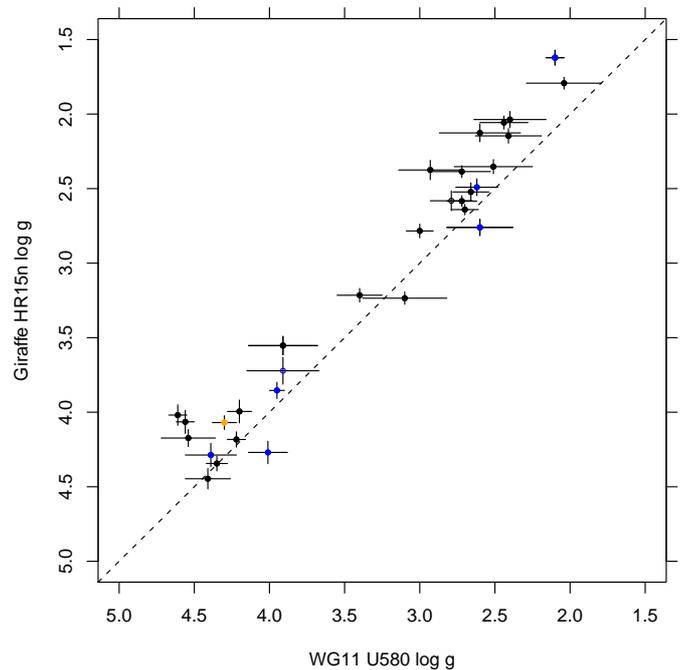}}
\caption{Same comparison as in Figure~\ref{uves-giraffe-teff}, for \logg.
\label{uves-giraffe-logg}}
\end{figure}

\begin{figure}
\resizebox{\hsize}{!}{
\includegraphics[bb=20 10 465 475]{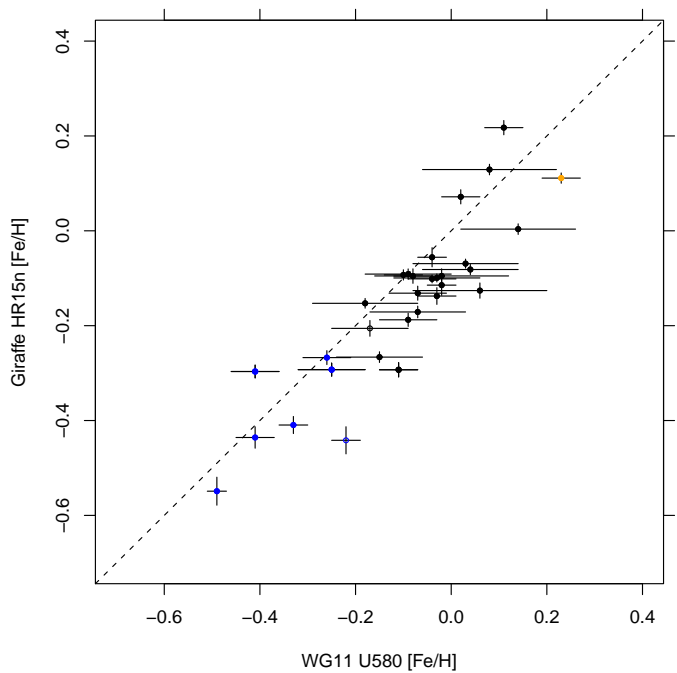}}
\caption{Same comparison as in Figure~\ref{uves-giraffe-teff}, for
metallicity $z$.
\label{uves-giraffe-feh}}
\end{figure}

More than 40 stars in the \gam\ dataset were targeted by Gaia-ESO
Survey observations with both UVES and Giraffe. Such common targets (not
necessarily cluster members) are very useful for testing the consistency
of results of independent analysis of UVES and Giraffe data, as explicitly
indicated in the Survey Management Plan.
We therefore compare here the results from the UVES data analysis made by
WG11 with our results for Giraffe HR15n spectra of the same stars
(not of their downgraded-resolution UVES spectra as in
Section~\ref{bench}). Since our
method was natively developed for Giraffe HR15n spectra, this test is
perhaps more significant than tests made in Section~\ref{bench} above.
For instance, residual instrumental signatures in the (degraded) UVES
data (e.g., imperfect order merging) might have slightly affected
results from those tests.
The number of common Giraffe-UVES target stars, on the other
hand, is significantly smaller than the total number of UVES spectra
used in the tests of Section~\ref{bench}.

We show in Figure~\ref{uves-giraffe-teff} a comparison between the \teff\
values derived from UVES data by WG11 (34 stars in {\rm GESiDR1Final} data
release) with those derived by
our method from HR15n data. Color coding indicates metallicity according to
the UVES results. Figures~\ref{uves-giraffe-logg}
and~\ref{uves-giraffe-feh} show analogous comparisons for \logg\ and
metallicity.
We find a mean \teff\ offset (UVES minus Giraffe) of -2~K, and a
std.dev. of differences of 92~K. This latter number is slightly smaller
than that obtained from benchmark stars (which however span a wider
range in all parameters), and well comparable to the std.dev. of fitting
residuals obtained in our \teff\ calibration in Sect.~\ref{calibr} in the same
\teff\ range (110~K for F to mid-G stars, 70~K for mid-G to late-K
stars). These uncertainties, consistent across a varied range of
\teff\ calibrations methods (direct measurements for benchmarks,
photometry for \gam\ MS/PMS stars, high-resolution UVES data) are
therefore robust estimates of the {\rm total} (statistical plus
systematic) errors on \teff\ delivered by our method.

For \logg, the UVES-Giraffe comparison yields a mean offset of 0.21~dex and
a std.dev. of differences of 0.21~dex. This latter is again slightly
lower than that obtained from metal-rich benchmark stars. Finally,
the \fe\ mean offset and std.dev. of UVES-Giraffe differences are respectively
0.056 and 0.084: this latter is significantly lower than the corresponding
std.dev. using benchmark stars (note however that some of the latter
have large errors in their literature metallicity, see
Fig.~\ref{bench-feh}).

Despite that the \logg\ and \fe\ mean offsets resulting from this
comparison are slightly different from zero, we choose not to revise for
the moment our calibrations for these quantities, both because of the
much smaller number of stars contributing to this test, compared to the
ELODIE spectra we used above as calibrators, and also since the Gaia-ESO
Survey analysis procedures are still evolving and consolidating.

\subsection{Tests on clusters as calibrators}
\label{clusters}

\begin{figure*}[t]
\resizebox{\hsize}{!}{
\includegraphics[bb=20 10 465 475]{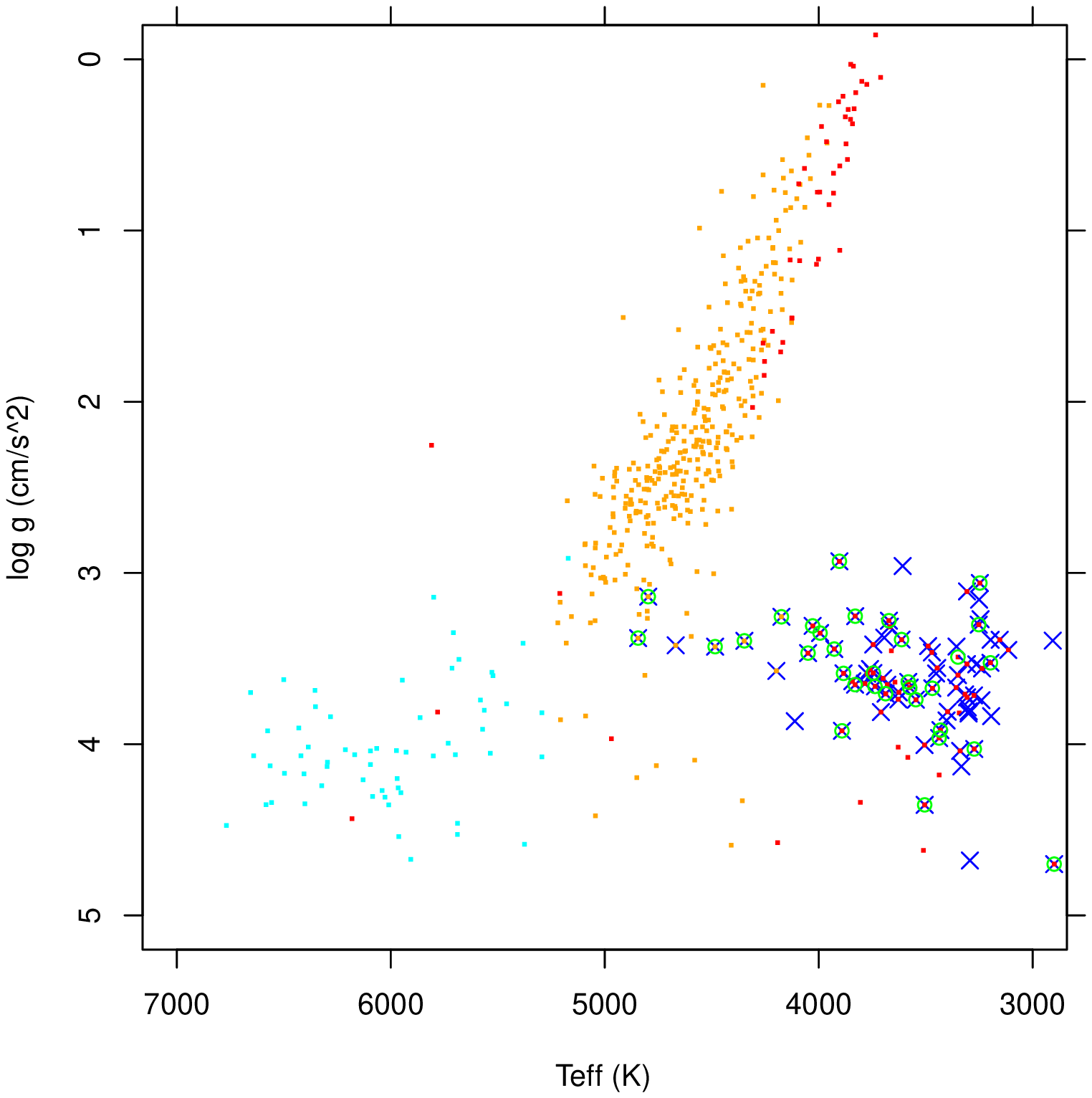}
\includegraphics[bb=20 10 465 475]{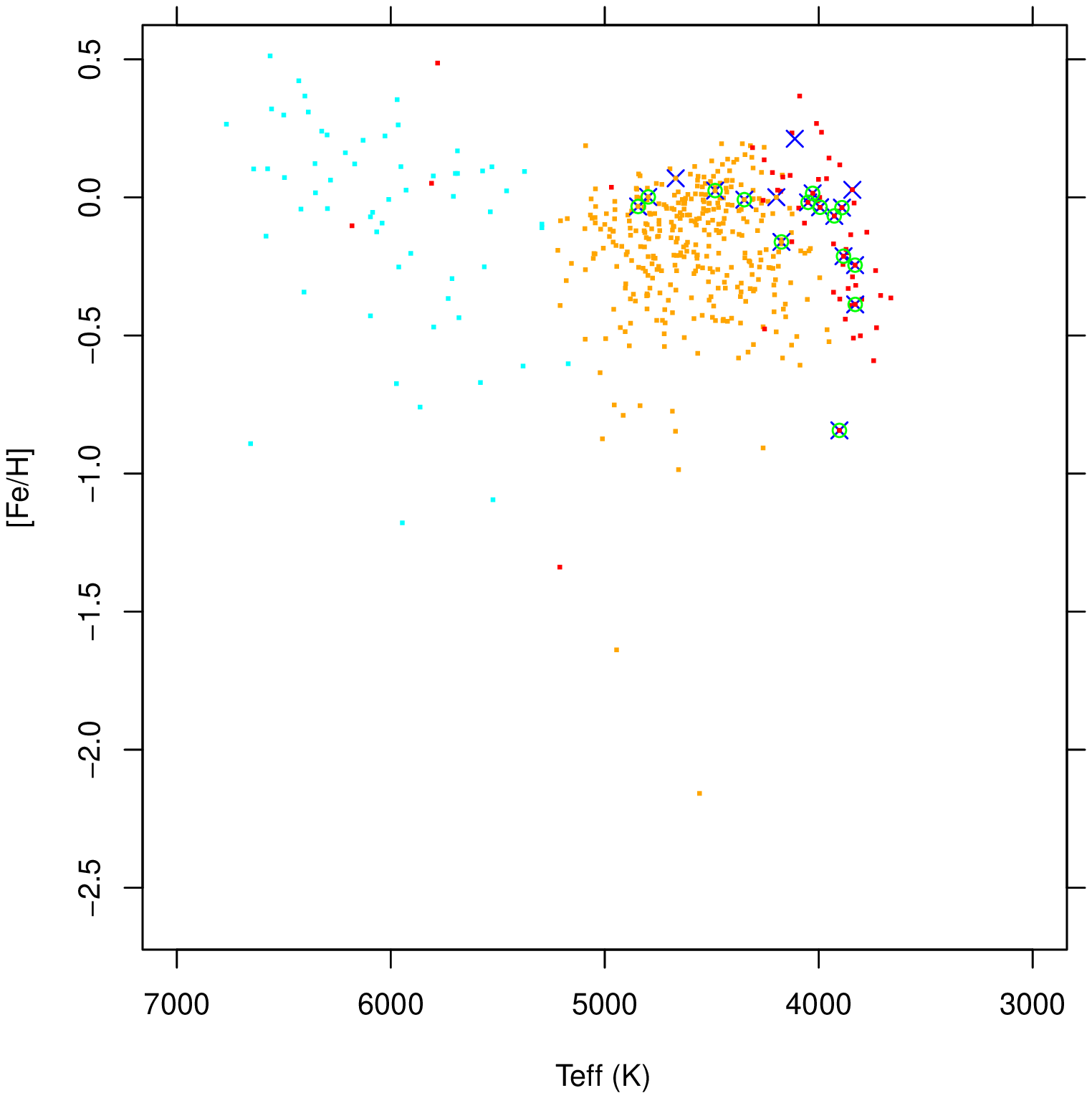}
}
\caption{Diagrams for Chamaeleon~I: {\rm Left:} ($T_{eff}, \log g$) diagram.
{\rm Right:} ($T_{eff}, [Fe/H]$) diagram.
Symbols have the same meaning as in Fig.~\ref{v-vi-2}.
Blue crosses are cluster members from Luhman (2007).
The few datapoints with plotted color not matching their derived \teff\
values are probable or confirmed peculiar stars.
\label{chaI}}
\end{figure*}

\begin{figure*}
\resizebox{\hsize}{!}{
\includegraphics[bb=20 10 465 475]{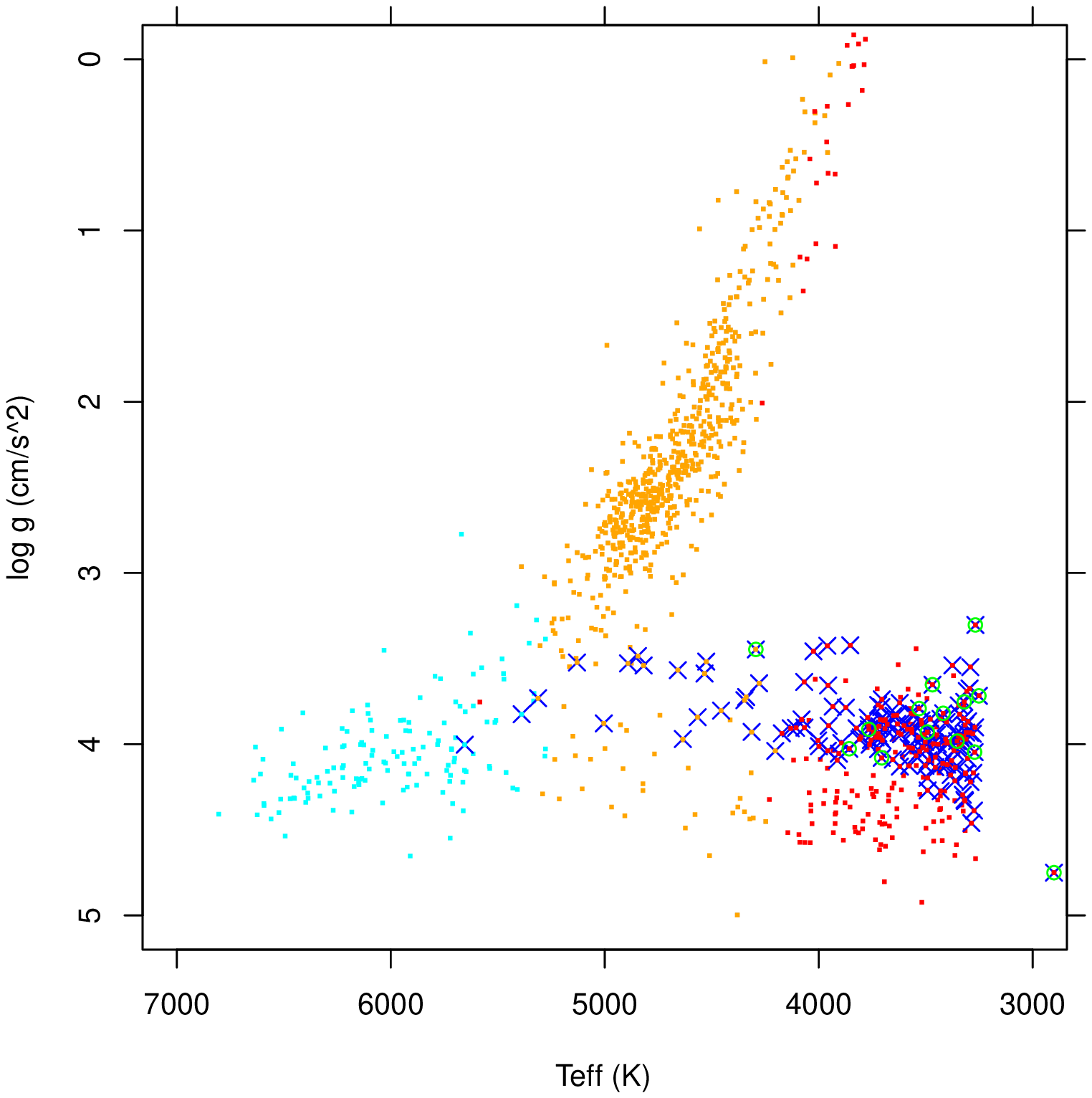}
\includegraphics[bb=20 10 465 475]{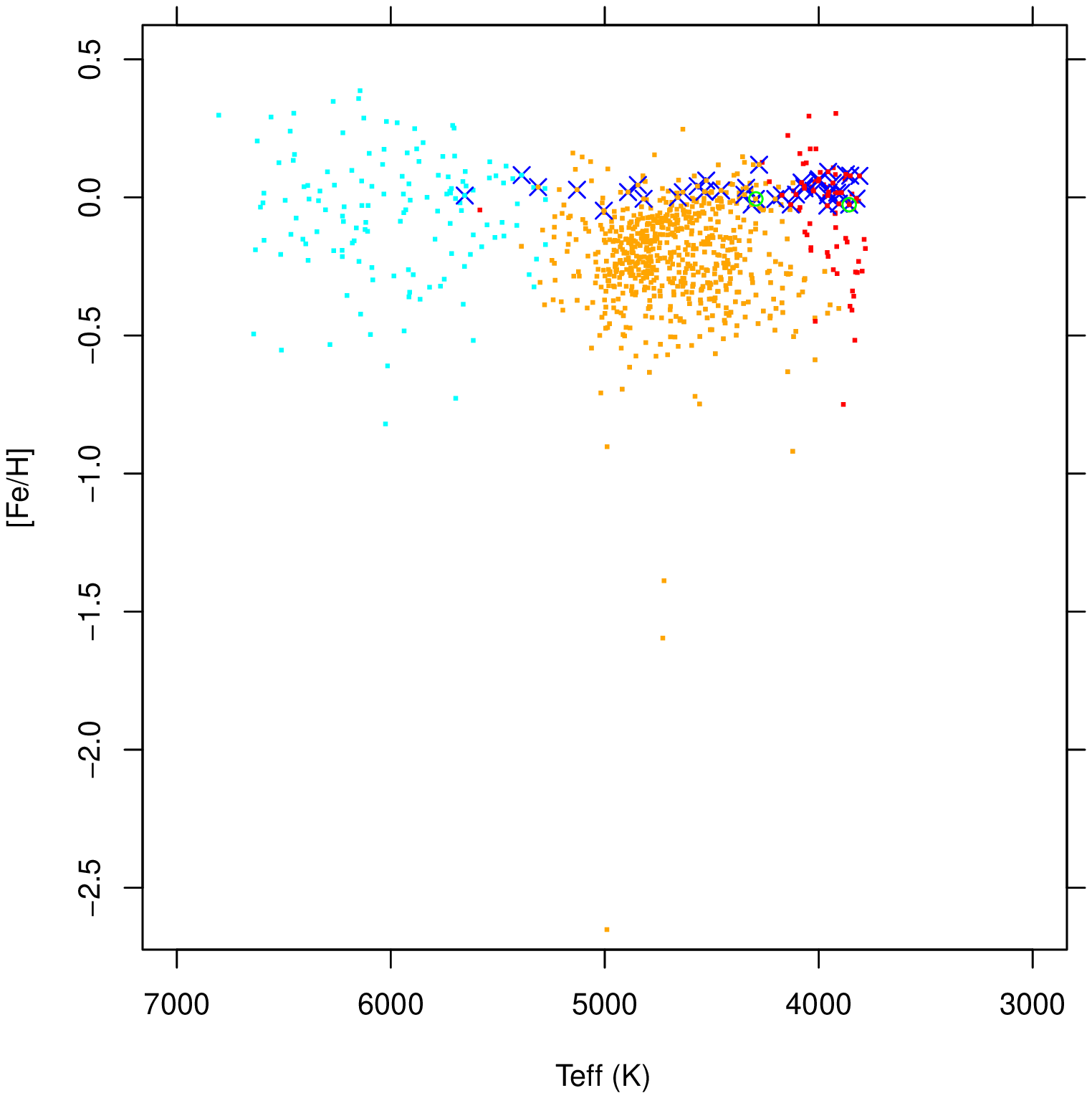}
}
\caption{Same as Figure~\ref{chaI}, for the \gam\ cluster.
Symbols have the same meaning as in Fig.~\ref{chaI}.
\label{gamma2Vel}}
\end{figure*}

Last, we show how the stellar parameters derived from our spectral
indices perform for Gaia-ESO Survey young clusters.
Classically, clusters have been used as calibrators for stellar
evolutionary models since they are composed of stars having all (nearly)
the same age and metallicity. Analogously, we need here to ensure that
the putative members of a given cluster define loci in both the
($T_{eff}, \log g$) and the ($T_{eff}, [Fe/H]$) diagrams, in
reasonable agreement with a single metallicity value, and also with
information available from photometry and model isochrones.
We therefore show these diagrams, derived using our method on Gaia-ESO Survey
data from release {\rm GESiDR1Final} for clusters \object{Chamaeleon~I}
and \gam\ (1-3~Myr and 5-10~Myr, respectively), in Figures~\ref{chaI}
and~\ref{gamma2Vel}.
Blue crosses represent members (from Sect.~\ref{obs-sample} for \gam,
from Luhman 2007 for Cha~I), and green circles CTTS; only stars with
$v \sin i <50$~km/s are plotted.

We observe from these Figures (left panels) that the median \logg\
of M-type Cha~I members is $\sim 3.7$, while that of M-type \gam\
members is distinctly higher, being $\sim 4.0$.
Therefore, our gravity determination does
indeed {\rm describe the gradual contraction of PMS stars toward the
MS}, confirming our findings in Sect.~\ref{tau-gam}. Moreover, the derived
metallicities (right panels) of putative cluster stars
(apart from very few outliers of
questionable membership) have a sharply peaked distribution, with no
(spurious) dependence on \teff.
Only a few CTTS members of Cha~I deviate from a flat dependence of
metallicity on \teff, but this is likely due to the strong veiling
affecting these stars' spectra, and weakening their absorption lines.
The $1-\sigma$ width of the derived metallicity distributions for these
clusters (robustly determined as the difference between the 84\% and
50\% quantiles, to minimize the effect of non-member outliers and of
veiled spectra) are 0.053~dex for Cha~I, and 0.042~dex for \gam.
These ranges are {\rm sensibly smaller} than the $1-\sigma$ scatter found above
(0.22~dex) for metal-rich benchmark stars, leading to at least a legitimate
suspicion that the benchmark metallicity scatter is dominated by errors in
their literature values rather than by uncertainties in our derivation.
Again, this is one more confirmation of the validity of our procedure.
The clusters' measured $1-\sigma$ metallicity scatter is also nearly
one-half that found in the UVES-Giraffe common-target comparison of
Sect.~\ref{gir-uves}, presumably because of the wider range of
parameter values covered by these latter stars.

\section{Discussion and future prospects}
\label{discuss}

In the previous sections we have defined a set of spectral indices,
which enabled us ultimately to place stars in an empirical
$(\gamma,\tau)$ diagram, representing a fairly regular
mapping of the $(T_{eff},\log g)$ plane for intermediate- and late-type stars,
for all gravities from the MS up to giants. For this approach to be
successful, the indices were carefully selected, on the basis of
suitable template spectra from the Gaia-ESO \gam\ dataset.
This approach is fundamentally different
from usual curve-of-growth studies of Fe~I/Fe~II lines, for which
insufficient spectral lines are available in the wavelength range
studied here. It is instead more similar to spectral-typing methods, but
made quantitative thanks to the use of spectral indices. These were
accurately defined to take also advantage of the higher resolution of our
spectra, compared to usual low-resolution classification spectra. This
compensates the lower number of available diagnostics in the relatively
limited wavelength range covered by the HR15n setup. In general, our
results falsify the often-made statement that this observational setup
is not adequate for spectral classification.

The results presented in this work were made possible in a fundamental
way by the extensive parameter-space coverage of the studied stellar
sample. Without a combined study of both MS/PMS and giant stars it would
have been probably impossible to select, e.g., line ratios most sensitive
to gravity.

Since \gam\ contains mostly low-mass, late-type stars, we have devoted
particular attention to devise a set of indices which are especially
suited to measure temperature and gravity in that mass range. Whether
the same indices are equally suited to discriminate satisfactorily, e.g.,
gravity (age) of more massive stars in a younger cluster, or to
discriminate between MS stars, and evolved subgiants and giants in a
much older cluster, remains to be studied on the basis of suitable
datasets, such as more massive cluster of different ages, some of which
are part of the Survey program.

Another little-explored problem is that of heavily-veiled CTTS, which
are basically absent in \gam. For these stars, we have discussed ways of
determining veiling and correct for its effect, but the effectiveness of
the procedure still needs to be tested on real veiled spectra, many of which
are found among, e.g., Chamaeleon~I member stars in the
Gaia-ESO {\rm GESiDR1Final} data release.
We should also note that in clusters immersed in an H$_{II}$ region, the
nebular (narrow) \ha\ emission often masks completely any narrow
(chromospheric) emission from the cluster stars. This probably prevents
the use of the $\alpha_c$ index as a tool for selection of
chromospherically active stars.
Actual tests on such type of clusters are needed to effectively
understand the useability of $\alpha_c$ under these circumstances.

One extremely interesting application of the present work is the study
of non-accreting stars in PMS clusters,
since the locus occupied by the PMS band in the
$(\gamma,\tau)$ diagram is expected to {\rm change as a function of age},
especially at the youngest ages. The $(\gamma,\tau)$ diagram might
therefore become a very useful, distance-independent {\rm age indicator},
complementing those already existing such as the Li abundance.
The comparison between the \gam\ and Cha~I $(\gamma,\tau)$ diagrams,
shown in Sect.~\ref{clusters}, agrees with their expected age dependence.

Observations of older clusters,
instead, are important to define as accurately as possible the MS locus
in the $(\gamma,\tau)$ diagram, without any possible contribution from PMS
stars. This would greatly help to define the region in the diagram where
MS stars are {\rm not} to be found, and only very young/PMS stars may be
found.

We have repeatedly noted the presence of outliers in several of our
index-index diagrams for the \gam\ dataset.
Their positions in the respective diagrams are not due to
statistical errors in the indices used (maximum errors are
carefully reported for each index), but are instead directly related to
some spectral peculiarities with respect to the bulk of other stars
(either cluster or field stars).
By comparison with index values for stars in the UVES-POP dataset, it is
likely that they are chemically peculiar stars, supergiants,
and carbon (or carbon-enhanced) stars. 
In this respect, spectral peculiarities involving an excess abundance of
barium (barium stars, but also several types of carbon-enhanced stars)
are easily spotted by our index $\beta$, sampling the
Ba~II $\lambda$6496.897 line.
A study of the performances of our indices based on a much more varied sample
of peculiar stars from, e.g., the Tomasella \e (2010) spectral atlas is being
made, and will be presented in a future work.

\section{Summary}
\label{concl}

We have studied a large set of stellar spectra in the \gam\ cluster
region, all taken within the Gaia-ESO Survey with VLT/Giraffe setup HR15n.
Our immediate aim was to derive basic parameters (\teff, \logg,
veiling, CTTS indicators) for low-mass cluster stars; at the same time,
we tried to address the more general issue of finding useful spectroscopic
indicators of fundamental parameters for stars observed with
the same resolution and wavelength range.

To this aim, we have defined a number of spectral indices, with the following
characteristics:
\begin{enumerate}
\item {\rm Molecular indices:}
seven indices $\mu_1-\mu_7$, sampling the most important TiO molecular-band
jumps in HR15n, plus an average TiO index $\mu$ derived from them, which
is a good \teff\ indicator for late-type stars.
\item {\rm \ha\ indices:}
two indices, $\alpha_w$ and $\alpha_c$, sampling respectively the \ha\
wings and core, either in absorption (for early-type \teff\
determination) or in emission as CTTS/accretion indicator ($\alpha_w$),
or chromospheric indicator ($\alpha_c$).
\item {\rm 6490-6500~\AA\ "quintet" indices:}
two indices, $\beta_t$ and $\beta_c$, sampling respectively the entire
group ("quintet") of strong lines near 6495\AA, and its "core"; a
difference index $\beta$ is then derived, highly sensitive to gravity
over a wide spectral-type range, while $\beta_t$ is a good \teff\
indicator for intermediate-type stars.
\item {\rm 6625-6635~\AA\ temperature index:}
an index $\zeta_1$ from the ratio of suitable Fe~I lines, useful as
additional \teff\ measure for intermediate-type stars.
\item {\rm 6760-6775~\AA\ gravity index:}
an index $\gamma_1$, sampling a set of lines strongly sensitive to
gravity for intermediate-type and late-type stars.
\end{enumerate}
Most notably, two {\rm global indices} were computed from the above indices:
index $\tau$, sensitive to temperature across the whole spectral-type
range spanned by our stars, and a gravity-sensitive index
$\gamma$. Using these indices we are able to place stars in a
$(\gamma,\tau)$ diagram, which enables us to {\rm discriminate rather clearly
MS from PMS stars (at the \gam\ cluster age)}, and even more strongly
from giant stars, through the intermediate- and low-mass range.
This diagram is a new, useful diagnostic tool when applied to young PMS low-mass
stars, both to assess membership, and as an {\rm empirical age indicator}.

We have initially calibrated our $\tau$ index on the basis of the widely used
KH95 calibrations and observed $V-I_c$ colors. For MS/PMS stars in the
\gam\ dataset, the best-fit
calibration yielded 1$\sigma$ errors on \teff\ on the order of 110~K for
early-type stars (hotter than $\sim$5500~K), 73~K for intermediate-type stars
(4300-5500~K), and 50~K for late-type stars (colder than 4300~K).

Moreover, comparison with the literature stellar parameters of
stars in the ELODIE dataset has enabled us to calibrate quantitatively the
temperature and gravity dependence of our indices in the general case,
and to introduce a new composite index $\zeta$, effective to measure
stellar metallicity $z$. Limitations posed by the stellar composition of
the ELODIE dataset remain for the low-mass stellar range, and the
problem of calibrating our gravity index in the late-K and M-star range was
therefore solved by using theoretical Siess \e (2000) isochrones for
\gam\ member stars.

Extensive tests using current Gaia-ESO Survey results from UVES
observations of a large number of stars spanning a wide range of
parameters, as well as using Gaia-ESO Survey benchmark stars, strengthen
the validity of our calibrations.
Typical 1$\sigma$ errors on \logg\ determinations were found to be
of 0.2~dex, and on metallicity $[Fe/H]$ of 0.08~dex, for solar-like
metallicities.
The consistency of our calibrations
was also checked against Gaia-ESO Survey observations of young
clusters with very good results: the rms metallicity scatter for members
is always found to be much less than the above 1$\sigma$ error, derived
using non-metallicity-homogeneous star samples as reference.  We find that our
method is able to provide gravity measurements for low-mass PMS stars,
of sufficient accuracy to infer gravity-based ages for PMS clusters (or
relative ages at least).

As a first test of the usefulness of our indices, we find that six stars
in the \gam\ dataset, showing PMS characteristics and falling much nearer
to the ZAMS in the
CMD than the bulk of cluster members, are also characterized by a higher
gravity than these latter according to our indices, supporting that
they are older PMS stars. This conclusion is also strengthened by their
systematically lower Li EW with
respect to the bulk of cluster member stars, well compatible with an older age.
We consider this a rather strong evidence of a previous generation of
stars, now dispersed in the field.

We have also discussed areas of possible improvement in the technique,
such as extension to higher-mass and
older clusters, and a better understanding of the effect of veiling in CTTS.

\begin{acknowledgements}
We wish to thank an anonymous referee for his/her helpful suggestions.
We acknowledge support from INAF and Ministero dell'Istruzione,
dell'Universit\`a e della Ricerca (MIUR) in the form of the grant
"Premiale VLT 2012".
The results presented here benefit from discussions held during the
Gaia-ESO workshops and conferences supported by the ESF (European
Science Foundation) through the GREAT Research Network Programme.
This work was also partly supported by the European Union FP7 program
through ERC grant number 320360, and by the Leverhulme Trust through
grant RPG-2012-541.
We also acknowledge financial support from "Programme National de Cosmologie
and Galaxies" (PNCG) of CNRS/INSU, France.
T.B.\ was funded by grant No.\ 621-2009-3911 from The Swedish Research Council.
This research has made use of the SIMBAD database,
operated at CDS, Strasbourg, France.
\end{acknowledgements}

\bibliographystyle{aa}

\newpage

\begin{table}
\caption{
Placeholder for CDS Table
\label{table2}}
\begin{tabular}{c}
\hline
"table3.dat" \\
\hline
\end{tabular}
\end{table}

\end{document}